\documentclass[epj,nopacs]{svjour}
\pdfoutput=1

\input{profpythia6-preamble}

\begin{document}

\title{Systematic event generator tuning for the LHC}
\author{Andy Buckley\inst{1} \and
  Hendrik Hoeth\inst{2} \and
  Heiko Lacker\inst{3} \and
  Holger Schulz\inst{3} \and
  Jan Eike von Seggern\inst{3}
}
\institute{
  Institute for Particle Physics Phenomenology, Durham University, UK \and
  Department of Theoretical Physics, Lund University, Sweden \and
  Physics Department, Berlin Humboldt University, Germany
}

\date{Received: date / Revised version: date}

\abstract{
  In this article we describe \professor, a new program for tuning model
  parameters of Monte Carlo event generators to experimental data by
  parameterising the per-bin generator response to parameter variations and
  numerically optimising the parameterised behaviour. Simulated experimental
  analysis data is obtained using the \rivet analysis toolkit. This paper
  presents the \professor procedure and implementation, illustrated with the
  application of the method to tunes of the \pythiasix event generator to data
  from the LEP/SLD and \tevatron{} experiments. These tunes are substantial
  improvements on existing standard choices, and are recommended as base tunes
  for LHC experiments, to be themselves systematically improved upon when early
  LHC data is available.
}

\maketitle

\section{Introduction}

It is an inevitable consequence of the physics approximations in Monte Carlo
event generators that there will be a number of relatively free parameters which
must be tweaked if the generator is to describe experimental data. Such
parameters may be found in most aspects of generator codes, from the
perturbative parton cascade to the non-perturbative hadronisation models, and
on the boundaries between such components. Since non-perturbative physics models are
by necessity deeply phenomenological, they typically account for the majority of
generator parameters: typical hadronisation models require parameters to
describe e.g.~the kinematic distribution of transverse momentum (\pT) in hadron
fragmentation, baryon/meson ratios, strangeness and $\{\eta, \eta'\}$
suppression, and distribution of orbital angular
momentum~\cite{Sjostrand:2006za,Corcella:2002jc,Bahr:2008tf,Gleisberg:2008ta}. The
result is a proliferation of parameters, of which between
\ofOrder{\text{10--30}} are of particular importance for collider physics
simulations.

Apart from rough arguments about their typical scale, these parameters are
freely-floating: they must be matched to experimental data for the generator to
perform well. Even parameters which appear fixed by experiment, such as
\LambdaQCD, should be treated in generator tuning as having some degree of
flexibility since the generator (unlike nature) can only apply them in a
fixed-order scheme with incomplete large log resummation. It is also important
that the experimental data to which parameters are tuned covers a wide range of
physics, to ensure that in fitting one distribution well, others do not suffer
unduly. Performing such a tune manually is slow, does not scale well, and cannot
be easily adapted to incorporate new results or generator models. In addition,
the results are always sub-optimal: a truly good tuning of a generator, which
can highlight deficiencies in the physics model as well as provide improved
simulations for experimentalists, requires a more systematic approach.

In this paper, we describe the \professor\footnote{Originally derived from the
  construction ``\textsc{pro}cedure \textsc{f}or \textsc{es}timating
  \textsc{s}ystematic err\textsc{or}s, but aesthetics compel us otherwise.}
tuning system, which eliminates the problems with manual
and brute-force tunings by parameterising a generator's response to parameter
shifts on a bin-by-bin basis, a technique introduced by \tasso and later
used by \aleph and \delphi \cite{Althoff:1984rf,Braunschweig:1988qm,Buskulic:1992hq,Barate:1996fi,Hamacher:1995df,Abreu:1996na}.
This parameterisation, unlike a brute-force method, is then amenable to numerical
minimisation within a timescale short enough to make explorations of tuning
criteria possible. Adding new data or generator models to the system is also
relatively simple. We then apply the \professor system to optimisations of the
\pythiasix event generator against $\eplus\eminus$ event shape and flavour
spectrum data from \lepi and SLD, and to minimum bias (MB) and underlying event (UE)
data from CDF. The resulting tunes (one for each of the two \pythiasix parton
shower/multiple parton interaction (MPI) models) are substantial improvements on
existing tunes, and demonstrate the \professor system as an important tool for
LHC event simulation both before data taking, and in response to early
measurements in the new energy regime.

The \professor system is based on simulated experimental analysis data, which is
conveniently provided by the \rivet analysis library\cite{Waugh:2006ip}. As
\professor and \rivet development have been closely linked, we first briefly
summarise \rivet; however, \professor is not limited to tuning on data from
\rivet --- any source of comparable histogram data is a valid input.

\section{\rivet}

The \rivet library is a successor to the \hera-oriented \hztool generator
analysis library~\cite{Bromley:1995np}. Like its predecessor, \rivet is both a
library of experimental analyses and of tools for calculating physical
observables from an event record. It is written in object-oriented \Cpp, and in
particular emphasises the separation of analysis from generator steering: the
analyses are performed on \hepmc~\cite{Dobbs:2001ck} event record objects with
no knowledge of or ability to influence the generator behaviour. The reference
data files for the $\mathcal{O}(40)$ included analyses are bundled with the
package and used to synchronise the Monte Carlo simulation (MC) and reference
data binnings. For efficiency, observable calculators cache their results
between analyses for each event, ensuring that there are no redundant expensive
computations. Standard tool libraries from inside and outside high energy
physics are used where prudent; for example, almost all jet algorithms are used
via the \fastjet~\cite{Cacciari:2006sm} package.

For the purposes of the tuning and validation studies presented here, we used
the \agile interface library to pass parameters to generators (primarily
\pythiasix, as will be seen) and to populate \hepmc events from the
\textsc{HepEvt} common block. \agile provides programmatic and command-line
interfaces to several generators, including the \fortran
\herwigsix~\cite{Corcella:2002jc} and \pythiasix~\cite{Sjostrand:2006za}
shower/hadro\-ni\-sa\-tion codes, optionally combined with the \alpgen multi-jet
merging generator~\cite{Mangano:2002ea}, the \charybdis black hole
generator~\cite{Harris:2003db}, and the \jimmy hard underlying event
generator~\cite{Butterworth:1996zw} for \herwig.

\section{Tuning methods}
While \rivet provides a system for comparing a given generator tuning to a wide
range of experimental data, it has no intrinsic mechanism for improving the
quality of that tune. Historically, the main methods of generator tuning have
been the purely manual ``by eye'' method, and a brute-force scan of the
parameter space.

\xparagraph{Manual tunes} Tuning any complex system by eye is evidently
non-optimal, and would barely be worth mentioning were it not the most widely
used method until now! Manual methods require significant insight into the
algorithmic response to parameter choices for even semi-reasonable results, and
are intrinsically slow since the procedure typically involves a lot of
iterations of parameter choices --- even with unhappily low statistics, the
turn-around time of a set of runs is a day or more. The scaling is also poor:
few humans can cope with manual optimisation of more than five or so parameters,
guided by a similar number of comparison plots. The responsiveness to new data
or models is similarly deficient, since tuning a different generator essentially
involves starting from scratch and, having done a tune once, few people are
enthusiastic to repeat the exercise! The prevalence of manual tunings, despite
their myriad shortcomings, is a major motivator for the development of
\professor.

\xparagraph{Brute force tunes} This label includes any direct approach which
involves running generators very many times. Na\"ively, one can think about
dividing a parameter space up into a grid and then sampling on the grid line
intersections. It will be readily seen that such an approach does not scale: a
comprehensive scan of 5 parameters, with 10 divisions in each parameter will
require 100,000 generator runs, each perhaps making 10~million events --- even
then, the sampling granularity will be insufficient for meaningful
results. Randomly sampling the space, looking for serendipitous best-\chisq
values has more merit, but is similarly bedevilled by scaling problems and a
lack of satisfying ways to either systematically improve the ``best'' point, or
to know whether the minimum that was stumbled into is local or global.

Finally, the approach of putting a generator code into a Markov Chain Monte
Carlo (MCMC) optimiser such as \minuit may be summarily dismissed. While the
approaches above have the benefit of being parallelisable, MCMC is an
intrinsically serial method: one must wait for the $n$th ``function'' evaluation
to decide where the $(n+1)$th will be. Since generator runs take days, and even
the burn-in periods of MCMC samplers may require thousands of samples, this
approach is clearly unrealistic.

\xparagraph{Parameterisation-based tunes}
The final approach, which has a lengthy
history~\cite{Althoff:1984rf,Braunschweig:1988qm,Buskulic:1992hq,Barate:1996fi,Hamacher:1995df,Abreu:1996na},
is to parameterise the generator
behaviour. Since the fit function itself is expected to be complicated and not
readily parameterisable, there is a layer of indirection: the polynomial is
actually fitted to the generator response, $\text{MC}_b$, of each observable
bin $b$ to the changes in the $P$-element parameter vector $\p=(p_1,\dots ,p_P)$.

Having determined, via means yet undetailed, a good parameterisation of the
generator response to the steering parameters for each observable bin, it
remains to construct a goodness of fit (\gof) function and minimise it. The
result is a predicted parameter vector, $\p_\text{tune}$, which should (modulo
checks of the technique's robustness) closely resemble the best description of
the tune data that the generator can provide.

In parameterisation-based tuning, the run time is dominated by the time taken to
run the generator and generate the reference data points.  Assuming that
sufficient CPU is available to run several hundred MC jobs in parallel, this is
at most a few days; the time taken to convert this to a predicted set of best
parameters is a few minutes (and can again be parallelised for different
configurations as a safety check.) Presuming the details elided above to be tractable,
this technique offers the possibility of systematic tuning on a timescale
compatible with rapid and exploratory re-tunings, ideal for responding to early
LHC measurements.

Parameterisation-based optimisation is the approach taken by the \professor
system. The following sections document the details of the \professor method and
implementation, and tests of its robustness.

\section{The \professor method}

To summarise, the rough formalism of systematic generator tuning is to define a
goodness of fit (\gof) function between the generated and reference data, and
then to minimise that function. The intrinsic problem is that the true fit
function will certainly not be analytic and any iterative approach to
minimisation will be doomed by the expense of evaluating the fit function at a
new parameter-space point. What we require is an optimisation method designed
for very computationally expensive functions whose form is not known \emph{a
  priori}. Parameterisation-based optimisation meets these criteria by using
numerical methods to mimic the behaviour of an expensive function by using
inexpensive ones, and by being amenable to parallelisation in the critical
stages. The details to be described in this section are: the choice of general
parameterisation function, the method for fitting the general function to the
specific response of a MC event generator, the goodness of fit function to be
used, and the method of maximising its quality.

\subsection{The parameterised response function}

As already mentioned, the function to be parameterised is not the overall
goodness of fit function between the simulation and the reference data, but the
large set of observable bin values for every bin, $b$, in every
distribution. Accordingly, the output of the first stage of \professor is a set
of functions $f^{(b)}(\p)$, which model the true MC response, $\text{MC}_b$, of each
observable bin to changes in the $P$-element parameter vector, \p.

This ensemble of parameterisations is useful in two ways: first (and most
importantly), it provides safety against deviations from the form of the
parameterising function, since such deviations are not likely to be correlated
between a majority of the bins in normal regions of parameter space. This
incoherence of failure to describe the bin-wise generator response ensures that
the aggregated measure of generator modelling is faithful to the true
behaviour. Second, by breaking the problem down to a fine-grained level, it is
possible to select particular regions of distributions as more interesting than
the rest --- say, the peak of the \PZ \pT spectrum or the thrust distribution,
which are particularly sensitive to QCD modelling.

To account for lowest-order parameter correlations, a polynomial of at least
second-order is used as the basis for bin parameterisation:
\begin{align}
  \label{eq:poly}
  \text{MC}_b(\p) 
  \approx f^{(b)}(\p)
  = \alpha^{(b)}_0 + \sum_i \beta^{(b)}_i \, p^\prime_i 
  + \sum_{i \le j} \gamma^{(b)}_{ij} \, p^\prime_i \, p^\prime_j
  ,
\end{align}
where the shifted parameter vector $\pprime \equiv \p - \pzero$.

The number of parameters and the order of the polynomial determine the number of
coefficients to be determined. For a second order polynomial in $P$ parameters,
the number of coefficients is 
\begin{align}
  \Nn{2}{P} = 1 + P + P(P+1)/2,
\end{align}
since only the
independent components of the matrix term are to be counted. For a general
polynomial of order $n$, the number of coefficients is
\begin{align}
  \Nn{n}{P} = 1 + \sum_{i=1}^{n} \, \frac{1}{i\,!} \, \prod_{j=0}^{i-1} (P+j).
\end{align}
How the number of parameters scales with $P$ for 2nd and 3rd order polynomials
is tabulated in \TabRef{tab:ncoeffs}.

\begin{table}[t]
  \centering
  \begin{tabular}[t]{lll}
    \toprule
    Num params, $P$ & \Nn{2}{P} (2nd order) & \Nn{3}{P} (3rd order) \\
    \midrule
    1   & 3         & 4   \\
    2   & 6         & 10  \\
    4   & 15        & 35  \\
    6   & 28        & 84  \\
    8   & 45        & 165 \\
    10  & 66        & 286 \\
    \bottomrule
  \end{tabular}
  \caption{Scaling of number of polynomial coefficients \Nn{n}{P} with dimensionality 
    (number of parameters) $P$, for polynomials of second order ($n=2$) and third order ($n=3$).}
  \label{tab:ncoeffs}
\end{table}

A useful feature of using a polynomial for the fit function, other than its
general-purpose robustness, is that the actual choice of \pzero is irrelevant: a
shift in the reference point simply redefines the $\{\alpha,\beta,\gamma\}$
coefficients, but the function remains the same. Hence we are free to choose a
numerically stable value within each parameter's chosen range without loss of
generality: we use the centre of the hypercube $[\p_{\text{min}},
\p_{\text{max}}]$, as will be defined in the next section.

\subsection{Fitting the response function}
\label{sec:parameterisation}

Given a general polynomial, we must now determine the coefficients
$\alpha,\beta,\gamma$ for each bin so as to best mimic the true generator
behaviour. This could be done by a Monte Carlo numerical minimisation method,
but there would be a danger of finding sub-optimal local minima, and
automatically determining convergence is a potential source of
problems. Fortunately, this problem can be cast in such a way that an efficient
and deterministic method can be applied.

One deterministic way to determine the polynomial coefficients would be to run the generator
at as many parameter points, $N$, as there are coefficients to be determined. A
square $N \times N$ matrix can then be constructed, mapping the appropriate
combinations of parameters on to the coefficients to be determined; a normal
matrix inversion can then be used to solve the system of simultaneous equations
and thus determine the coefficients. Since there is no reason for the matrix to
be singular, this method will always give an ``exact'' fit of the polynomial to
the generator behaviour. However, this suggestion fails to acknowledge the true complexity of
the generator response: we have engineered the exact fit by restricting the
number of samples on which our interpolation is based, and it is safe to assume
that taking a larger number of samples would show deviations from what a
polynomial can describe, both because of intrinsic complexity in the true
response function and because of the statistical sampling error that comes from
running the generator for a finite number of events. What we would like is to
find a set of coefficients (for each bin) which average out these effects and
are a least-squares best fit to the oversampled generator points. As it happens,
there is a generalisation of matrix inversion to non-square matrices ---~the
\emph{pseudoinverse} \cite{nla.cat-vn441566}~--- with exactly this property.

As in our matrix inversion example, the set of ``anchor'' points for each bin are determined
by randomly sampling the generator from $N$ parameter space points in a
$P$-dimensional parameter hypercube $[\,\p_{\text{min}},
\p_{\text{max}}]$ defined by the user. This definition requires physics
input --- each parameter $p_i$ should have its upper and lower sampling
limits $p_{\text{min,max}}$ chosen so as to encompass all reasonable
values. In practice, we find that generosity in this definition is sensible, as
\professor may suggest tunes which lie outside conservatively chosen
ranges, forcing a repeat of the procedure. On the other hand, the
parameter range should not be too large, to keep the volume of
the parameter space small and to make sure that the parabolic
approximation gives a good fit to the true Monte Carlo response. Each
sampled point may actually consist of many (parallel) generator runs, which are
then merged into a single collection of simulation histograms. The
simultaneous equations solution described above is possible if the
number of sampled points is the same as the number of coefficients
between the $P$ parameters, i.e.  $N = \Nmin{P} = \Nn{n}{P}$. The more
robust pseudoinverse method applies when $N > \Nmin{P}$: we prefer to
oversample by at least a factor of 2.

The numerical implementation of the pseudoinverse uses a standard
singular value decomposition (SVD) \cite{1480176}. First, the polynomial
is cast into the form of a scalar product,
\begin{align}
  \text{MC}_b(\p) 
  \approx f^{(b)}(\p)
  = \sum_{i=1}^{\Nmin{P}} c^{(b)}_i \, \ptilde_i ,
\end{align}
where the $c^{(b)}_i$ coefficients are the independent components of
$\alpha^{(b)}_0$, $\beta^{(b)}_i$, and $\gamma^{(b)}_{ij}$ in
\EqRef{eq:poly}, and \vptilde is an \emph{extended parameter vector}
containing all the corresponding combinations of the parameter vector
components. Given sets of sampled points $\{\p\}$ and generator values
$\{\text{MC}^{(b)}\}$, the above implies the matrix equation,
\begin{align}
  \label{eq:matrixfwd}
  \vvalb = \Ptilde \, \vcoeffb \ ,
\end{align}
where \vvalb contains the generated bin values at the sample points, and the
rows of \Ptilde are composed of extended parameter vectors like \vptilde.  

For a two parameter case with parameters $x$ and $y$, the above may be
explicitly written as
\newcommand{\columnfill}{%
  \begin{pmatrix}
    \alpha_0 \\ \beta_x \\ \beta_y \\ \gamma_{xx} \\ \gamma_{xy} \\ \gamma_{yy}
  \end{pmatrix}
}
\begin{align}
  \underbrace{
    \vphantom{\columnfill}
    \begin{pmatrix}
      v_1 \\ v_2 \\ \vdots \\ v_N
    \end{pmatrix}
    }_{\vval\text{ (values)}}
  =
  \underbrace{
    \vphantom{\columnfill}
    \begin{pmatrix}
      1 & x_1 & y_1 & x^2_1 & x_1y_1 & y^2_1 \\
      1 & x_2 & y_2 & x^2_2 & x_2y_2 & y^2_2 \\
      &     &     &  \vdots   &   &          \\
      1 & x_N & y_N & x^2_N & x_Ny_N & y^2_N
    \end{pmatrix}
    }_{\Ptilde\text{ (sampled parameter sets)}}
  \underbrace{
    \begin{pmatrix}
      \alpha_0 \\ \beta_x \\ \beta_y \\ \gamma_{xx} \\ \gamma_{xy} \\ \gamma_{yy}
    \end{pmatrix}
  }_{\vcoeff\text{ (coeffs)}}
\end{align}
where the numerical subscripts indicate the $N$ generator runs. Note that
the columns of \Ptilde include all $\Nmin{2} = 6$ combinations of
parameters in the polynomial, and that \Ptilde is square (i.e.
minimally pseudoinvertible) when $N = \Nmin{P}$.

The \coeffb{i} can then be determined by pseudoinversion of~\Ptilde,
\begin{align}
  \label{eq:matrixbackwd}
  \vcoeffb = \Ipseudo[\Ptilde] \, \vvalb ,
\end{align}
where \Ipseudo is the pseudoinverse operator.

Except for demanding more sample points than can be computed in reasonable time
on the available batching facilities, the order of the polynomial has no
influence on the functioning of the parameterisation. Hence the method may be
extended in accuracy of the fitting function as required. In practice, a 2nd
order polynomial suffices for almost every MC generator distribution studied to
date, i.e. there is no correlated failure of the fitted description across a
majority of bins in the vicinity of best generator behaviour.

It is worth justifying a little more our apparent obsession with polynomial
parameterisations, other than their general ubiquity and robustness. The key
point is that a translational free parameter like \pzero is hard to express as
part of the linear combination of parameters required by this inversion method;
the invariance of the fitted polynomial function under shifts in the reference
point is one simple way to neatly avoid this problem.

\subsection{Goodness of fit function}

We choose a heuristic \chisq function, but other goodness of fit (\gof) measures
could certainly be used instead. Since the relative importance of various
distributions in the observable set is a subjective thing --- for example, given
20 event shape distributions and one charged multiplicity, it would certainly be
sensible to weight up the multiplicity by a factor of at least 10 or so to
maintain its relevance to the \gof measure --- we include per-observable
weights, \wO for each observable $\mathcal{O}$, in our \chisq definition:
\begin{align}
  \chisq(\p) = 
  \sum_{\mathcal{O}} \wO \sum_{b \, \in \, \mathcal{O}} 
  \frac{ (f^{(b)}(\p) - \mathcal{R}_b)^2 }{ \Delta^2_b },
  \label{eq:chi2}
\end{align}
where $\mathcal{R}_b$ is the reference value for bin $b$, and the error
$\Delta_b$ is the total uncertainty of the reference for bin $b$.  In practice
we attempt to generate sufficient events at each sampled parameter point that
the statistical MC error is much smaller than the reference error for all
bins. For future tuning studies on \Cpp MC generators, we will include a
``theory'' error corresponding to the degree of disbelief ($\approx 10\%$) that
the generator authors feel is appropriate to avoid any single observable biasing
the \gof heavily (overtuning). In computing the number of degrees of freedom,
the weights again enter:
\begin{align}
  N_{\text{df}} = \sum_{\mathcal{O}} w_{\mathcal{O}} \, |\{ b \in \mathcal{O} \}|.
  \label{eq:ndf}
\end{align}

It should be noted that there is unavoidable subjectivity in the choice of these
weights, and a choice of equal weights is no more sensible than a choice of
uniform priors in a Bayesian analysis. Physics input is necessary in both
choosing the admixture of observable weights according to the criteria of the
generator audience --- for example, a \Pbottom-physics experiment may prioritise
distributions that a general-purpose detector collaboration would have little
interest in --- and to ensure that the end result is not overly sensitive to the
choice of weights.

\subsection{Maximising the total \gof}

The final stage of our procedure is to minimise the parameterised \chisq
function. It is tempting to think that there is scope for an analytic global
minimisation at this order of polynomial, but not enough Hessian matrix elements
may be calculated to constrain all the parameters, and hence we must finally
resort to a numerical minimisation. This is the numerically weakest point in the
method, since the weighted quadratic sum of hundreds of polynomials is a very
complex function and there is scope for getting stuck in a non-global
minimum. Hence the choice of minimiser is important.

The output from the minimisation is a vector of parameter values which, if the
parameterisation and minimisation stages are faithful, should be the optimal
tune according to the (subjective) criterion defined by the choice of observable
weights.

\subsection{Final checks}

\begin{figure}[t]
  \centering
  \includegraphics[width=0.5\textwidth]{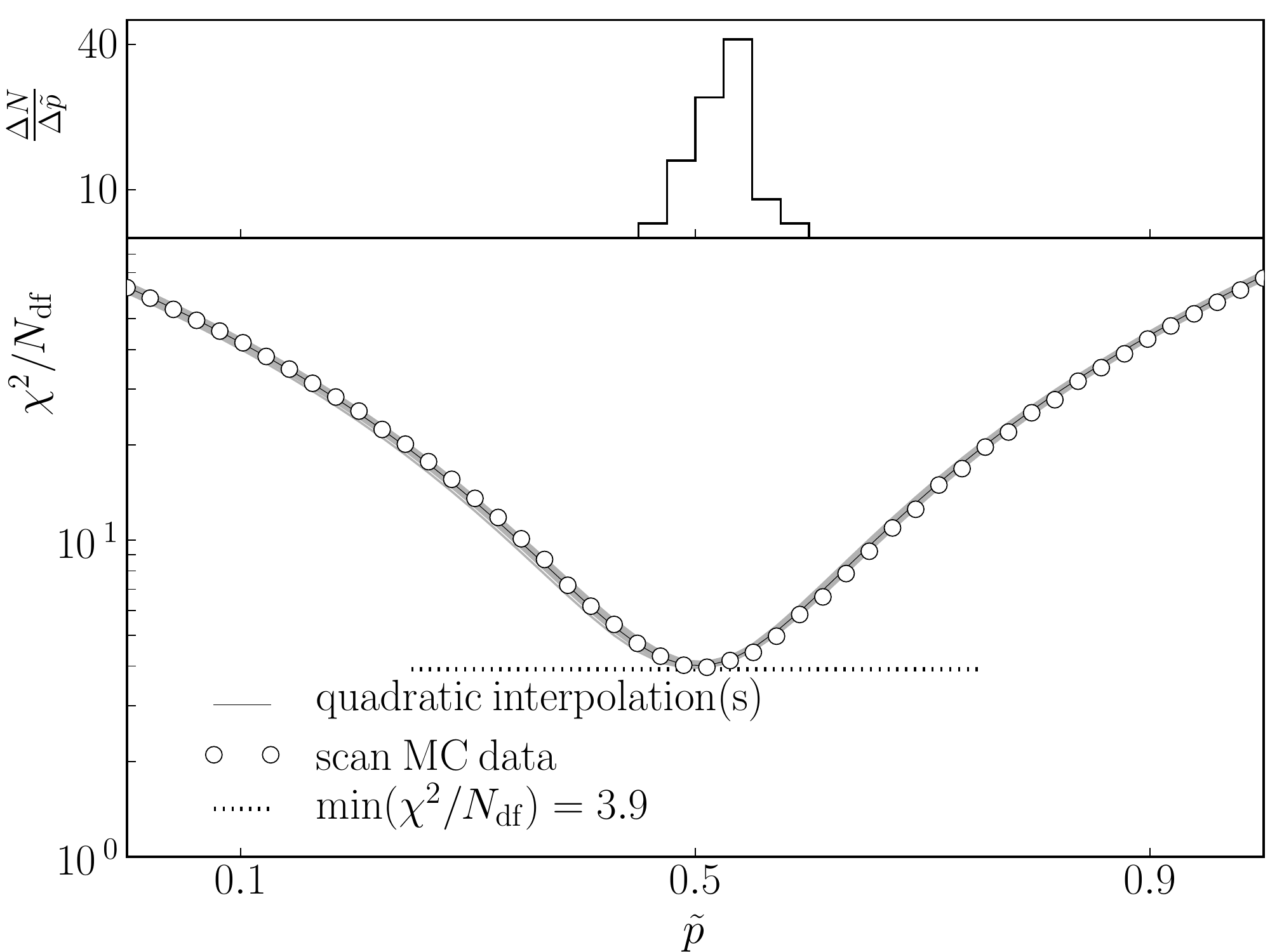}
  \caption{%
    The results of an example line-scan in \chisqNdf through nine-dimensional
    (flavour-) parameter space, demonstrating the agreement between
    \professor{}'s predicted values (lines, interpolation versus data) and the
    true values (white dots, MC-scan versus data).  Shown are 100
    interpolations that use 299 out of 399 available runs (grey band) and the
    interpolation that uses all available runs which fits in perfectly (black
    line).  The histogram displays the distribution of minimisation results
    derived from these interpolations.  The scan line is chosen such that it
    pierces a hypercube that is symmetric around one of the predicted minima,
    such that in the linear line-scan parameter $\tilde{p}$, the predicted
    minimum is at $\tilde{p}=0.5$, and the hypercube boundaries are at
    $\tilde{p} = \{0,1\}$.  Its volume is about $0.03\%$ of the total volume of
    the parameter sampling hypercube, illustrating the futility of attacking
    this optimisation problem via a grid scan of the space.%
  }
  \label{fig:profscan}
\end{figure}

On obtaining a predicted best tune point from \professor, it is prudent to check
the result by running the generator again at the predicted tune: this can be
done directly with \rivet. It is also useful to verify that the generator behaves in
the vicinity of the predicted point as predicted by the parameterisation by
scanning the generator along a line which passes through the ``best'' point
and comparing to the \professor prediction of how the \chisq will change.
This also enables explicit comparisons of default/alternative tunes to
\professor's predictions, by making the scan line intersect both points
and plotting the slice of the \gof function along the line. Such a line scan
can be seen in \FigRef{fig:profscan}.

A final important point of procedure remains: so far we have spoken entirely of
the procedure as a single set of runs entering the parameterisation
and minimisation procedure. However, this is rather dangerous: it may be that we
are picking an inappropriate set of runs, or that a subset of points are skewing
the fit and minimisation away from the true behaviour. Even if this is not the
case, the lack of any alternative to which we can compare means that we have
little knowledge of the procedure's systematic uncertainties. Hence, we have
also found it to be useful to oversample by a considerable fraction, $N \gg
\Nmin{P}$, and then to perform the parameterisation and \chisq minimisation
for a large number of distinct run-combinations, $N_{\text{tune}}$, where $\Nmin{P} \ll N_{\text{tune}}
\le N$. The set of different predicted responses and tunes from all the
different $N_{\text{tune}}$-run combinations provides a systematic control. It is important
that the tuning run combinations need a significant degree of independence from
one another if their indication of systematics is to be believed,
i.e. $N_{\text{tune}} \ll N$ for most of the tune run-combinations. In practice,
using $N_{\text{tune}} = N$ typically gives good results, but not necessarily the best
possible.

\section{Implementation}

In this section we describe the implementation of the \professor method as a
class library and a set of programs which use it. The majority of the code
is written in Python, but makes use of the \numpy numerical
library~\cite{numpy} and the \scipy scientific library~\cite{scipy}, in
which most of the core functions are implemented in CPU-efficient C code.
Additionally, the \scipy ``weave'' functionality can be used to automatically generate C
code which is compiled and cached for later use automatically at run-time,
with a corresponding speed-up factor of $\sim 5$.  The \professor user interface
is a set of distinct Python scripts which produce tunes based on input data,
explore line scans in the parameter space, and so on. Many parts of the
process are designed to be parallelisable on a batch system, and a standard
templating system~\cite{cheetah} is used to build batch scripts for this
purpose.

\subsection{Data generation}

Having decided what parameters to tune to what observables, the random
parameter points need to be sampled. This is done using the
\kbd{prof-scanparams} script which only needs a file of parameter ranges and
the desired number of parameter points as input. We are using \python's uniform
random generator \kbd{random.uniform} to independently sample values in all
dimensions of a hypercube defined by the input file.  To deal with parameters 
whose effect is non-linear, it is recommended that the parameters themselves be 
invertibly transformed to a more linear form, rather than sampling the space 
non-uniformly: this can be easily added by modifying the script, but is not an 
intrinsic feature.

The default output format is a list of simple (name, value) pairs, suitable for
use with the AGILe generator interfaces, but more complex templating can also be
used for native use with generators, such as \herwigpp, which have a more complex
configuration system.

The number of runs must be at least \Nmin{P},\footnote{To clarify, $P$ here is the
  number of \emph{independent} parameters to be sampled, i.e. the dimensionality
  of the sample space. The constrained parameters do not contribute to this parameter
  count.} but we typically use several times this number, so that
the parameterised function is not artificially anchored to the sampled values
but may float away from them to exploit the least-squares property of the
pseudoinverse. This is important independently of the considerations about
considering many independent run combinations --- this higher-level bet hedging
is less than useful if all of the available combinations are intrinsically
untrustworthy.

It is worth noting that despite the scaling of \Nmin{P}, the volume of the
hypercube still scales exponentially with $P$ and that the number of samples had
better keep approximate pace with this scaling, especially in wide scans where
many different generator behaviour regimes may be encountered. The power law
scaling of the polynomial does not obviate the responsibility to ensure that the
fitting method sees a representative sample of the space to be fitted. It is
also wise to ensure that the sample ranges are chosen so as to include the
default tune, at least in the first phase of a tuning: hopefully this would
automatically be the case, since a first set of sampling ranges
should at least include all ``reasonable'' values of each parameter.

The job of running the generator and \rivet (and of merging the output
histograms from different kinematic regions, if required) is mainly left up to
the user. This is because different system configurations, the variety of
batching systems, the choice of contributing Rivet analyses, etc. effectively
mandate some user customisation of run scripts. Attempting to automate this
process would likely lead to disastrously algorithmic tuning efforts, but we
note that our choice of observables for \pythiasix, as described in
\SecRef{sec:py6}, is as good a template as any currently available.

For most of the \professor procedure, the analysis data is defined by a
directory structure containing a reference directory and a set of run
directories, each of which contains histogram files from \rivet. The same
structure may be conveniently used to store output from different tunes. It is
also possible for analysis programs other than \rivet to provide input for
\professor tuning, provided that their data format is in a format which can be
used by \professor, or can be converted to such a format. The currently
most-used data format is the ``AIDA'' XML format, as this is the main \rivet
output format. When, as planned, \rivet's data format is upgraded to use the
simpler ``YODA'' data files, \professor will also support this format. Yet more 
formats can also be supported in response to demand.

Loading of the data files is currently ``eager'', i.e. all data files are read
in and stored in memory during processing. For large data sets, e.g. $\sim 1000$
sampled parameter points with distributions amounting to $\sim 10^4$ bins per
point, this produces a lead time of $\sim 1$ minute on a typical workstation and
large memory occupancy. For larger input sets, where this lead time may be less
tolerable, a ``lazy'' loading and pro-active garbage collection on unused bins
will be explored.

\subsection{Tuning}

The main tuning stage is accessed via the \kbd{prof-tune} program. This performs
the combination of parameterisation and optimisation against reference data for
each of a set of MC run combinations, based on the runs found in the input
directory. The run combinations can either be uniquely and randomly generated at
run-time by \kbd{prof-tune}, or can be supplied via a plain text file in which
each line is a white-space separated list of run names. This latter method is
most useful for parallelising the tuning for a large number of run combinations.

\subsubsection*{Parameterisation and fitting}

\professor currently supports second- and third-order polynomials for
parameterisation --- as previously discussed, these are robust against
origin-translation in a way necessary for the pseudoinverse method to work, and
our experience is that a second-order polynomial has so far been sufficient for
almost all purposes in generator tuning.

For the numerical evaluation of the pseudoinverse procedure, \numpy's
implementation of the singular value decomposition is used.

\subsubsection*{\gof optimisation}

To have an intuitive way of excluding single bins from the \gof calculation,
the implemented \chisq function differs slightly from \EqRef{eq:chi2} in
that weights are not applied on a per-observable but on a per-bin scope. The
\Ndf definition is changed accordingly. By this, single bins can be left out
of the \chisq calculation by setting their respective weights to zero. We
use this to veto bins with zero error as this usually indicates that these
bins were not filled during the data generation.  After applying the weights
and vetoes all bins with zero weight are filtered out. From the resulting
bins the \Ndf is computed and a \chisq function $\chisq(\p)$ is constructed,
which is passed to the minimiser.

The optimisation of the heuristic \chisqNdf function is implemented using
minimisers from \scipy and also \pyminuit~\cite{pyminuit}, a Python
interface to the CERN Minuit package~\cite{James:1975dr}.  As Minuit uses a Markov chain method,
which copes with high dimensional problems better than the \scipy Nelder-Mead
simplex minimiser, and also offers error estimates and covariance calculations,
it is the preferred and default choice.  \professor is also able to apply
limits to each parameter in minimisation, to exclude unphysical results. The
limits used in such cases should not just be the sampling limits, unless
those were determined by physical restrictions, since a minimisation falling
outside the sample limits is actually a useful result which should not be
obscured.

By default the starting point for the minimiser is the centre of the
parameter space defined by our parameter sampling ranges. It is also
possible to specify a manual or random starting point. \minuit evaluates
the parameter uncertainties by calculating those parameter points at which
the \chisqNdf value exceeds that of the minimum by 1: for a truly
\chisq-distributed test statistic this should correspond to a ``$1\sigma$'' 
68\% confidence limit estimate.

A successful minimisation will write out its details to file, including the
optimal parameters and their correlations, a file of parameterised histograms
for each of the observables included in the fit (based on the parameterisation at
the tune point), and information about the number of parameters, optimal \gof
value(s), etc. These can then be studied and plotted as described in the next
section.

\subsubsection*{Tuning output and visualisation}

The result of the tuning stage, in the form of the \kbd{prof-tune} program, is
a file of tune points, including their \gof scores. If the tuning has been
parallelised, there will be several such files, which can be merged together
if desired. The tunes can be visualised either textually or graphically.

Graphical visualisation is particularly useful, and comes in several different forms:

\xparagraph{\gof vs. parameter value} Each tuning parameter produces a plot of
\gof vs. parameter value, with parameter sampling boundaries indicated. Run
combinations of different size are represented in different colours, and points
which lie outside \emph{any} of the sampling boundaries are indicated by lighter
shades of their point colour: this makes it easy to see how predicted tunes fit
into the high-dimensional sample space without having to perform feats of mental
gymnastics. Clearly, if a cluster of points falls outside one or more sampling
boundaries, their \gof values are less than trustworthy and a re-run of the
generator sampling, with expanded boundaries is recommended.

\xparagraph{\gof vs. weight combination} If several combinations of weights have
been tested, they can be graphically displayed side-by-side to verify that the
tune is robust against reasonable shifts of \gof definition.

\xparagraph{Correlation display} For each minimisation result we also store the
covariance matrix between parameters and values, as calculated by the
minimiser. The \professor{}-system provides the user with the possibility to
calculate the coefficients of correlation $\rho_{ij}$ for each pair of
parameters $(i,j)$ from the (symmetric and real) $P$-dimensional covariance matrix
$\mathcal{C}$:
\begin{align}
    \rho_{ij} = \frac{\mathcal{C}_{ij}}{\sqrt{\mathcal{C}_{ii}\mathcal{C}_{jj}}}
\end{align}
The resulting parameter-parameter correlations can be displayed as colour-map
plots or as tables such as in \TabRef{tab:correlations-flavour}.

\begin{figure}[t]
  \centering
  \includegraphics[width=0.5\textwidth]{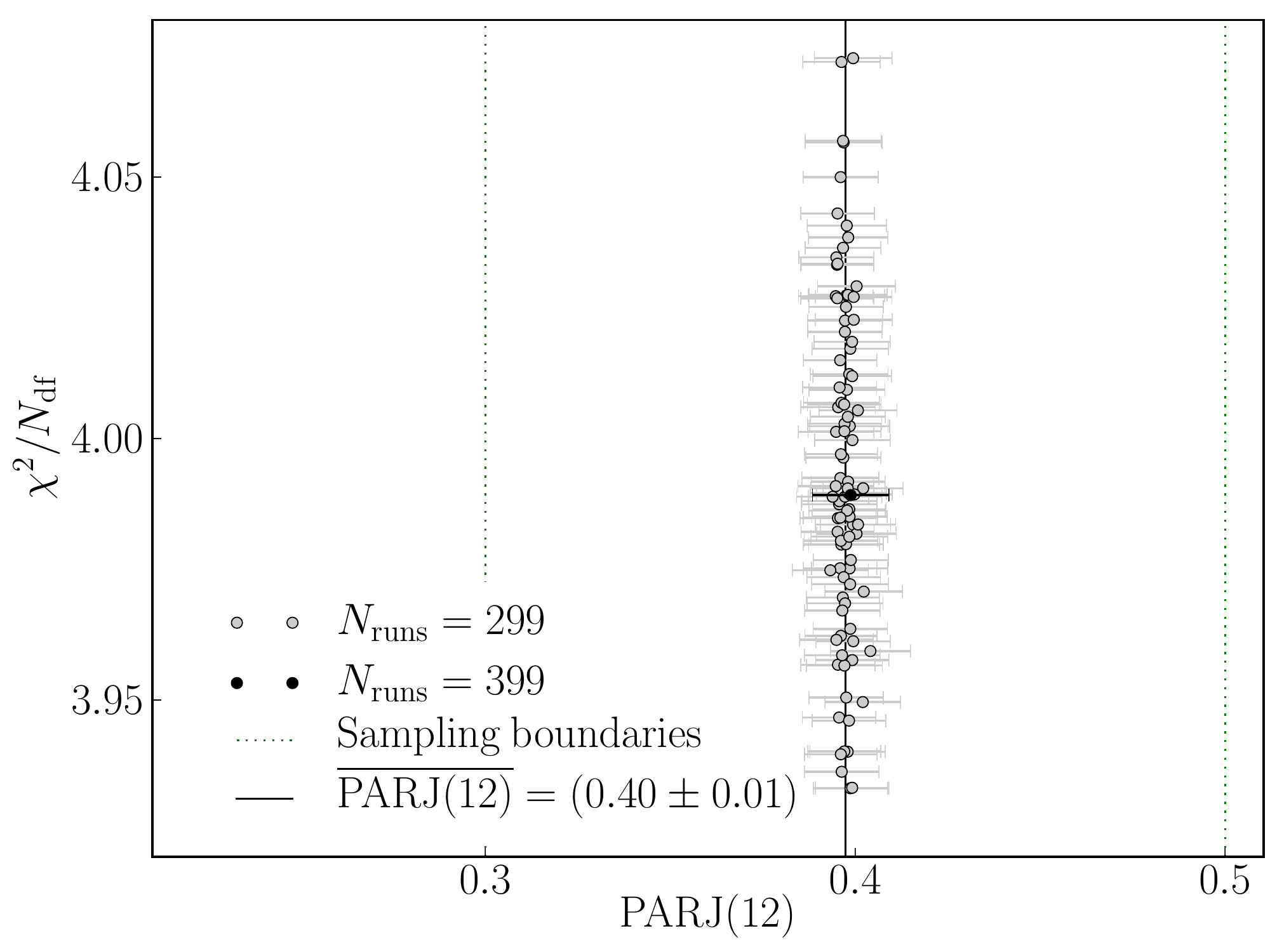}
  \caption{%
    Two dimensional \chisqNdf versus parameter value plot
    showing minimisation results derived from 100 interpolations that use 299
    out of 399 available runs and one result that uses all available runs for
    the probability that a strange meson has spin 1. The observable-weight
    combination used can be found in \TabRef{tab:obsweight-flavour}. 
    The  parameter errors are estimated by the minimiser by going up one unit in
    \chisq.%
  }
  \label{fig:minparj12}
\end{figure}

\xparagraph{Sensitivities}
It is desirable to tune to those observables most sensitive to parameter
changes. Clearly, if a parameter has no effect on an observable at all the
minimiser is very likely to yield useless results or it might even fail to
converge.  The \professor package offers the calculation of a bins sensitivity
to changes of the values of the parameters in question for tuning based on the
parameterisation:
\begin{align}
    S^{(b)}_i(\p) \approx \frac{f^{(b)}(\p + \vec{\varepsilon})-f^{(b)}(\p) }{f^{(b)}(\p)}\cdot\frac{\p}{\p + \vec{\varepsilon}}
\end{align}
where the $\vec{\varepsilon}_i$ are conveniently set to one percent of the interval
$[\,\p_i^{\text{min}}, \p_i^{\text{max}}]$.

A side-by-side comparison of an observables sensitivity to all parameters
included in the parameterisation is available as colour-map plots.  These plots
may be used to identify and remove those observables that have little or no
impact at all. They could also be seen as an \emph{a posteriori} justification for the
choice of observables included in a tuning.

\xparagraph{Interactive parameterisation explorer}%
The script \kbd{prof-I} can be used to interactively explore the effect of
shifts in parameter space on the shapes of observables. All parameter values can
be adjusted via sliders in a graphical user interface. The resulting shapes of
the observables, calculated from the parameterisations, are updated in real time
as the sliders are moved, and static data or MC histograms can also be shown for
comparison. As the goodness of fit is also displayed, \kbd{prof-I} can be also
used as an aid in manual tunings.

\xparagraph{Histogram prediction}%
\kbd{prof-tune} also helpfully produces a directory of histogram files, one for
each tuning, which makes it possible to see how each distribution is predicted
to behave at that point without running the generator and incurring the usual
(typically multi-day) delay. This is particularly useful when choosing how to
weight distributions to achieve the desired quality of fits --- a subjective
prioritising of particular physics which cannot be avoided and usually requires
some iteration.
\bigskip

These visualisations of tunes are extremely useful, not least for iterating the
choice of sample boundaries in the early stages of a tune. At this point, the
sampling boundaries must be wide enough to include the predicted tunes --- if
the prediction consistently falls outside the boundaries, it is probably
indicative of a problem with the generator physics model --- but also narrow
enough that the sampling is representative of the space. A too-wide initial scan
may be too coarse to yield stable results. Care should also be taken
when tightening boundaries for secondary tune stages, since in the case of
strong parameter--parameter correlations the optimum may unexpectedly appear
once again outside the boundaries due to the improved description of the
correlations. The main rule is that there \emph{are} no truly reliable rules,
and some iteration will certainly be required.

Once the boundaries have been chosen for a final stage of tuning, if there are
observed cases where the polynomial does not sufficiently describe the
generator, it is worth trying the third-order polynomial.

\subsection{Validation}
\label{sec:validation}

Before parameterising real MC-generated data, the parameterisation algorithm was
tested for robustness against the distribution of the anchor points and its
behaviour when dealing with data which cannot fit exactly to the parameterising
polynomial: such troublesome data was simulated by smeared sampling from
polynomials of various orders. This round of validation also sought to verify
that the \gof calculated from the parameterised bins resembles that obtained
directly from the MC-generated data, and is described briefly in this section.

\subsubsection*{Robustness of the parameterisation algorithm}

\begin{figure}[t]
    \begin{center}
      \includegraphics[width=0.48\linewidth]{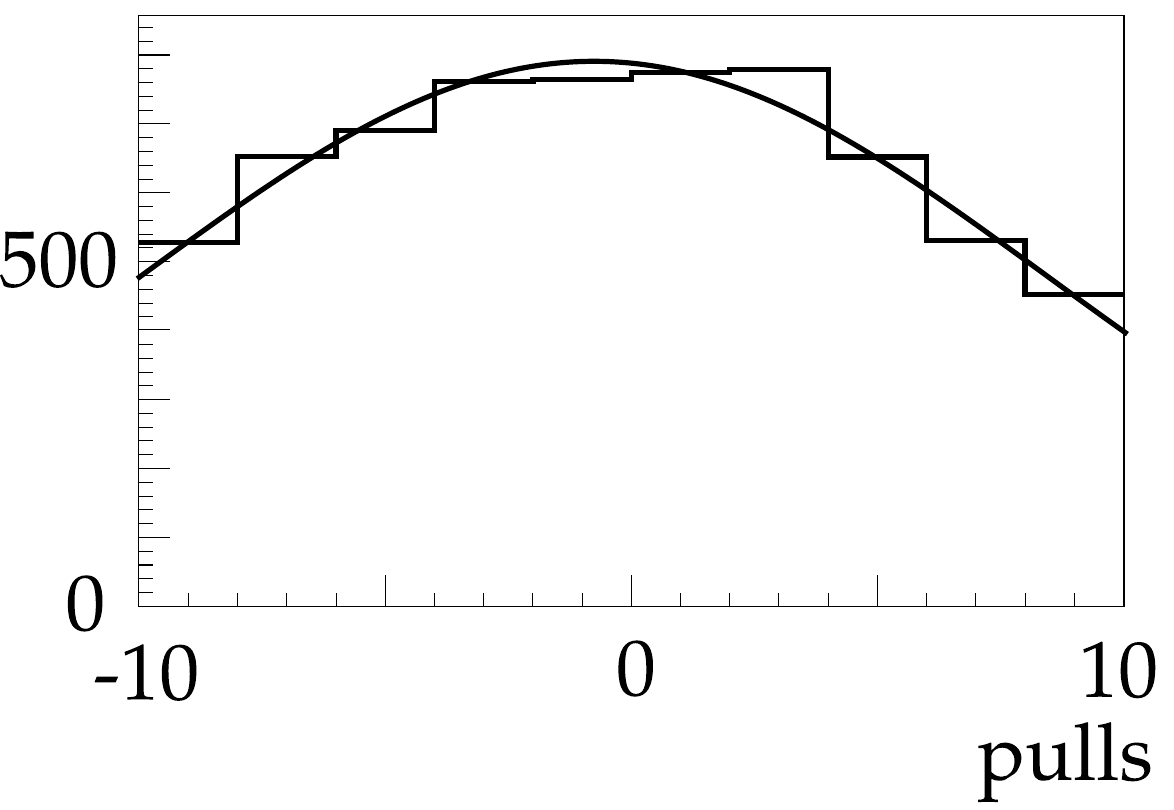}
      \includegraphics[width=0.50\linewidth]{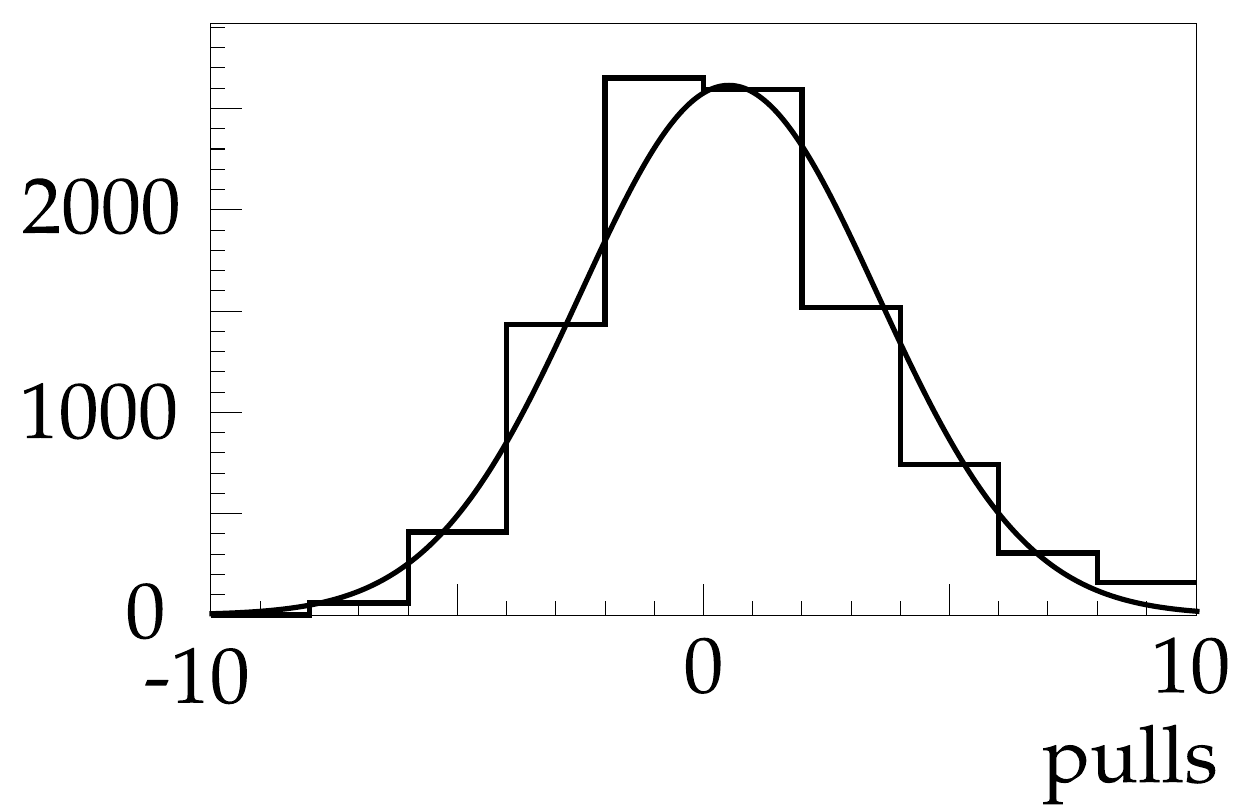}
    \end{center}
    \caption[Examples of pull distributions]{Example pull distributions:
    Parameterisations of data generated with a smeared fourth-order
    polynomial (see \EqRef{eq:skeweddistribution}) in 7 dimensions are
    compared to the unsmeared polynomial.  The parameterisations have been
    created (a) using the minimal number of anchor points $\Nmin{7} = 37$
    and (b) using $\Nmin{7} + 6 = 43$ anchor points. One can clearly observe
    that the pull distribution narrows when using additional sample points.
    }
    \label{fig:pulls}
\end{figure}

The \professor \gof function can be influenced by several observable weight
combinations, ${\wO}$, and also by the number of runs in each random run
combination used for the parameterisation. This offers possibilities to check
the predicted minima for systematic errors due to improper parameterisation or
overtuning to a specific set of observable weights.


In addition, the minimisation results obtained from quadratic
interpolations were compared to those obtained from cubic
interpolations. We did not find a significant difference between the
predictions, although the cubic interpolation describes the generator
response better in regions that are far away from the minimum.

%
The basic functioning of the polynomial parameterisation was tested with
input data generated with a second-order polynomial with random coefficients
--- the {\em known} coefficients were compared to those of the resulting
parameterisations. The robustness of parameterising error-smeared and data from
non-second-order distributions was also tested. Example input data were
generated using second- to fourth-order polynomials, especially polynomials
of the form
\begin{align}
  \label{eq:skeweddistribution}
    f(\p) = (\p-\vec{m}_1)^2(\p - \vec{m}_2)^2 +\vec{a} \cdot \p,
\end{align}
and were smeared using an Gaussian error. Then, the \emph{un}smeared original
polynomial and the parameterisation were evaluated at 10000 randomly located
points and a simple \chisqNdf and ``pull'' statistics were calculated as \gof measures,
where the pulls were calculated as follows:
\begin{align}
    p = \sum_{i=1}^{10000} \frac{f_\text{unsmeared}(\vec{x}_i) -
    f_\text{param}(\vec{x}_i)}{\sigma},
\end{align}
with the $\vec{x}_i$ being the test points, $f_\text{unsmeared}$ and
$f_\text{param}$ the unsmeared polynomial and parameterisation respectively,
and $\sigma$ the width of the Gaussian error distribution.
A Gaussian distribution was then fitted to the pull histogram. This procedure was
followed for several different dimensions of parameter space $P$ and different numbers
of sample points $N = \Nmin{P}, \Nmin{P} + 2, \ldots$. 

Using the minimal \Nmin{P} sample points resulted in wide variation of fit
quality, with observed \chisqNdf varying across several orders of magnitude, and
broad --- in the low dimensional case even biased --- pull distributions. Using
additional sample points reduced all this unwanted behaviour, e.g.~in the case
of a 7-dimensional parameter space and data generated from a fourth-order
polynomial, the average width of pull distribution fell from 7.9 for \Nmin{7}
samples to 3.2 for $\Nmin{7} + 6$ samples, and the corresponding ranges of
observed \chisqNdf fell from \ofOrder{10\text{--}10^3} to
\ofOrder{1\text{--}10}.  Consequently, we discourage use of parameterisations
based on the minimal number of sample points: the evidence is simply that they
do not provide reliable descriptions of the parameter space. Examples of pull
distributions are shown in \FigRef{fig:pulls}.

Finally, the influence of the distribution of the sample points in the parameter
hypercube on the parameterisation quality was tested. We performed 5000
parameterisations based on error-smeared data. \chisqNdf values were computed
as above, along with several different measures of the anchor point distributions
\begin{itemize}
    \item average and minimal Cartesian distance
    \item average and minimal distance of the projections on the parameter axis.
\end{itemize}
These were studied as 2D histograms. Analysing the low-dimensional cases revealed
a dependence of the \gof on the averaged distances for infrequent anchor
point samples, which again could be eliminated by oversampling. The more
relevant high-dimensional cases did not show this dependence; however
oversampling narrowed the range of observed \chisqNdf by several orders
of magnitude.  The parameter space dimensionality for these latter tests ranged
from $P=1$ to $10$, and the number of anchor points from $N = \Nmin{P}$ to
$\Nmin{P} + 10$.


\subsubsection*{Tune verification}

As mentioned, it is useful to visualise \professor tunes along lines in
parameter space, particularly lines which intersect both the predicted tune and
an alternative or default configuration. \professor provides a program,
\kbd{prof-scanchi2}, to perform this scan for saved parameterisations and/or to
generate generator/\rivet batch scripts from supplied templates. This enables
verification that the \gof really behaves as parameterised and that the chosen
point really is close to a \gof extremum.

To reduce the risk that a minimum returned by the numerical minimisation is a
local minimum, \kbd{prof-tune} can perform several minimisations with different
starting points.

A tighter tune, either \professor or grid-scan based, could be performed based
on the correspondence between the true and parameterised \gof in the tune
region, although in practice the deviations are small enough that we have not
risked overtuning by attempting to do so.

\subsubsection*{Tuning stability}
The \professor system offers two different ways to get an estimate of how
well a predicted minimum is defined. 

We can benefit from oversampling the parameter space with respect to \Nmin in
such a way that numerous run combinations\footnote{We usually do about 100
  minimisations. The run combinations are chosen in a randomised procedure, and
  it is explicitly checked that there are no duplicates.} may be chosen for
different parameterisations simply by omitting a fraction of all the available
Monte Carlo runs.  In order to reduce the correlation between run combinations
we usually choose this fraction to be about one-third. This is clearly a
compromise between the quality of the parameterisation and the degree of
correlation introduced by choosing several run combinations.

The outcome of all minimisations can be displayed parameter-wise such as in
\FigRef{fig:minparj12}. We observe that the minimisation result derived from all
available Monte Carlo runs always lies within the distribution of \chisqNdf
from random run combination tunes, illustrating that certain interpolations fit
the data better than others but that using all the information available always
gives a good average description.

Instead of varying the parameterisation it is also possible to influence the \gof
function. This can be done by independently applying a weight to each observable
included in the tuning. This more-or-less subjective approach is justified
by two facts. Firstly, we certainly do not expect the generator's response 
function to be a simple polynomial, and secondly we know that most models
should not be expected to be capable of describing all observables: for example, 
all generators fail to describe the $\pT^\text{out}$ distribution in $e^+e^-$ data.


\subsection{Performance and comments}

The focus in testing and commissioning the \professor system has until recently
been focused on \pythiasix tunes against \lep, \sld, and \tevatron data. Here we were able to
interpolate and minimise up to 10 parameters at a time for roughly 100
distributions, but beyond this the minimisation time became large and we were
less satisfied with the resulting minima. The latter effect probably represents the fact that
\Nmin{P} only specifies the minimum number of samples needed to algebraically
constrain a curve in $P$-space, but tells us nothing about the number of points
needed to adequately represent the space --- this depends on more complex things
like the rate of change of the function, the extent to which it oscillates, and
the degree of correlation between parameters: in general it will rise much
faster than the algebraic constraint. 

We advise that a maximum of 10 parameters be observed whenever possible, and
less than that is advisable. Since generators usually have many more parameters
than this, some approximate factorisations into semi-independent groups must be
found. As usual, this requires appreciation of the generator physics, and
ideally input from the generator authors.

\section{Complete tuning of \pythiasix{}}
\label{sec:py6}

For the first production tuning we chose the \pythiasix event
generator, as this is a well-known generator which has been tuned before
and which we expected to behave well. Additionally, the tuning of \pythiasix
was arguably more pressing than that of any other generator as it is used for 
the majority of LHC experiment MC simulation, and the newer parton shower/MPI
model had never before been tuned in detail. All results in this paper are
based on the version 6.418.

Our multi-stage approach to tuning was to fix the flavour composition and
the fragmentation parameters to the precision data from \lep{} and \sld{} before
continuing with the parameters related to hadron collisions, for which we use
data from the \tevatron{}.

\subsection{Parameter factorisation strategy}

In \pythia{}, the parameters for flavour composition decouple well from the
non-flavour hadronisation parameters such as the Lund string parameters $a$,
$b$, $\sigma_q$, and from the shower parameters (\alphaS{}, cut-off). Parameters
related to the underlying event and multiple parton interactions are decoupled
from the flavour and fragmentation parameters. In order to keep the number of
simultaneously tuned parameters small, we decided to follow a three-stage
strategy. In the first step the flavour parameters were optimised, keeping
almost everything else at its default values (including using the
virtuality-ordered shower). In the second step the non-flavour hadronisation and
shower parameters were tuned --- using the optimised flavour parameters obtained
in the first step. The final step was tuning the underlying event and multiple
parton interaction parameters to data from the two \tevatron{} experiments
\cdf{} and \dzero{}.

\subsection{Flavour parameter optimisation}

The observables used in the flavour tune were hadron multiplicities and
their ratios with respect to the \Ppiplus multiplicity measured at
\lepi and \sld~\cite{Amsler:2008zz}, as well as the \Pbottom-quark
fragmentation function measured by the \delphi{}
collaboration~\cite{delphi-2002}, and flavour-specific mean charged
multiplicities as measured by the \opal{}
collaboration~\cite{Ackerstaff:1998hz}. For this first production we
chose to use the Lund-Bowler fragmentation function
for \Pbottom-quarks (invoked in \pythiasix{} by setting
$\MSTJ{11}=5$) with a fixed value of $r_b = 0.8$
(\PARJ{47}), as first tests during the validation phase of the
\professor{} framework showed that this setting yields a better
agreement with data than the default common Lund-Bowler parameters
for \Pcharm{} and \Pbottom{} quarks.

\begin{figure}[t]
  \centering
  \includegraphics[width=0.5\textwidth]{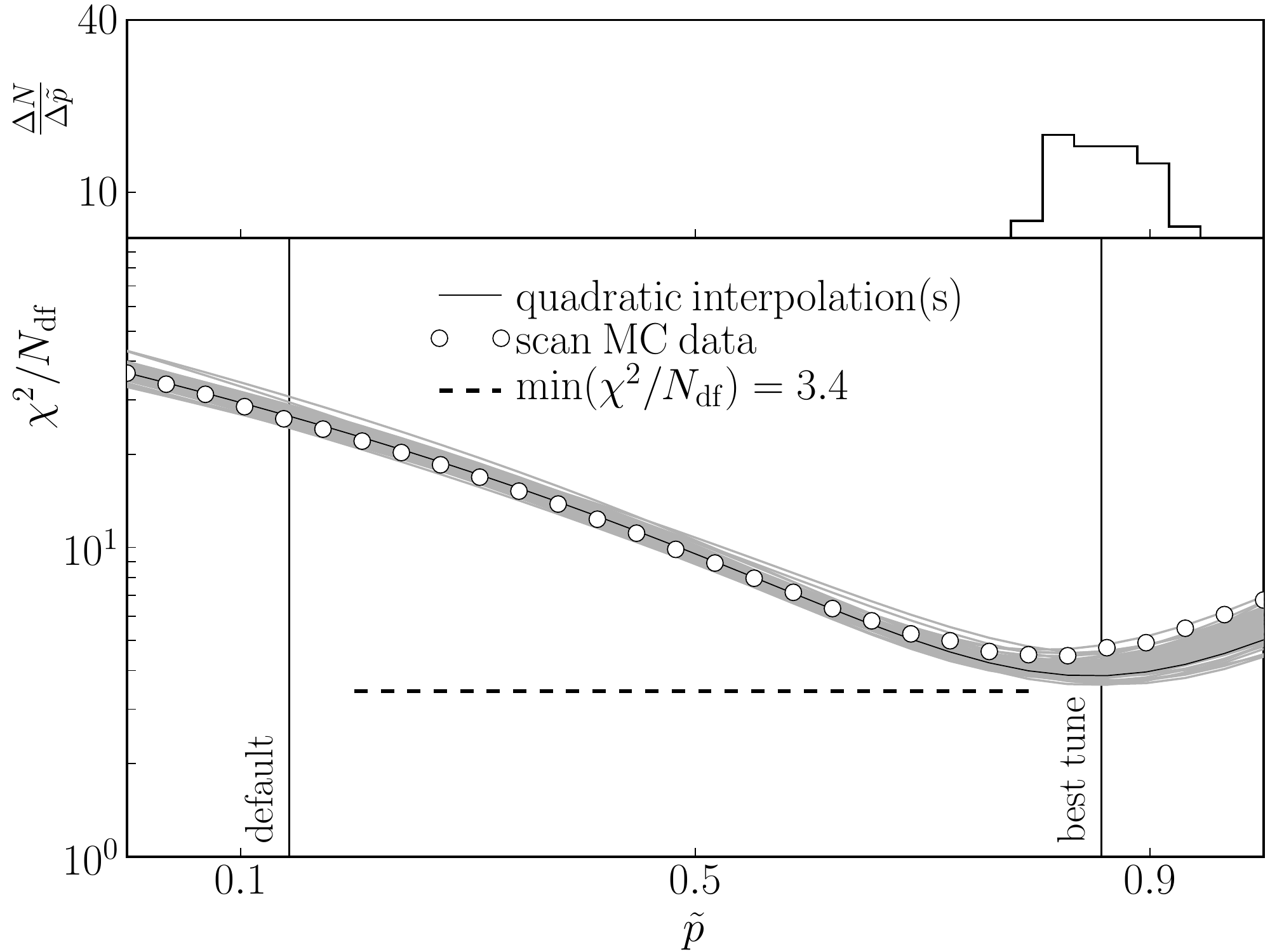}
  \caption{%
    Pythia 6 ($Q^2$ shower) \chisqNdf variation along a line in the 9D
    flavour-hyperspace, linearly parameterised by $\tilde{p} \in [0,1]$.  The
    line shown runs between the default and the Professor flavour tuning. The
    white dots are the true generator \chisqNdf values, and the grey lines an
    ensemble of parameterisations from the Professor procedure that use about
    two-thirds of the available MC-runs. The black line represents the
    interpolation calculated from all available runs. The histogram displays the
    distribution of minimisation results (if projected on the scan-line) derived
    from these interpolations.%
  }
  \label{fig:profscan_flavour}
\end{figure}

For the tuning we generated 500,000 events at each of 180~parameter points.  The
tuned parameters are the basic flavour parameters like diquark suppression,
strange suppression, or spin-1 meson rates. All parameters are listed in
\TabRef{tab:tune-flavour} together with the tuning results and a \chisqNdf
line-scan plot comparing the default with tuned parameter sets in
\FigRef{fig:profscan_flavour}. The physics observables and their weights for the
tuning are listed in \TabRef{tab:obsweight-flavour}.

Since the virtuality-ordered shower was used for tuning the flavour
parameters, we tested our results also with the \pT-ordered shower in
order to check if a separate tuning was necessary. Turning on the
\pT-ordered shower and setting $\Lambda_\text{QCD} = 0.23$ (the
recommended setting before our tuning effort) we obtained virtually the
same multiplicity ratios as with the virtuality-ordered shower. This
confirms the decoupling of the flavour and the fragmentation parameters
and no re-tuning of the flavour parameters with the \pT-ordered shower
is needed.

In \TabRef{tab:multiplicities}, we compare several measured mean hadron
multiplicities in $\eplus\eminus$ collisions at \unit{91}{\GeV} to \pythia{}
predictions with default settings and with our tune. In particular, the strange
sector is significantly improved, although there is a slight degradation for
charm and bottom mesons.

\subsection{Fragmentation optimisation}
\label{sec:fragmentationtune}

Based on the new flavour parameter settings, the non-flavour
hadronisation and shower parameters were tuned separately for the
virtuality-ordered and for the \pT-ordered shower. The observables
used in this step of the tuning were event shape variables, momentum
spectra, and the mean charged multiplicity measured by the \delphi{}
collaboration~\cite{Abreu:1996na}, momentum spectra and flavour-specific
mean charged multiplicities measured by the \opal{}
collaboration~\cite{Ackerstaff:1998hz}, and the \Pbottom-quark fragmentation
function measured by the \delphi collaboration~\cite{delphi-2002}.
All observables and their weights are listed in
\TabRef{tab:obsweight-frag}.

We tuned the same set of parameters for both shower types
(\TabRef{tab:tune-frag}). To turn on the \pT-ordered shower, \MSTJ{41}
was set to 12. In the case of the virtuality-ordered shower, this
parameter stayed at its default value. For both tunes, we generated
1~million events at each of 100~parameter points.

During the tuning of the \pT-ordered shower it transpired that the fit
prefers uncomfortably low values of the shower cut-off \PARJ{82}. Since
this value needs to be at least $2 \cdot \LambdaQCD$, and
preferably higher, it was manually fixed to 0.8 to keep the parameters
in a physically meaningful regime. Then the fit was repeated with the
remaining five parameters.

The second issue we encountered with the \pT-ordered shower was that
the polynomial parameterisation $f^{(b)}$ for the mean charged
multiplicity differed from the real Monte Carlo response by about 0.2
particles. This discrepancy was accounted for during the \chisq
minimisation, so that the final result does not suffer from a bias in
this observable.

\FigRef{fig:tune-frag-q2} shows some comparison plots between the \pythia{}
default and our new tune of the virtuality-ordered shower.  Even though this
shower has been around for many years, and \pythia{} has been tuned before in
this mode, there was still clearly room for improvement in the default settings.

\FigRef{fig:tune-frag-pt} shows comparisons of the \pT{}-ordered shower.  This
shower is a new option in \pythia{} and has not been tuned systematically
before. Nevertheless, the \pythia{} manual recommends to set \LambdaQCD to
0.23. The \ATLAS collaboration in their 2008 production tune chose to leave
this parameter as set for the $Q^2$ shower, so for a full breadth of comparison
our plots show our new tune, the default with $\LambdaQCD = 0.23$, and the
settings used by \ATLAS~\cite{moraes-tuneATLAS}.

\subsection{Underlying event and multiple parton interactions}

For the third step we tuned the parameters relevant to the underlying event,
again both for the virtuality-ordered shower and the old MPI model, and
for the \pT-ordered shower with the interleaved MPI model. This was
based on various Drell-Yan, jet physics, and minimum bias measurements
performed by \cdf{} and \dzero{} in \runi{} and
\runii{}~\cite{Affolder:1999jh,Affolder:2001xt,Acosta:2001rm,%
  Aaltonen:2009ne,cdf-note9351,cdf-leadingjet,Abazov:2004hm}.
\TabRef{tab:obsweight-ue} lists all observables and their
corresponding weights used in the tuning.

The new MPI model differs significantly from the old one, hence we had to tune
different sets of parameters for these two cases. For the virtuality-ordered
shower and old MPI model we took Rick Field's tune~DW~\cite{field-tuneDW} as
guideline. In the case of the new model we consulted Peter Skands, author of the
new MPI model, and used a setup similar to his
tune~S\O~\cite{Sandhoff:2005jh,Skands:2007zg} as starting point. All switches
and parameters for the UE/MPI tune, and our results, are listed in
\TabsRef{tab:params-ueq2} and~\ref{tab:params-uept}.

One of the main differences we observed between the models is their behaviour in
Drell-Yan physics. The old model had difficulties describing the \PZ \pT{}
spectrum~\cite{Affolder:1999jh} and we had to assign a high weight to that
observable in order to force the Monte Carlo to get the peak region of the
distribution right (note that this is the only observable to which we assigned
different weights for the tunes of the old and the new MPI model). The new model
on the other hand gets the \PZ \pT{} correct almost out of the box, but
underestimates the underlying event activity in Drell-Yan events as measured
in~\cite{cdf-note9351}.  The same behaviour can be observed in Peter Skands'
tunes~\cite{Skands:2009zm}. We are currently investigating this issue
together with the generator authors.

Another (albeit smaller) difference shows in the hump of the turn-on in many of
the UE distributions in jet physics. This hump is described by the new model,
but mostly missing in the old model. Although the origin of this hump is thought
to be understood as an ambiguity in defining the event direction for
events with only little and soft activity, the model differences
responsible for its presence/absence in
the two Pythia models is not yet known in any detail.

It should be noted that the parameters \PARP{71} and \PARP{79} could not be
constrained very well with the observables tuned to.

\FigsRef{fig:tune-ue-1} to \ref{fig:tune-ue-5} show some comparisons between our
new tune and various other tunes.  For the virtuality-ordered shower with the
old MPI model we show Rick Field's tunes~A~\cite{field-tuneA} and
DW~\cite{field-tuneDW} as references, since they are well-known and widely
used. For the \pT{}-ordered shower and the new MPI framework we compare to Peter
Skands' new ``Perugia 0'' tune \cite{Skands:2009zm}. We also include the 2008
\ATLAS tune~\cite{moraes-tuneATLAS} in our comparison, since it is widely used
at the \lhc{}, but note that the energy scaling is strongly disfavoured by the
existing data, as well as the issues discussed in \SecRef{sec:fragmentationtune}.

\subsection{Tune verification}
\begin{figure}
  \centering
  \subfigure[Determining directions of largest/smallest uncertainty]{
  \label{fig:linescan-schematic}
    \scalebox{.55}{
    \input{linescan-ellipsis}
    }
  }
  \subfigure[Line-scan along direction of smallest uncertainty]{
  \label{fig:linescan-steep}
  \includegraphics[width=0.45\textwidth]{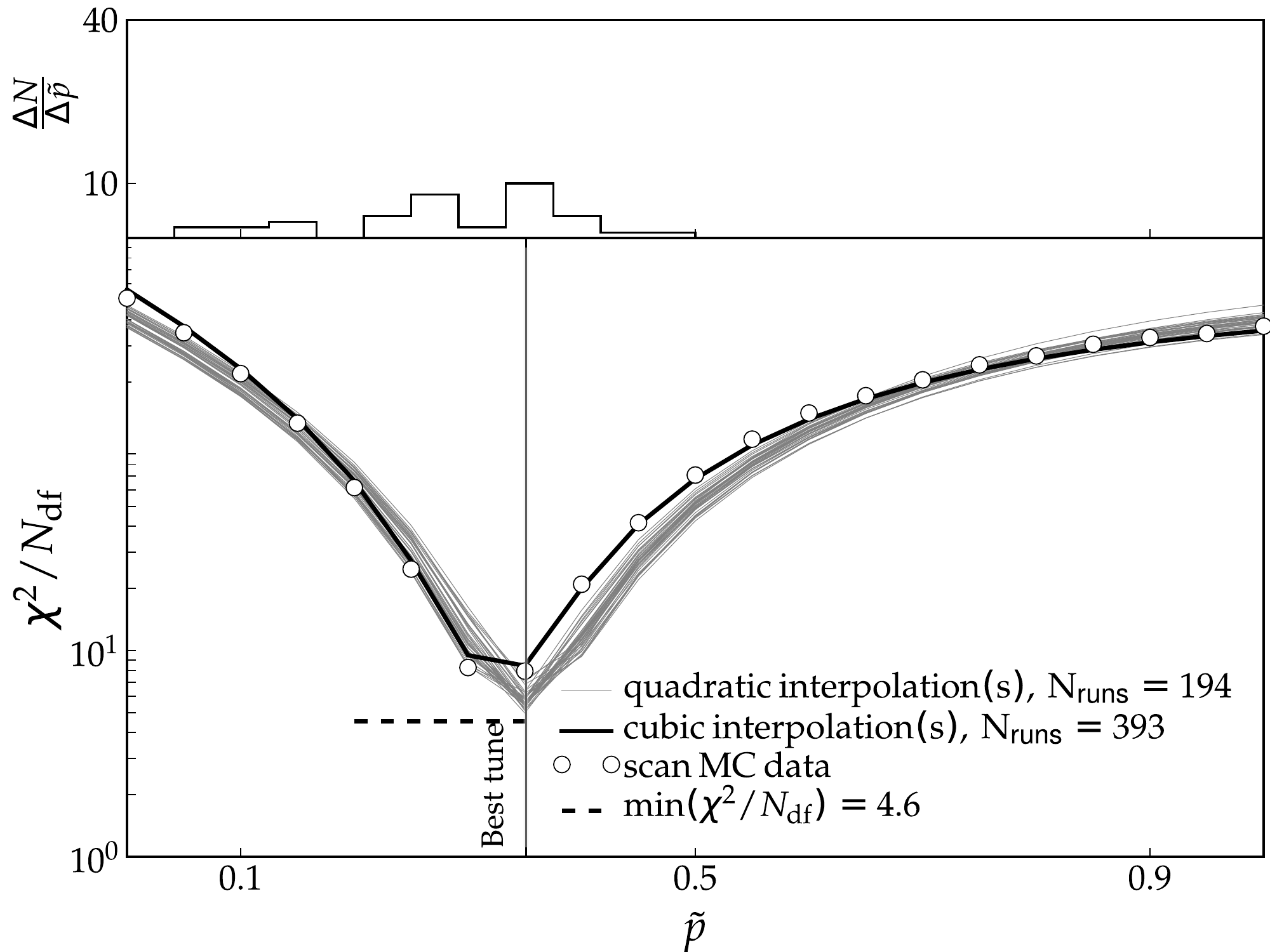}
  }
  \subfigure[Line-scan along direction of largest uncertainty]{
  \label{fig:linescan-shallow}
  \includegraphics[width=0.45\textwidth]{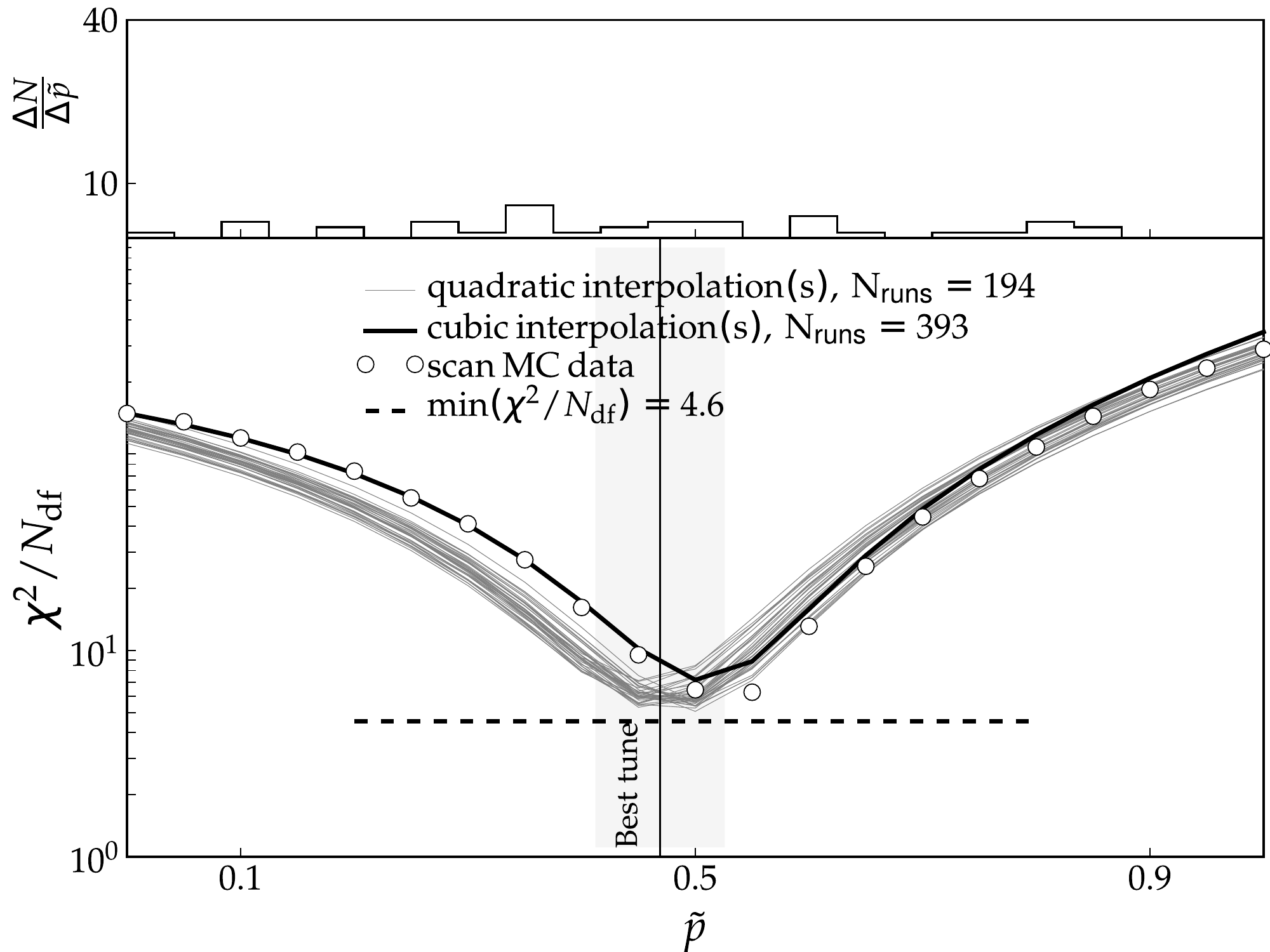}
  }
  \caption{%
    Line-scan validation of the best tuning estimate obtained with \professor
    for the underlying event (\pT-ordered shower).  The sketch in
     \subref{fig:linescan-schematic} illustrates how the directions of largest
    and smallest uncertainty are found based on a best tune's covariance matrix
    for a toy two dimensional case --- the real UE tune has 10 parameters.  The
    line-scans are done along the direction of the smallest uncertainty in
    \subref{fig:linescan-steep} and the largest uncertainty in
    \subref{fig:linescan-shallow}, parameterised by $\ptilde\in\left[0, 1\right]$.  The histograms on top of the line-scans show
    the distribution of minimisation results obtained with the corresponding
    parameterisations. The gray band in \subref{fig:linescan-shallow} indicates
    the parameter uncertainties (estimated by the minimiser) of the best tune estimate as quoted by
    \minuit. The corresponding band in \subref{fig:linescan-steep} is a thin
    line, since the best tune value is very well-defined.%
  }
  \label{fig:ueptlinescans}
\end{figure}

Much effort has been put into the verification of the new tune.  We have
performed line-scan validations along the directions in parameter space that
correspond to the largest and the smallest uncertainty based on the covariance
matrix, $\mathcal{C}$, of the new tune's minimisation result. These directions
should coincide with those where the \gof-function is very flat or
very steep, respectively.  The eigen-decomposition of $\mathcal{C}$ can be
written as 
\begin{align}
    \mathcal{C} = T^{\mathsf T} \Sigma T
\end{align} 
where $T$ is a rotation matrix and $\Sigma$ a diagonal matrix.  The eigenvalues
of $\Sigma$ are related to the axes of the rotated hyper-ellipsoid of
$\mathcal{C}$ . The eigenvectors $\vec{\sigma}_\text{max, min}$ of the largest
and the smallest (absolute) eigenvalues are rotated back into the original
system and used to define the scan-lines: 
\begin{align}
        \vec{d}_\text{max, min} =  T\vec{\sigma}_\text{max, min}
\end{align}
A schematic illustration of this procedure for a two dimensional case can be
found in \FigRef{fig:linescan-schematic}. It was checked that the parameter
points sampled from these lines are kept in the region of interpolation.  From
the parameterisations available, only those where the obtained minima project
onto the according scan line were chosen for the line-scan.  The line-scans can
be found in \FigRef{fig:linescan-steep} and \FigRef{fig:linescan-shallow}.  It
shows that the parameterisations and the actual generator response are in very
good agreement, especially in the region of the minimum.  However, in the case
of \FigRef{fig:linescan-shallow} the quadratic interpolations seem to shift the
line-scan a little to the left but the deviation is inside the parameter
uncertainties indicated by the grey vertical band. In both cases we observe an
even better description of the generator response by the cubic interpolation,
especially in regions further away from the minimum.

\subsection{Tuning stability}
\begin{figure}
  \centering
  \includegraphics[width=0.45\textwidth]{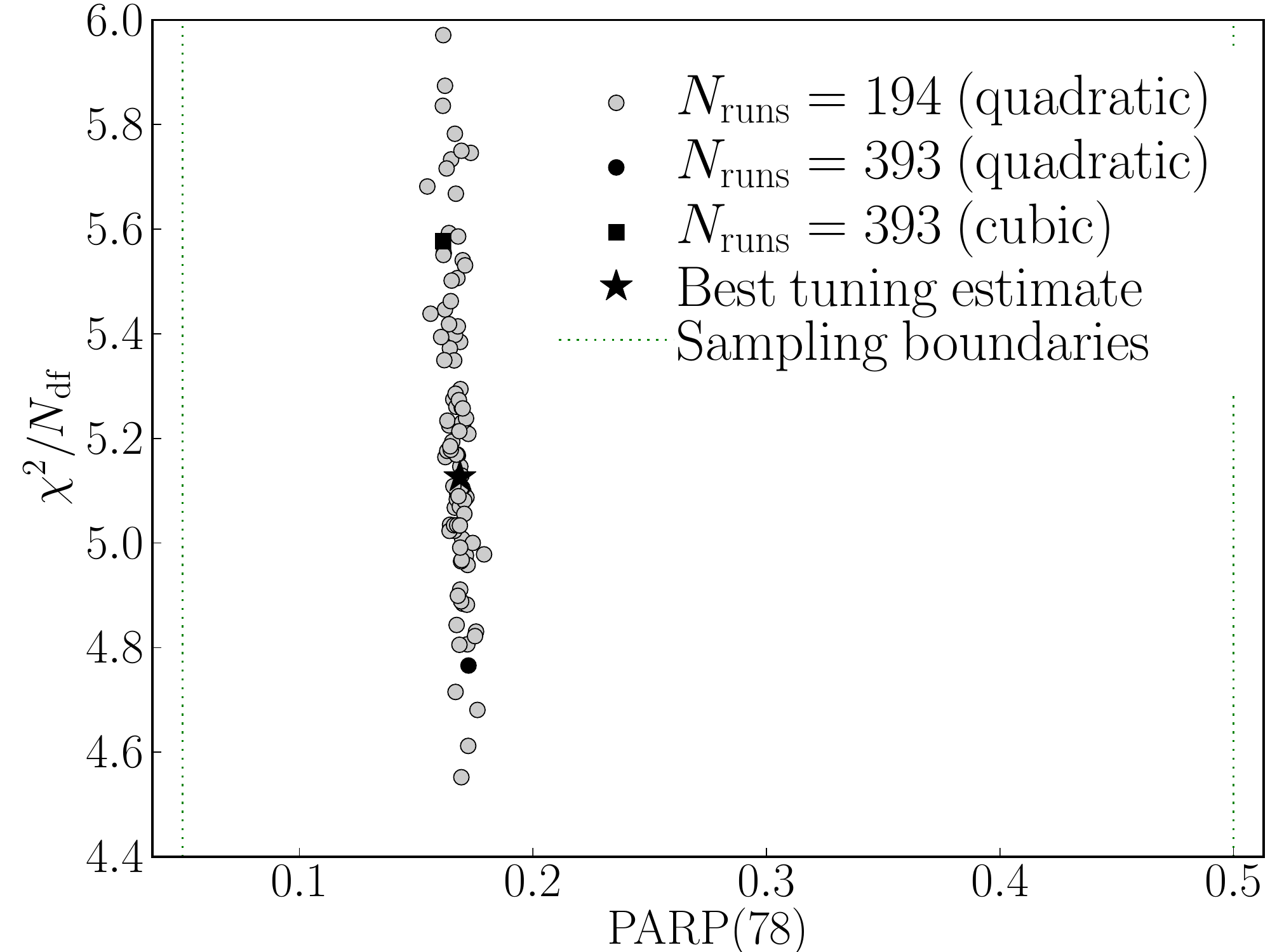}
  \includegraphics[width=0.45\textwidth]{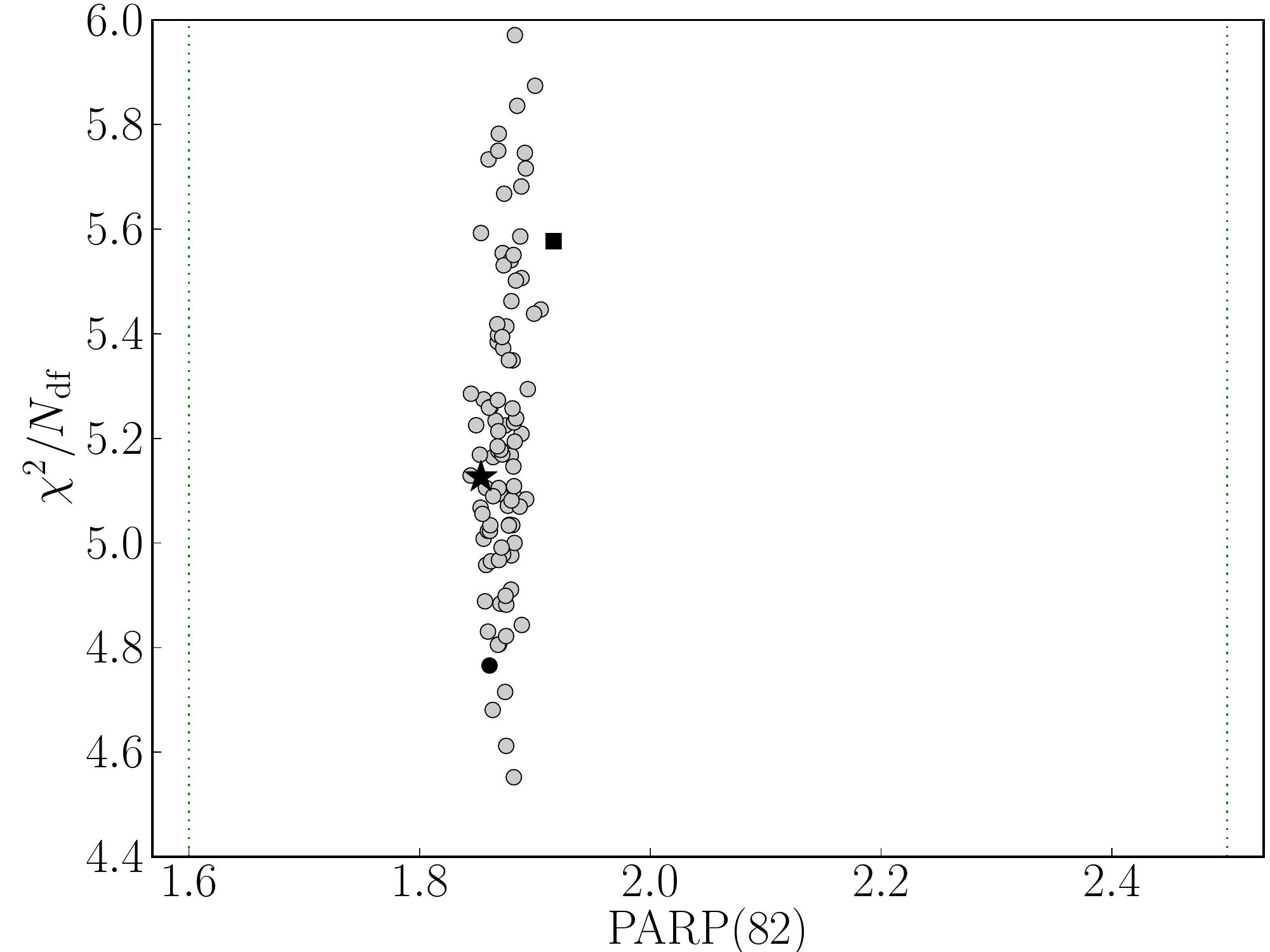}
  \includegraphics[width=0.45\textwidth]{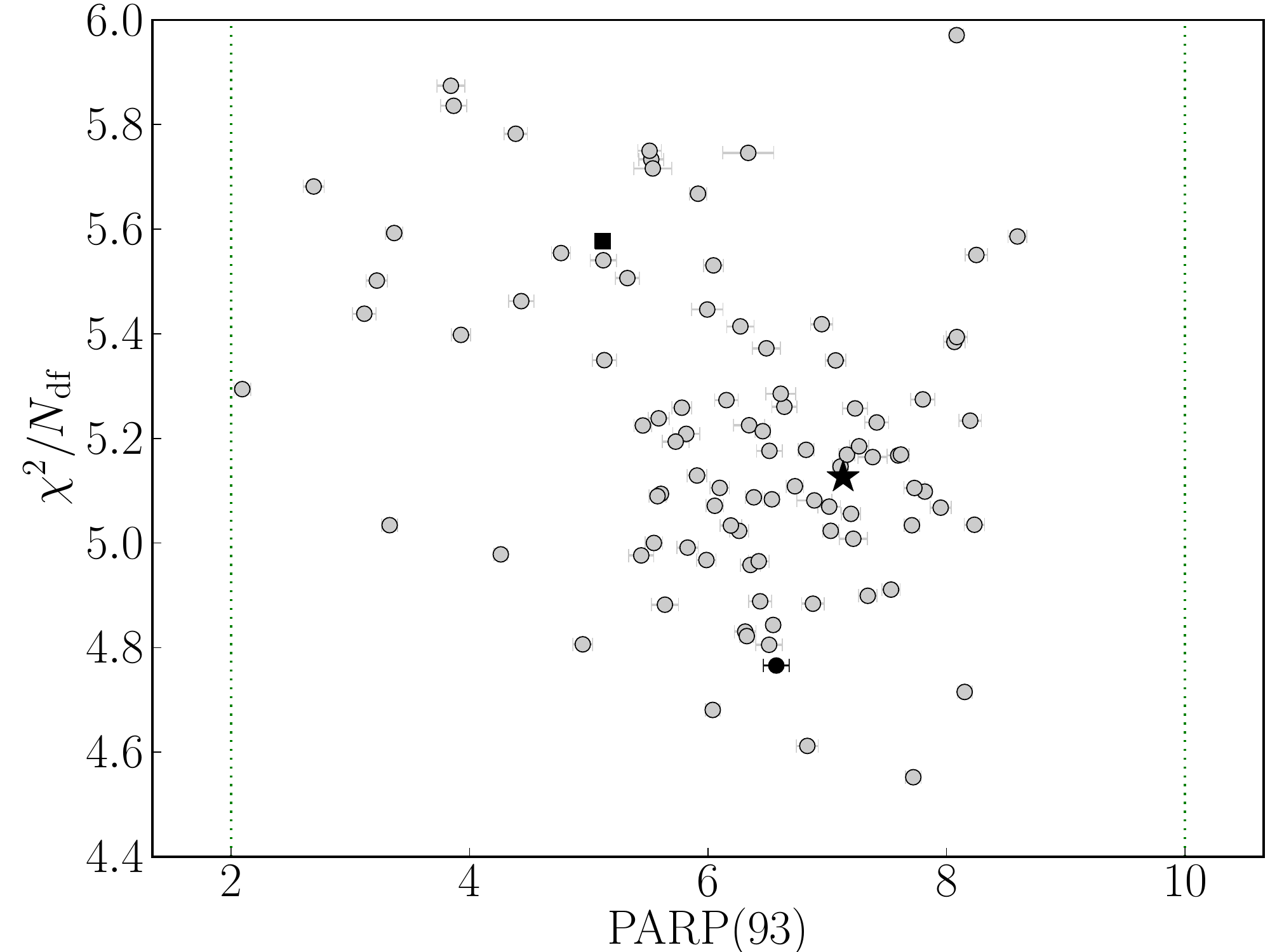}
  \caption{%
    Examples of parameter-wise distributions of minimisation results obtained
    for the underlying event (\pT-ordered shower) with the weights found in
    \TabRef{tab:obsweight-ue} and various $N_{\text{tune}}$ and polynomial orders. 
    For each parameter the goodness of fit of each minimisation result is projected
    on to the corresponding parameter axis. The minimum is clearly
    well-defined for some parameters like \PARP{78} or PARP{82}; others are
    scattered over the whole parameter range, e.g. PARP(93), indicating that we
    are not sensitive to these parameters, i.e. they are not very important for
    a good description of the data we tuned to.
    The best tune estimate, indicated by a star, was derived from a quadratic
    parameterisation that uses 194 runs.}
  \label{fig:ueptchi2vsparams}
\end{figure}

An example parameter-wise comparison of the goodness of fit of minimisation
results obtained from 100 quadratic parameterisations that use 194 runs as well
as from two parameterisations (one quadratic, one cubic) that use 393 runs can
be found for the three parameters \PARP{78}, \PARP{82} and \PARP{93} in
\FigRef{fig:ueptchi2vsparams}.  In the case of \PARP{78} we find a very well
defined minimum, indicating that the used observables are sensitive to this
parameter. A similar picture with a somewhat larger spread is found for
\PARP{82}. Here, the cubic parameterisation predicts a slightly larger value
than the quadratic ones do. \PARP{93} does not show a well defined minimum,
indicating that the observables we tuned to are not very sensitive to this
parameter, i.e., that \PARP{93} is not very important for a good description of
the data.  However the goodness of fit values suggest that a large value ($> 5~
\GeV$) is preferred. 

It has been observed that the scattering of the minimisation results among each
other is \emph{larger} compared to the errors calculated by \minuit, indicating
that for the tuning of the underlying event the choice of run-combinations is
accompanied by a systematic error. This stands in opposition to the tunings of
flavour- and fragmentation parameters.

\section{Conclusions}

MC generator tuning is currently in something of a boom era: there has been much
activity in systematising tuning and validation in the past years, driven by the
growing realisation that MPI effects will be a highly significant effect at the
LHC and that existing data places only relatively weak constraints on their
scale.  In the last year, interest has been gradually converging on the \rivet{}
and \professor{} tools; as demonstrated in this paper, these are now in a state
where they can be used to achieve real physics goals and the \pythiasix{} tunes
described in \SecRef{sec:py6} have been a significant success, bringing new
accuracy, speed and systematic control to this previously vague topic.

Our development plans in the near future are very much aimed at tuning of the
newer \Cpp generator codes (\pythiaeight, \sherpa \& \herwigpp) to
$\eplus\eminus$ and hadron collider data as we have done here for \pythiasix. As
well as the hadron collider data shown here, we intend to extend \rivet's
coverage of analyses to include low energy data from UA1, UA5, RHIC and other
experiments. \PB-factory data, if it can be obtained, will help to improve
further the final state radiation and hadronisation tunes and challenge existing
models. We have also been able to use Professor in the setup described here for
similar tunings of \pythiasix to alternative PDFs, to be described in a separate
note.

We also anticipate using Professor within the LHC collaborations to provide
re-tunes of MC generators to early QCD-dominated data: this is a crucial step
for understanding the underlying event at LHC energies, since low energy data
provides little constraint on the evolution of the total $\Pproton\Pproton$
cross-section. We are collaborating with both \ATLAS and \cms to ensure that
pre-publication data can be best used to rapidly improve the simulation of
backgrounds to new physics searches.

There is clearly more potential for exploration and innovation in connection
with the \professor method and tools. One idea currently being investigated is
the provision of representative ``error tunes'' (cf. ``error PDFs'') and
uncertainty bands to give a quantitative estimate on how much models and tunes
may be expected to deviate from data; this is particularly relevant for
extrapolations such as the UE energy evolution. Other ideas include the use of
Professor to make fast predictions of generator distributions for e.g. SUSY
parameter space scans, optimisation of parameterised observables such as jet
measures against user-defined goodness functions, or even
multidimensional fits in experimental analyses like a simultaneous fit
of top mass and jet energy scale using MC templates. We look forward to
challenging current MC models with the combination of LHC data and the
new trend for statistically robust parameter exploration.

\section*{Acknowledgements}

We would like to thank Frank Krauss for convening the \professor collaboration
and for myriad useful discussions. This work was supported in part by the MCnet
European Union Marie Curie Research Training Network (contract
MRTN-CT-2006-035606), which provided funding for collaboration meetings and
attendance at research workshops such as ACAT08 and MPI@LHC.

Andy Buckley has been principally supported by a Special Project Grant from the
UK Science \& Technology Funding Council. Hendrik Hoeth acknowledges a MCnet
postdoctoral fellowship.

\bibliographystyle{h-physrev3}
{\raggedright
  \bibliography{profpythia6}
}
\begin{appendix}
  \onecolumn
  \clearpage
  \section{Tables}
  
  \vspace*{2em}
\begin{center}
  \begin{tabular}{lrrl}
    \toprule
    Parameter  & Pythia 6.418 default & Final tune  &            \\
    \midrule
    \PARJ{1}   & 0.1       & 0.073 & diquark suppression         \\ 
    \PARJ{2}   & 0.3       & 0.2   & strange suppression         \\ 
    \PARJ{3}   & 0.4       & 0.94  & strange diquark suppression \\ 
    \PARJ{4}   & 0.05      & 0.032 & spin-1 diquark suppression  \\ 
    \PARJ{11}  & 0.5       & 0.31  & spin-1 light meson          \\ 
    \PARJ{12}  & 0.6       & 0.4   & spin-1 strange meson        \\ 
    \PARJ{13}  & 0.75      & 0.54  & spin-1 heavy meson          \\ 
    \PARJ{25}  & 1         & 0.63  & \Peta suppression           \\ 
    \PARJ{26}  & 0.4       & 0.12  & \Petaprime suppression      \\
    \bottomrule
  \end{tabular}
  \captionof{table}{Tuned flavour parameters and their defaults.}
  \label{tab:tune-flavour}
\end{center}

  \vspace*{2em}
\begin{center}
  \begin{tabular}{lrrrl}
    \toprule
    Parameter & Pythia 6.418 default & Final tune ($Q^2$) & Final tune ($\pT$) & \\
    \midrule
    \MSTJ{11}  & 4    & 5     & 5     & frag. function         \\
    \PARJ{21}  & 0.36 & 0.325 & 0.313 & $\sigma_q$             \\ 
    \PARJ{41}  & 0.3  & 0.5   & 0.49  & $a$                    \\ 
    \PARJ{42}  & 0.58 & 0.6   & 1.2   & $b$                    \\ 
    \PARJ{47}  & 1    & 0.67  & 1.0   & $r_b$                  \\ 
    \PARJ{81}  & 0.29 & 0.29  & 0.257 & $\Lambda_\text{QCD}$   \\
    \PARJ{82}  & 1    & 1.65  & 0.8   & shower cut-off         \\
    \bottomrule
  \end{tabular}
  \captionof{table}{Tuned fragmentation parameters and their defaults for the
    virtuality and \pT-ordered showers.}
  \label{tab:tune-frag}
\end{center}

  \vspace*{2em}
\begin{center}
  \begin{tabular}{lrrl}
    \toprule
    Parameter & \multicolumn{1}{l}{Pythia 6.418 default} & \multicolumn{1}{l}{Final tune} &  \\
    \midrule
    PARP(62)  & 1.0 & 2.9 & ISR cut-off \\ 
    PARP(64)  & 1.0 & 0.14 & ISR scale factor for $\alpha_S$ \\ 
    PARP(67)  & 4.0 & 2.65 & max. virtuality \\ 
    PARP(82)  & 2.0 & 1.9 & $\pT^0$ at reference $E_\text{cm}$ \\ 
    PARP(83)  & 0.5 & 0.83 & matter distribution \\ 
    PARP(84)  & 0.4 & 0.6 & matter distribution \\ 
    PARP(85)  & 0.9 & 0.86 & colour connection \\ 
    PARP(86)  & 1.0 & 0.93 & colour connection \\ 
    PARP(90)  & 0.2 & 0.22 & $\pT^0$ energy evolution \\ 
    PARP(91)  & 2.0 & 2.1 & intrinsic $k_\perp$ \\ 
    PARP(93)  & 5.0 & 5.0 & intrinsic $k_\perp$ cut-off \\ 
    \bottomrule
  \end{tabular}
  \captionof{table}{Tuned parameters for the underlying event using the
    virtuality-ordered shower}
  \label{tab:params-ueq2}
\end{center}

  \clearpage
  \vspace*{2em}
\begin{center}
  \begin{tabular}{lrrl}
    \toprule
    Parameter & \multicolumn{1}{l}{Pythia 6.418 default} & \multicolumn{1}{l}{Final tune} &  \\
    \midrule
    PARP(64)  & 1.0 & 1.3 & ISR scale factor for $\alpha_S$ \\ 
    PARP(71)  & 4.0 & 2.0 & max. virtuality (non-s-channel) \\ 
    PARP(78)  & 0.03 & 0.17 & colour reconnection in FSR \\ 
    PARP(79)  & 2.0 & 1.18 & beam remnant x enhancement \\ 
    PARP(80)  & 0.1 & 0.01 & beam remnant breakup suppression \\ 
    PARP(82)  & 2.0 & 1.85 & $\pT^0$ at reference $E_\text{cm}$ \\ 
    PARP(83)  & 1.8 & 1.8 & matter distribution \\ 
    PARP(90)  & 0.16 & 0.22 & $\pT^0$ energy evolution \\ 
    PARP(91)  & 2.0 & 2.0 & intrinsic $k_\perp$ \\ 
    PARP(93)  & 5.0 & 7.0 & intrinsic $k_\perp$ cut-off \\ 
    \bottomrule
  \end{tabular}

  \bigskip
  \begin{tabular}{lrl}
    \toprule
    Switch  & Value & Effect \\
    \midrule
    MSTJ(41)  & 12 & switch on $\pT$-ordered shower \\ 
    MSTP(51)  & 7 & use CTEQ5L \\ 
    MSTP(52)  & 1 & use internal PDF set \\ 
    MSTP(70)  & 2 & model for smooth $\pT^0$ \\ 
    MSTP(72)  & 0 & FSR model \\ 
    MSTP(81)  & 21 & turn on multiple interactions (new model) \\ 
    MSTP(82)  & 5 & model of hadronic matter overlap \\ 
    MSTP(88)  & 0 & quark junctions $\to$ diquark/Baryon model \\ 
    MSTP(95)  & 6 & colour reconnection \\
    \bottomrule
  \end{tabular}
  \captionof{table}{Tuned parameters (upper table) and switches (lower table)
    for the underlying event using the $\pT$-ordered shower.}
  \label{tab:params-uept}
\end{center}

  \clearpage
  \vspace*{2em}
\begin{center}
  \begin{tabular}{lccc}
    \toprule
    Observable & \multicolumn{1}{l}{Weight} \\
    \midrule
    \Pbottom quark frag. function $f(x_{\PB}^{\text{weak}})$         & 1 \\ 
    Mean of \Pbottom quark frag. function $f(x_{\PB}^{\text{weak}})$ & 1 \\ 
    $\Pup\Pdown\Pstrange$ events mean charged multiplicity           & 1 \\ 
    $\Pcharm$ events mean charged multiplicity                       & 1 \\ 
    $\Pbottom$ events mean charged multiplicity                      & 1 \\ 
    All events mean charged multiplicity                             & 1 \\
    \midrule
    \Ppipm multiplicity                                              & 1 \\ 
    \Ppizero  multiplicity                                           & 1 \\ 
    $\Ppizero/\Ppipm$ multiplicity ratio                             & 6 \\ 
    $\PKplus/\Ppipm$ multiplicity ratio                              & 6 \\ 
    $\PKzero/\Ppipm$ multiplicity ratio                              & 6 \\ 
    $\Peta/\Ppipm$ multiplicity ratio                                & 2 \\ 
    $\Petaprime(958)/\Ppipm$ multiplicity ratio                      & 1 \\ 
    $\PDplus/\Ppipm$ multiplicity ratio                              & 1 \\ 
    $\PDzero/\Ppipm$ multiplicity ratio                              & 1 \\ 
    $\PDsplus/\Ppipm$ multiplicity ratio                             & 2 \\ 
    $(\PBplus\!, \PBd)/\Ppipm$ multiplicity ratio                    & 1 \\ 
    $\PBu/\Ppipm$ multiplicity ratio                                 & 1 \\ 
    $\PBs/\Ppipm$ multiplicity ratio                                 & 2 \\ 
    \midrule
    $\Prhozero(770)/\Ppipm$ multiplicity ratio                       & 9 \\ 
    $\Prhoplus(770)/\Ppipm$ multiplicity ratio                       & 9 \\ 
    $\Pomega(782)/\Ppipm$ multiplicity ratio                         & 9 \\ 
    $\PKstar^+(892)/\Ppipm$ multiplicity ratio                       & 2 \\ 
    $\PKstar^0(892)/\Ppipm$ multiplicity ratio                       & 2 \\ 
    $\Pphi(1020)/\Ppipm$ multiplicity ratio                          & 1 \\ 
    $\PDstar^+(2010)/\Ppipm$ multiplicity ratio                      & 1 \\ 
    $\PDsstar^+(2112)/\Ppipm$ multiplicity ratio                     & 1 \\ 
    $\PBstar/\Ppipm$ multiplicity ratio                              & 1 \\ 
    \midrule
    $\Pproton/\Ppipm$ multiplicity ratio                             & 3 \\ 
    $\PLambda/\Ppipm$ multiplicity ratio                             & 4 \\ 
    $\PSigmazero/\Ppipm$ multiplicity ratio                          & 2 \\ 
    $\PSigmapm/\Ppipm$ multiplicity ratio                            & 2 \\ 
    $\PXiminus/\Ppipm$ multiplicity ratio                            & 1 \\ 
    $\Delta^{++}(1232)/\Ppipm$ multiplicity ratio                    & 1 \\ 
    $\PSigmapm(1385)/\Ppipm$ multiplicity ratio                      & 1 \\ 
    \bottomrule
  \end{tabular}
  \captionof{table}{Observables and weights included in the flavour tune}
  \label{tab:obsweight-flavour}
\end{center}

  \clearpage      
  \vspace*{2em}
\begin{center}
  \begin{tabular}{lcc}
    \toprule
    Observable                                       & Weight ($Q^2$) & Weight (\pT) \\
    \midrule
    $\pT^\text{in}$  w.r.t. Thrust axes                         &   1 &   2 \\
    $\pT^\text{out}$ w.r.t. Thrust axes                         &   1 &   1 \\
    $\pT^\text{in}$  w.r.t. Sphericity axes                     &   1 &   2 \\
    $\pT^\text{out}$ w.r.t. Sphericity axes                     &   1 &   1 \\
    Scaled momentum, $x_p = |p|/|p_\text{beam}|$                     &   1 &   3 \\
    Log of scaled momentum, $\log{1/x_p}$                       &   1 &   3 \\
    Mean $\pT^\text{out}$ vs $x_p$                              &     &   1 \\
    Mean $\pT$ vs $x_p$                                         &     &   1 \\
    $1-\text{Thrust}$, $1-T$                                    &   1 &   6 \\
    Thrust major, $M$                                           &   1 &   4 \\
    Thrust minor, $m$                                           &   1 &   4 \\
    Oblateness = $M - m$                                        &   1 &   1 \\
    Sphericity, $S$                                             &   1 &   1 \\
    Aplanarity, $A$                                             &   1 &   1 \\
    Planarity, $P$                                              &   1 &   1 \\
    $C$ parameter                                               &   1 &   1 \\
    $D$ parameter                                               &   1 &   4 \\
    Energy-energy correlation, EEC                              &     &   1 \\
    Mean charged multiplicity                                   & 160 & 181 \\
    \midrule
    \Pbottom quark frag. function $f(x_{\PB}^{\text{weak}})$             &   1 &   2 \\
    Mean of \Pbottom quark frag. function $f(x_{\PB}^{\text{weak}})$     &   1 &   4 \\
    $\Pup\Pdown\Pstrange$ events mean charged multiplicity        &  20 &  10 \\
    \Pcharm events mean charged multiplicity                      &  20 &  10 \\
    \Pbottom events mean charged multiplicity                     &  20 &  10 \\
    $\Pup\Pdown\Pstrange$ events scaled momentum, $x_p = |p|/|p_\text{beam}|$   &     &   1 \\
    \Pcharm events scaled momentum, $x_p = |p|/|p_\text{beam}|$         &     &   1 \\
    \Pbottom events scaled momentum, $x_p = |p|/|p_\text{beam}|$        &     &   1 \\
    $\Pup\Pdown\Pstrange$ events log of scaled momentum, $x_p = |p|/|p_{beam}|$ &     &   1 \\
    \Pcharm events log of scaled momentum, $x_p = |p|/|p_\text{beam}|$  &     &   1 \\
    \Pbottom events log of scaled momentum, $x_p = |p|/|p_\text{beam}|$ &     &   1 \\
    \bottomrule
  \end{tabular}
  \captionof{table}{Observables and weights included in the fragmentation tune.}
  \label{tab:obsweight-frag}
\end{center}

  \clearpage
  \enlargethispage{\baselineskip}
  \vspace*{-1em}
\begin{center}
  \begin{tabular}{lc}
    \toprule
    Observable                                                    & Weight \\
    \midrule
    \textit{\cdf{} underlying event in min-bias events:}            &    \\
    $\pT(\PZ)$                                     & 40 ($Q^2$) / 10 (\pT) \\
    $N_\text{ch}$ density vs leading jet \pT (toward), min-bias     &  1 \\
    $N_\text{ch}$ density vs leading jet \pT (transverse), min-bias &  1 \\
    $N_\text{ch}$ density vs leading jet \pT (away), min-bias       &  1 \\
    $\sum \pT$ density vs leading jet \pT (toward), min-bias        &  1 \\
    $\sum \pT$ density vs leading jet \pT (transverse), min-bias    &  1 \\
    $\sum \pT$ density vs leading jet \pT (away), min-bias          &  1 \\
    $\sum \pT$ density vs leading jet \pT (toward), JET20           &  1 \\
    $\sum \pT$ density vs leading jet \pT (transverse), JET20       &  1 \\
    $\sum \pT$ density vs leading jet \pT (away), JET20             &  1 \\
    \pT distribution (transverse), leading $\pT > \unit{30}{\GeV}$  &  1 \\
    \midrule
    \textit{\cdf multiplicity measurement:}                         &    \\
    $N_\text{ch}$ distribution at \unit{630}{\GeV}                  &  2 \\
    $N_\text{ch}$ distribution at \unit{1800}{\GeV}                 &  2 \\
    \midrule
    \textit{\cdf underlying event in leading jet events:}           &    \\
    $N_\text{ch}$ density vs leading jet \pT (transverse)           &  1 \\
    $N_\text{ch}$ density vs leading jet \pT (transMAX)             &  1 \\
    $N_\text{ch}$ density vs leading jet \pT (transMIN)             &  1 \\
    $N_\text{ch}$ density vs leading jet \pT (transDIF)             &  1 \\
    $\sum \pT$ density vs leading jet \pT (transverse)              &  1 \\
    $\sum \pT$ density vs leading jet \pT (transMAX)                &  1 \\
    $\sum \pT$ density vs leading jet \pT (transMIN)                &  1 \\
    $\sum \pT$ density vs leading jet \pT (transDIF)                &  1 \\
    $\langle \pT \rangle$ (transverse)                              &  1 \\
    \midrule
    \textit{\cdf min-bias:}                                         &    \\
    $\langle \pT \rangle$ vs $N_\text{ch}$                          &  2 \\
    \midrule
    \textit{\cdf underlying event in Drell-Yan events analysis:}    &    \\
    $N_\text{ch}$ density vs lepton pair \pT (toward)               &  1 \\
    $N_\text{ch}$ density vs lepton pair \pT (transverse)           &  1 \\
    $N_\text{ch}$ density vs lepton pair \pT (transMAX)             &  1 \\
    $N_\text{ch}$ density vs lepton pair \pT (transMIN)             &  1 \\
    $N_\text{ch}$ density vs lepton pair \pT (transDIF)             &  1 \\
    $N_\text{ch}$ density vs lepton pair \pT (away)                 &  1 \\
    $\sum \pT$ density vs lepton pair \pT (toward)                  &  1 \\
    $\sum \pT$ density vs lepton pair \pT (transverse)              &  1 \\
    $\sum \pT$ density vs lepton pair \pT (transMAX)                &  1 \\
    $\sum \pT$ density vs lepton pair \pT (transMIN)                &  1 \\
    $\sum \pT$ density vs lepton pair \pT (transDIF)                &  1 \\
    $\sum \pT$ density vs lepton pair \pT (away)                    &  1 \\
    $\langle \pT \rangle$ (toward)                                  &  1 \\
    $\langle \pT \rangle$ (transverse)                              &  1 \\
    $\langle \pT \rangle$ (away)                                    &  1 \\
    $\pT^\text{max}$ (toward)                                       &  1 \\
    $\pT^\text{max}$ (transverse)                                   &  1 \\
    $\pT^\text{max}$ (away)                                         &  1 \\
    $\langle \pT(\text{lepton pair}) \rangle$ vs $N_\text{ch}$      &  1 \\
    $\langle \pT \rangle$ vs $N_\text{ch}$                          &  1 \\
    $\langle \pT \rangle$ vs $N_\text{ch}$, $\pT(\PZ) < \unit{10}{\GeV}$ &  1 \\
    \midrule
    \textit{\dzero{} dijet angular correlations:}                        &    \\
    dijet azimuthal angle, $\pT^\text{max} \in \unit{[75, 100]}{\GeV}$   &  2 \\
    dijet azimuthal angle, $\pT^\text{max} \in \unit{[100, 130]}{\GeV}$  &  2 \\
    dijet azimuthal angle, $\pT^\text{max} \in \unit{[130, 180]}{\GeV}$  &  2 \\
    dijet azimuthal angle, $\pT^\text{max} > \unit{180}{\GeV}$           &  2 \\
    \bottomrule
  \end{tabular}
  \samepage
  \captionof{table}{Observables and weights used for the tuning of the underlying event}
  \label{tab:obsweight-ue}
\end{center}

  \clearpage
  \vspace*{2em}
\begin{center}
  \begin{tabular}{lrrrrr}
    \toprule
    & PARJ(1)  & PARJ(2)  & PARJ(3)  & PARJ(4)  & PARJ(11) \\
    \midrule
    PARJ(1) & $1$& $0.32$ & $-0.75$ & $-0.34$ & $0.41 $ \\
    PARJ(2) &    & $1$    & $-0.39$ & $-0.26$ & $0.71 $ \\
    PARJ(3) &    &        & $1 $    & $0.63 $ & $-0.35$ \\
    PARJ(4) &    &        &         & $1 $    & $-0.33$ \\
    PARJ(11)&    &        &         &         & $1 $    \\
    
    \addlinespace
    & &  PARJ(12)     & PARJ(13)         & PARJ(25)      & PARJ(26)\\
    \midrule
    PARJ(1) & &  $0.05 $ & $0.05 $ & $0.20 $  & $0.31 $  \\
    PARJ(2) & &  $-0.08$ & $0.13 $ & $0.44 $  & $0.43 $  \\
    PARJ(3) & &  $0.04 $ & $-0.05$ & $-0.15$  & $-0.33$  \\
    PARJ(4) & &  $-0.04$ & $-0.08$ & $-0.50$  & $-0.27$  \\
    PARJ(11)& &  $0.01$  & $0.07 $ & $0.41 $  & $0.28 $  \\
    PARJ(12)& &  $1    $ & $0.02 $ & $0.09 $  & $-0.04$  \\
    PARJ(13)& &          & $1 $    & $-0.01$  & $0.08 $  \\
    PARJ(25)& &          &         & $1 $     & $0.04 $  \\
    PARJ(26)& &          &         &          & $1    $  \\
    \bottomrule
  \end{tabular}
  \captionof{table}{Correlation coefficients for flavour parameters as calculated by Minuit.}
  \label{tab:correlations-flavour}
\end{center}
    
  \vspace*{2em}
\begin{center}
  \begin{tabular}{lrrrrrr}
    \toprule
    & PARJ(21)& PARJ(41)& PARJ(42)& PARJ(47)& PARJ(81)& PARJ(82)\\
    \midrule
    PARJ(21) & $1$ & $0.55$ & $0.40$ & $0.22$ & $-0.33$ & $0.50$ \\
    PARJ(41) &     & $1$    & $0.95$ & $0.40$ & $0.15$  & $0.74$ \\
    PARJ(42) &     &        & $1$    & $0.47$ & $0.31$  & $0.52$ \\
    PARJ(47) &     &        &        & $1$    & $0.07$  & $0.18$ \\
    PARJ(81) & & & & & $1$                              & $0.04$ \\
    PARJ(82) & & & & &                                  & $1$    \\
    \bottomrule
  \end{tabular}
  \captionof{table}{Correlation coefficients for fragmentation parameters as calculated by Minuit.}
  \label{tab:correlations-frag}
\end{center}

  \vspace*{2em}
\begin{center}
  \begin{tabular}{lrrrrr}
    \toprule
    & PARP(64)     & PARP(71)     & PARP(78)      & PARP(79)      & PARP(82)    \\  
    \midrule
    PARP(64) & $1$ & $0.26$ & $-0.17$  & $-0.15$  & $-0.65$ \\
    PARP(71) &        & $1$ & $ 0.39$  & $ 0.04$  & $-0.26$ \\
    PARP(78) &        &        & $ 1$  & $-0.45$  & $-0.17$ \\
    PARP(79) &        &        &          & $ 1$  & $ 0.15$ \\
    PARP(82) &        &        &          &          & $ 1$ \\
    
    \addlinespace
    
    & & PARP(83)      & PARP(90)      & PARP(91)      & PARP(93)      \\
    \midrule
    PARP(64) & & $ 0.48$  & $-0.18$  & $-0.18$  & $-0.19$  \\
    PARP(71) & & $ 0.42$  & $-0.18$  & $-0.24$  & $-0.38$  \\
    PARP(78) & & $ 0.44$  & $-0.12$  & $-0.36$  & $ 0.20$  \\
    PARP(79) & & $-0.18$  & $ 0.17$  & $ 0.00$  & $-0.26$  \\
    PARP(82) & & $-0.58$  & $ 0.65$  & $ 0.15$  & $ 0.10$  \\
    PARP(83) & & $ 1$  & $-0.21$  & $-0.18$  & $ 0.12$  \\
    PARP(90) & &          & $ 1$  & $-0.09$  & $-0.08$  \\
    PARP(91) & &          &          & $ 1$  & $ 0.27$  \\ 
    PARP(93) & &          &          &          & $ 1$  \\
    \bottomrule
  \end{tabular}
  
  \captionof{table}{Correlation coefficients for underlying event parameters (\pT-ordered shower)}
  \label{tab:correlations-uept}
\end{center}

  \clearpage
  \section{Comparisons}

  \vspace*{2em}
\begin{center}

  \begin{tabular}{ll@{${}\pm{}$}lll}
    \toprule
    Particle   & \multicolumn{2}{l}{Data} & \pythia{}~6.418 default & Final tune \\
    \midrule
    $\pi^+$             & 17.02  & 0.19   & 17.10  & 17.23  \\
    $\pi^0$             & 9.42   & 0.32   & 9.69   & 9.78   \\
    $K^+$               & 2.228  & 0.059  & 2.311  & 2.126  \\
    $K^0$               & 2.049  & 0.026  & 2.211  & 2.068  \\
    $\eta$              & 1.049  & 0.08   & 1.012  & 1.014  \\
    $\eta'(958)$        & 0.152  & 0.02   & 0.301  & 0.171  \\
    $D^+$               & 0.175  & 0.016  & 0.166  & 0.219  \\
    $D^0$               & 0.454  & 0.03   & 0.494  & 0.490  \\
    $D^+_s$              & 0.131  & 0.021  & 0.128  & 0.101  \\
    $B^+, B^0_d$        & 0.33   & 0.052  & 0.346  & 0.373  \\
    $B^+_u$             & 0.178  & 0.006  & 0.173  & 0.186  \\
    $B^0_s$             & 0.057  & 0.013  & 0.052  & 0.037  \\
    \midrule
    $\rho^0(770)$       & 1.231  & 0.098  & 1.523  & 1.267  \\
    $\rho^+(770)$       & 2.4    & 0.43   & 2.86   & 2.40   \\
    $\omega(782)$       & 1.016  & 0.065  & 1.367  & 1.150  \\
    $K^{*+}(892)$       & 0.715  & 0.059  & 1.111  & 0.740  \\
    $K^{*0}(892)$       & 0.738  & 0.024  & 1.106  & 0.743  \\
    $\phi(1020)$        & 0.0963 & 0.0032 & 0.1942 & 0.1006 \\
    $D^{*+}(2010)$      & 0.1937 & 0.0057 & 0.2395 & 0.1974 \\
    $D^{*+}_s(2112)$    & 0.101  & 0.048  & 0.090  & 0.058  \\
    $B^*$               & 0.288  & 0.026  & 0.299  & 0.221  \\
    \midrule
    $p$                 & 1.05   & 0.032  & 1.221  & 1.117  \\
    $\Lambda$           & 0.3915 & 0.0065 & 0.3922 & 0.3507 \\
    $\Sigma^0$          & 0.076  & 0.011  & 0.075  & 0.096  \\
    $\Sigma^-$          & 0.081  & 0.01   & 0.069  & 0.091  \\
    $\Sigma^+$          & 0.107  & 0.011  & 0.074  & 0.095  \\
    $\Sigma^\pm$        & 0.174  & 0.009  & 0.143  & 0.186  \\
    $\Xi^-$             & 0.0258 & 0.001  & 0.0278 & 0.0282 \\
    $\Delta^{++}(1232)$ & 0.085  & 0.014  & 0.193  & 0.147  \\
    $\Sigma^-(1385)$    & 0.024  & 0.0017 & 0.037  & 0.025  \\
    $\Sigma^+(1385)$    & 0.0239 & 0.0015 & 0.0389 & 0.0266 \\
    $\Sigma^\pm(1385)$  & 0.0462 & 0.0028 & 0.0757 & 0.0516 \\
    \bottomrule
  \end{tabular}

  \samepage
  \captionof{table}{Mean hadron multiplicities in $e^+e^-$ collisions at
                    \unit{91}{\GeV} for data~\cite{Amsler:2008zz},
                    \pythia{}~6.418 default and our tune using the
                    virtuality-ordered shower. While there is a slight
                    degradation in charm and bottom mesons, the strange
                    sector is significantly improved (mesons and
                    baryons), and also particles like $\rho$ and
                    $\omega$ clearly benefit from the tuning.}
  \label{tab:multiplicities}
\end{center}

  \clearpage
  \enlargethispage{2\baselineskip}
  \vspace*{-1em}
  \begin{center}
    \vspace*{2em}
    \includegraphics[width=0.45\textwidth]{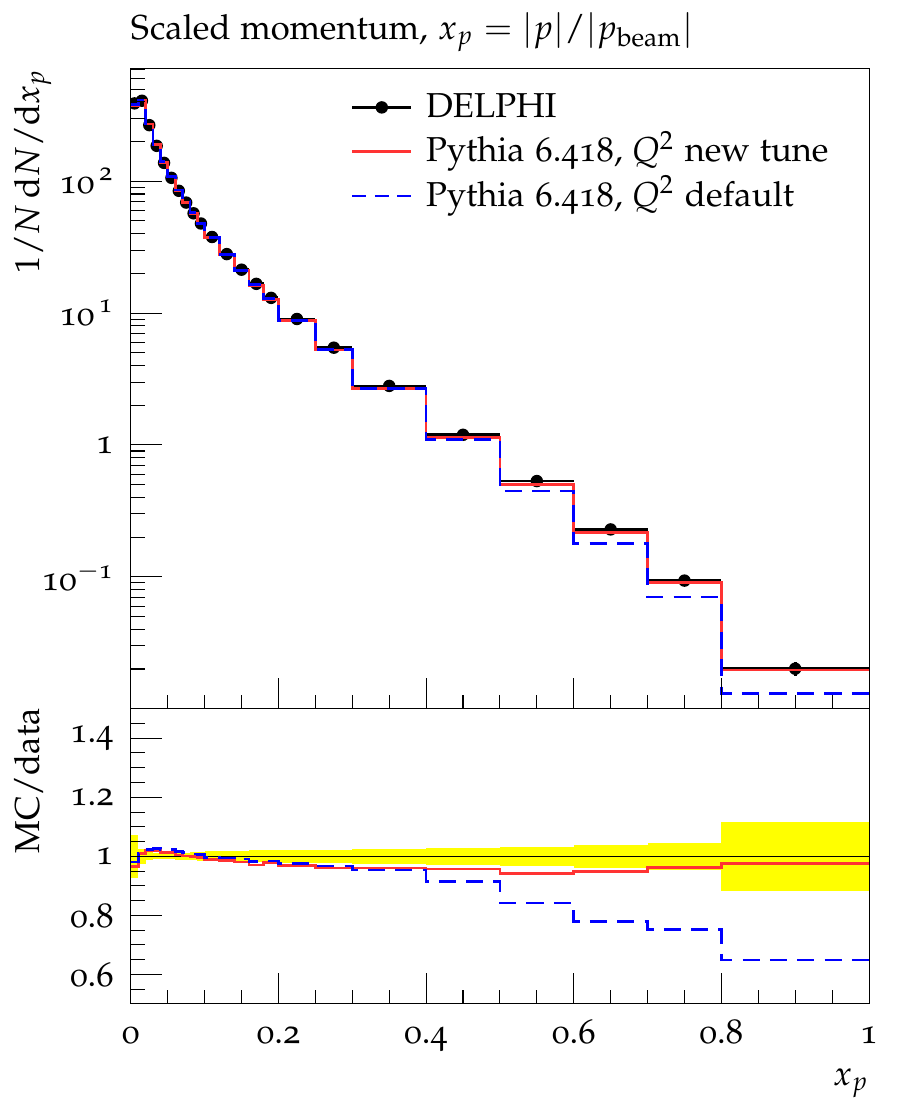}
    \includegraphics[width=0.45\textwidth]{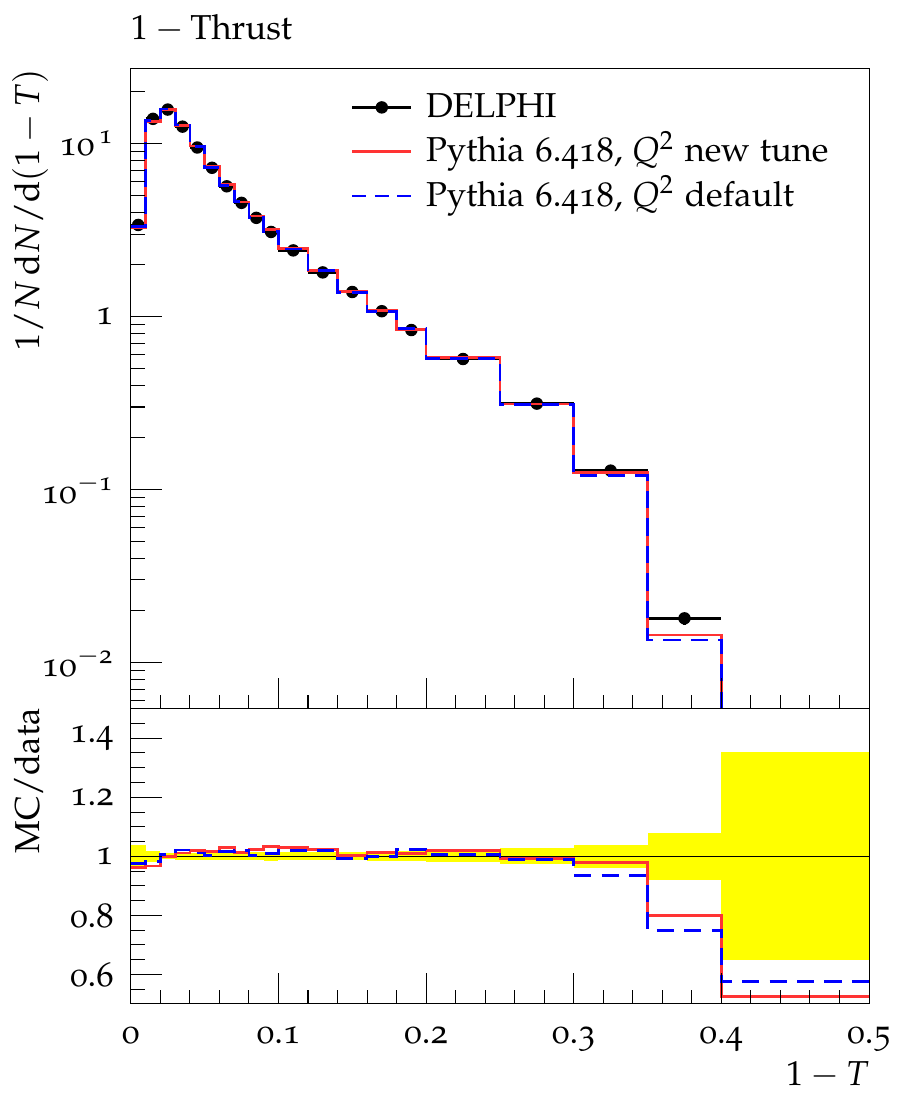}\\
    \vspace*{2em}
    \includegraphics[width=0.45\textwidth]{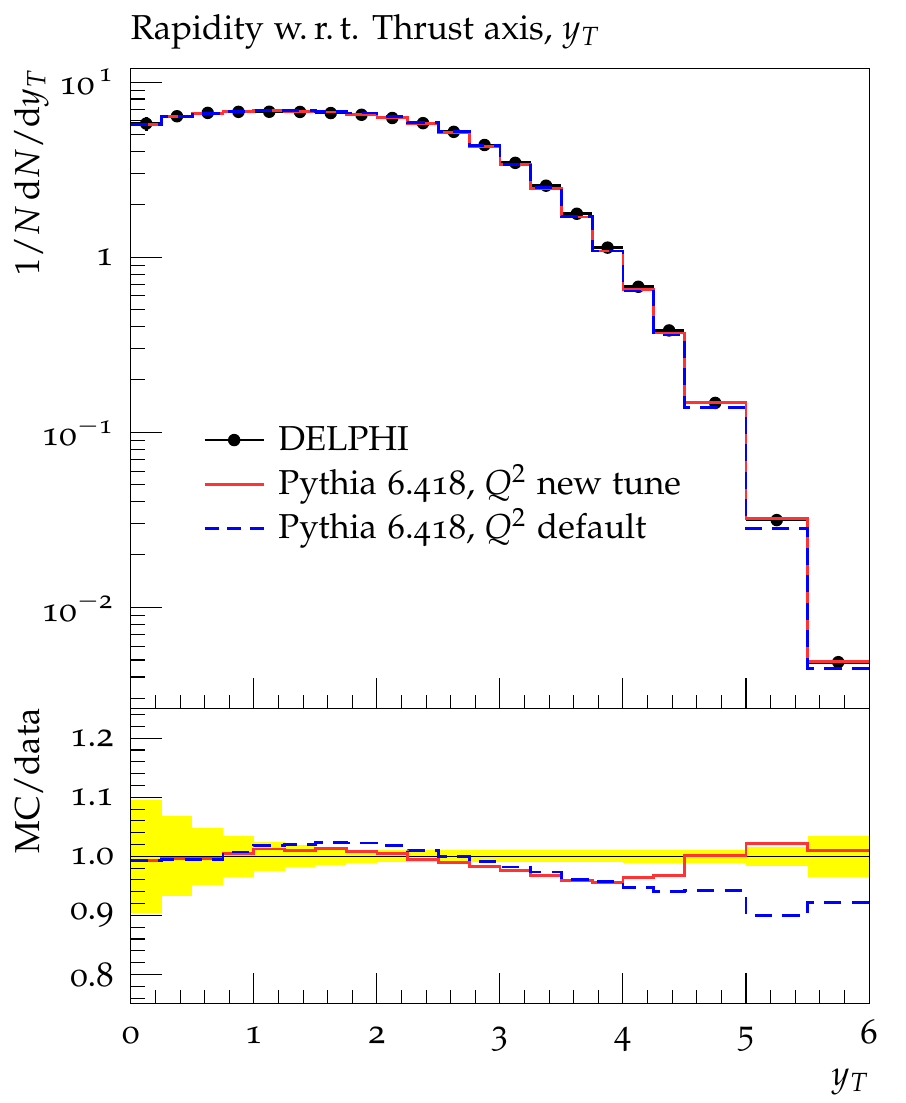}
    \includegraphics[width=0.45\textwidth]{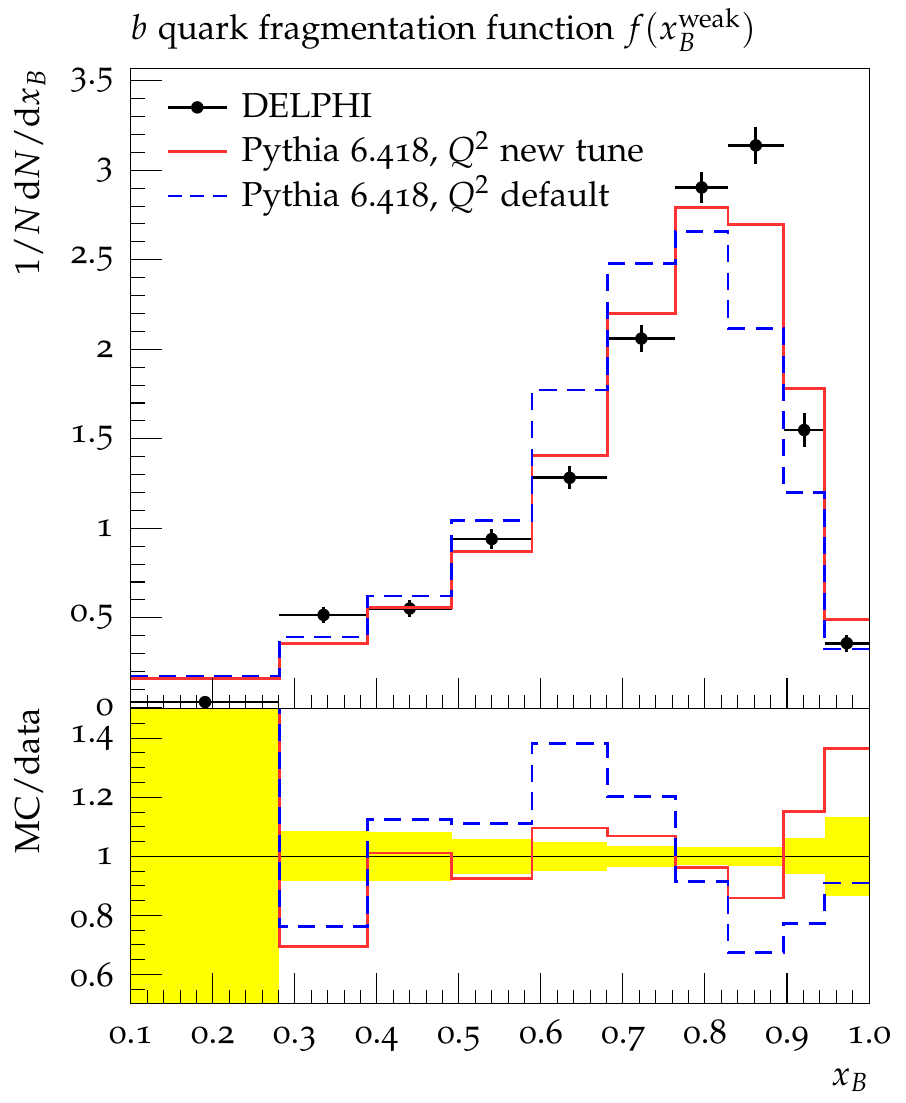}
    \captionof{figure}{Some example distributions for $\eplus\eminus$
      collisions using the virtuality-ordered shower. The solid line shows the
      new tune, the dashed line is the default. Even though the
      virtuality-ordered shower is well-tested and \pythia{} has been tuned
      several times, especially by the \lep{} collaborations, there is still
      room for improvement in the default settings.  Note the different scale
      in the ratio plot of the rapidity distribution.  The data in these plots
      has been published by \delphi{}~\cite{Abreu:1996na,delphi-2002}.}
    \label{fig:tune-frag-q2}
  \end{center}

  \clearpage
  \enlargethispage{2\baselineskip}
  \vspace*{-1em}
  \begin{center}
    \vspace*{2em}
    \includegraphics[width=0.45\textwidth]{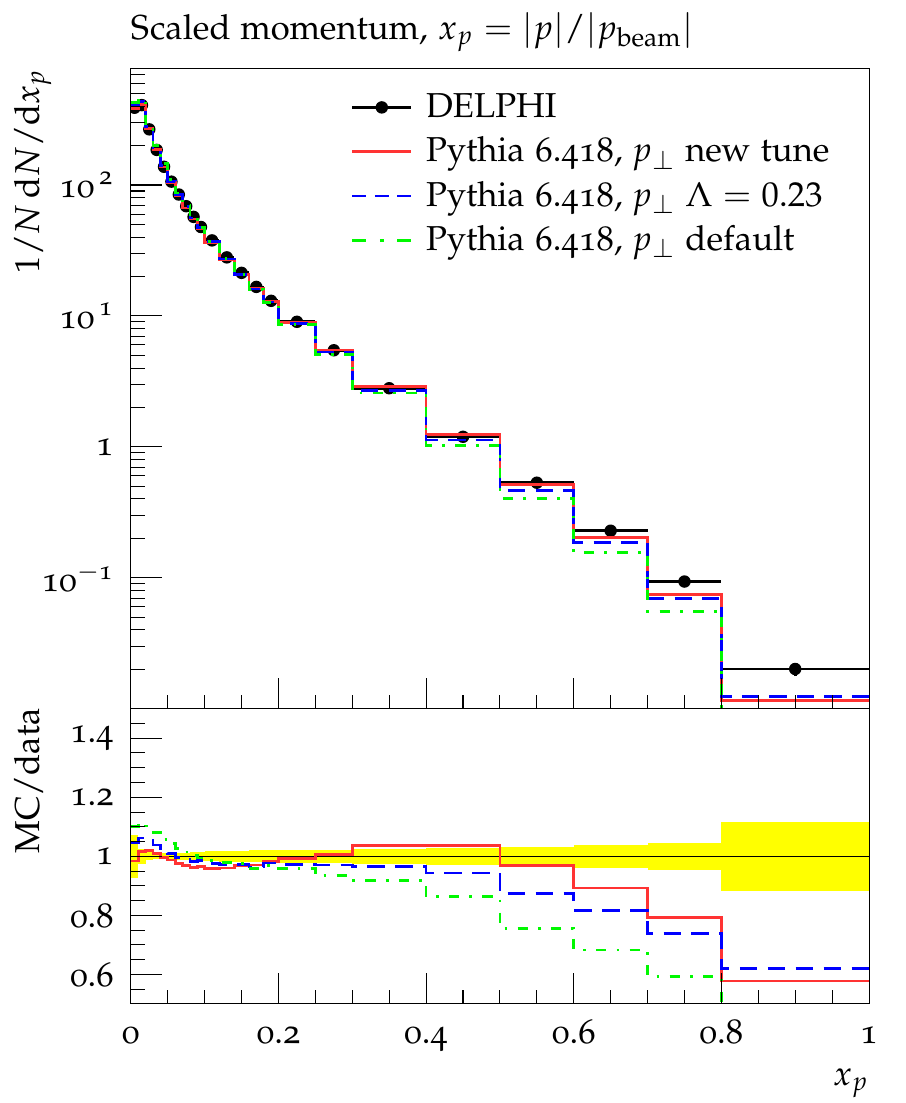}
    \includegraphics[width=0.45\textwidth]{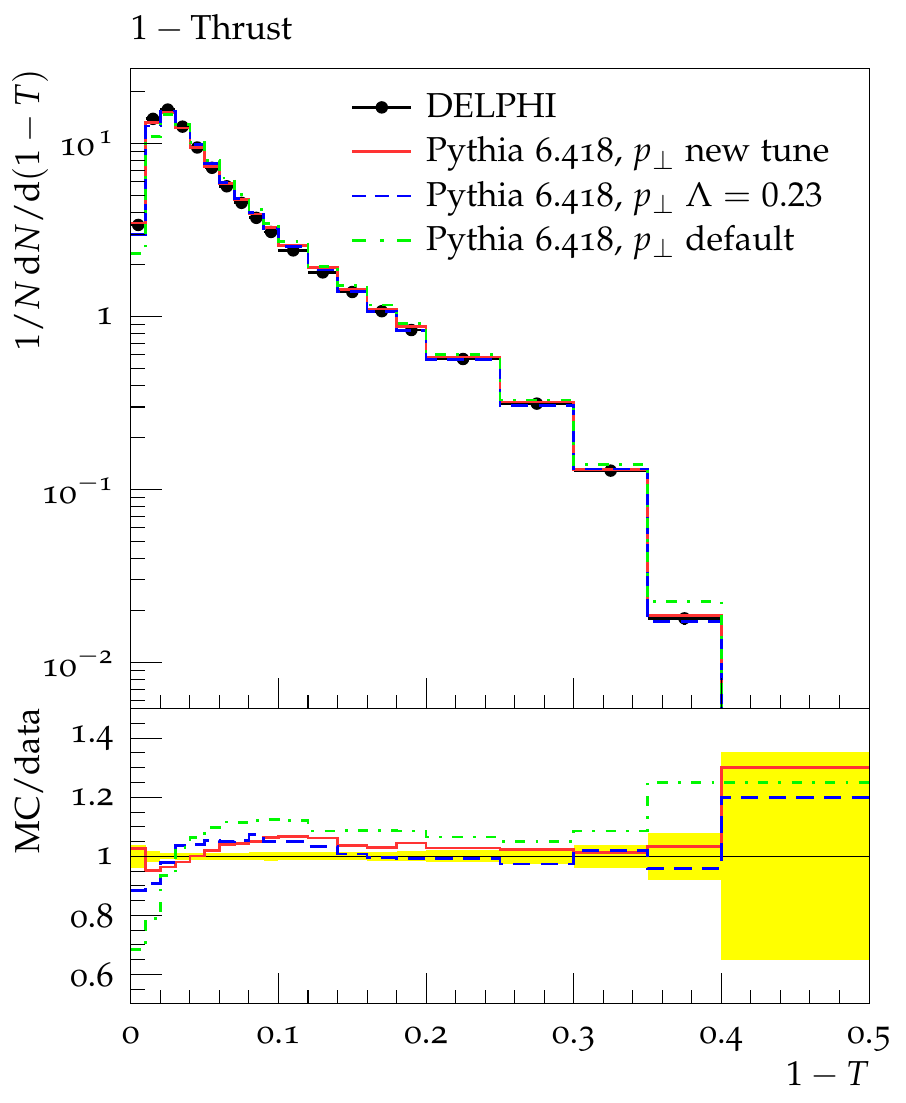}\\
    \vspace*{2em}
    \includegraphics[width=0.45\textwidth]{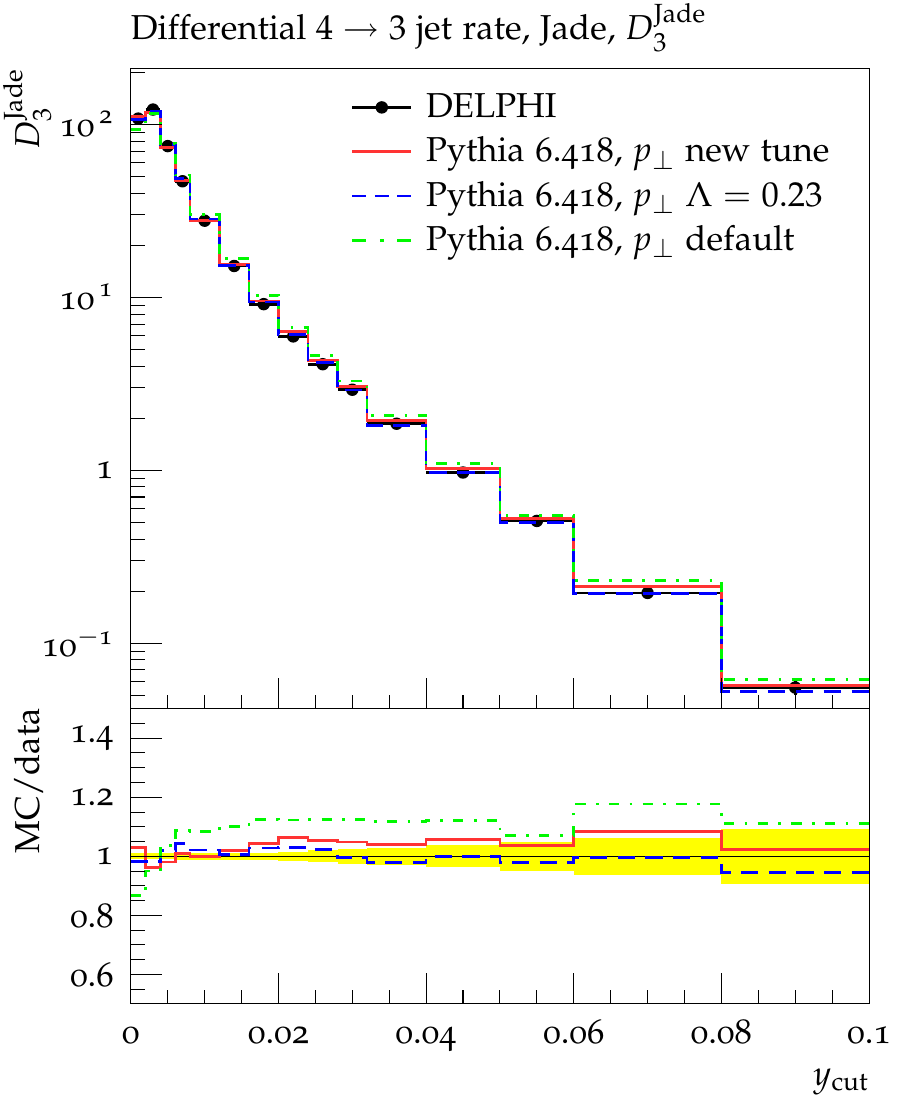}
    \includegraphics[width=0.45\textwidth]{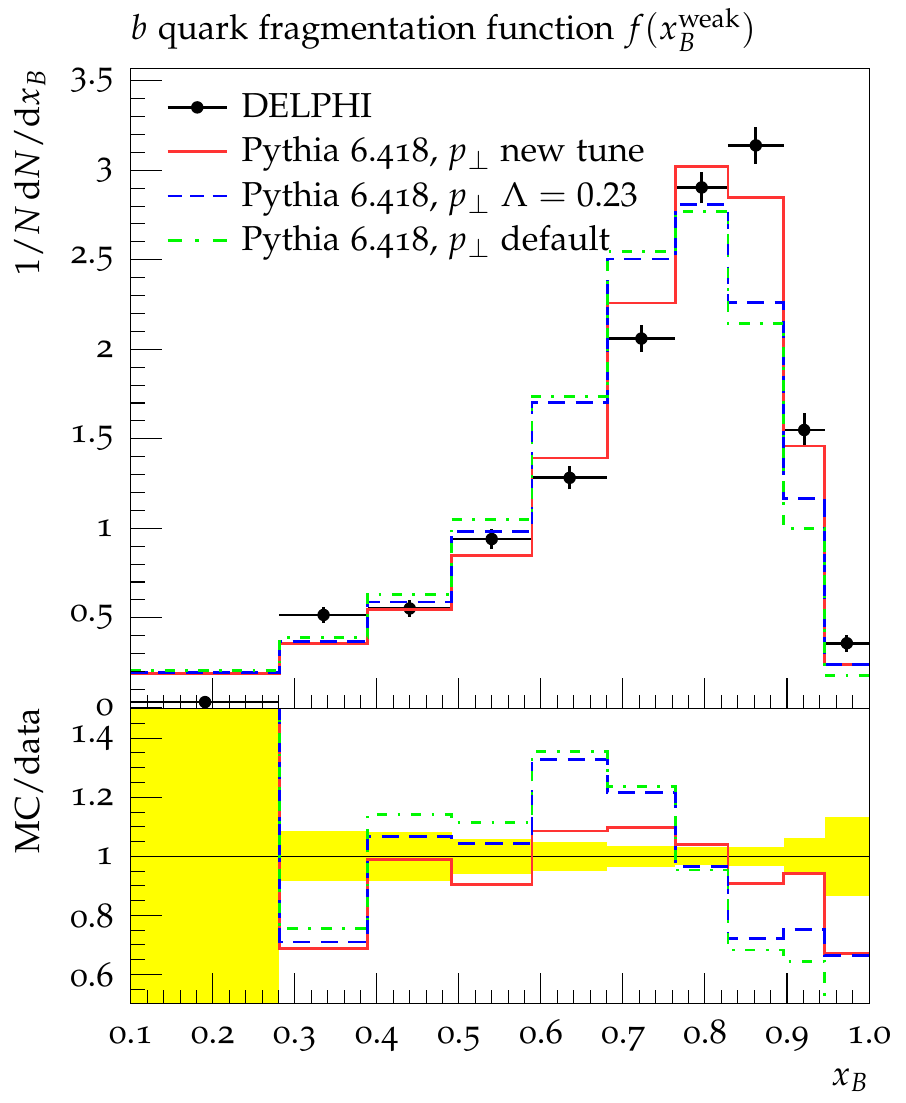}
    \captionof{figure}{Some example distributions for $\eplus\eminus$
      collisions using the \pT{}-ordered shower. The solid line shows the new
      tune, the dashed line is the old recommendation for using the
      \pT{}-ordered shower (i.\,e. changing \LambdaQCD to 0.23), and the
      dashed-dotted line is produced by switching on the \pT{}-ordered shower
      leaving everything else at its default (the choice made for the \ATLAS tune.)
      The data has been published by \delphi{}~\cite{Abreu:1996na,delphi-2002}.}
    \label{fig:tune-frag-pt}
  \end{center}

  \clearpage
  \enlargethispage{2\baselineskip}
  \vspace*{-1em}
  \begin{center}
    \vspace*{2em}
    \includegraphics[width=0.45\textwidth]{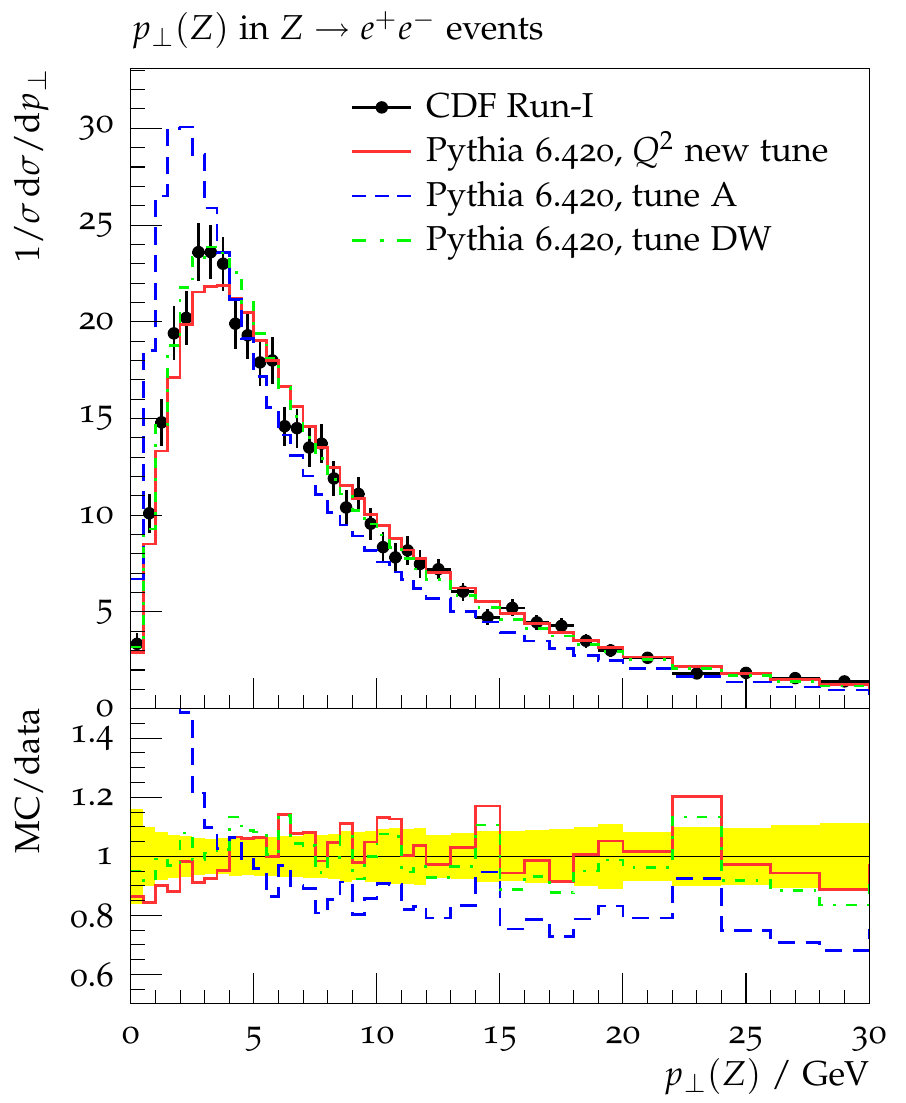}
    \includegraphics[width=0.45\textwidth]{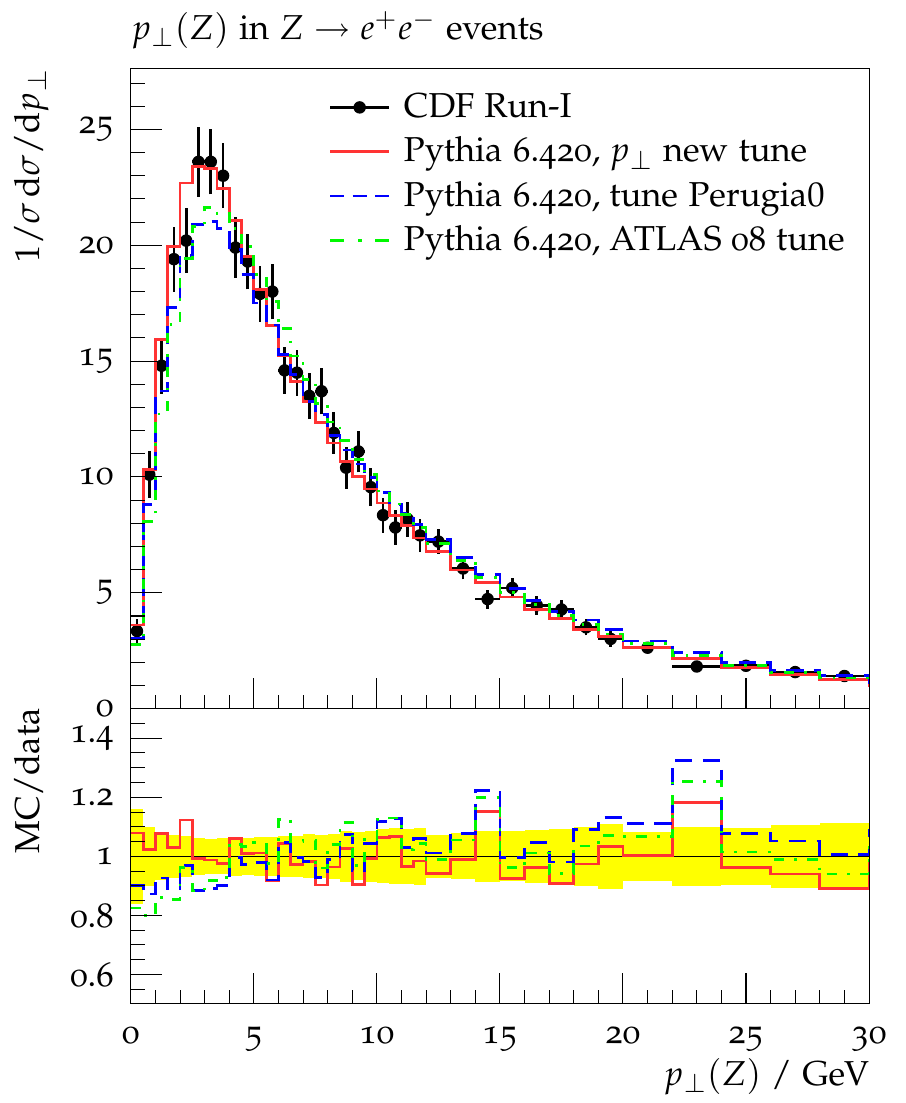}\\
    \vspace*{2em}
    \includegraphics[width=0.45\textwidth]{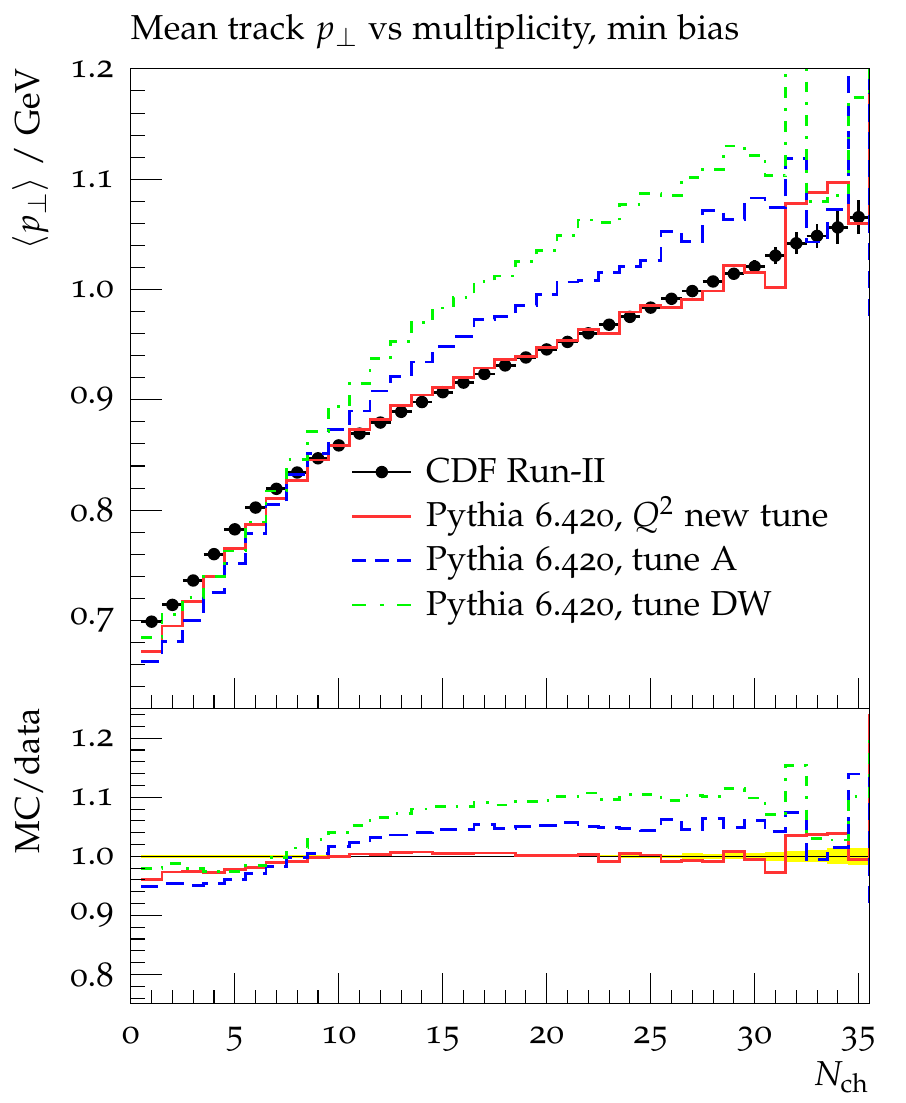}
    \includegraphics[width=0.45\textwidth]{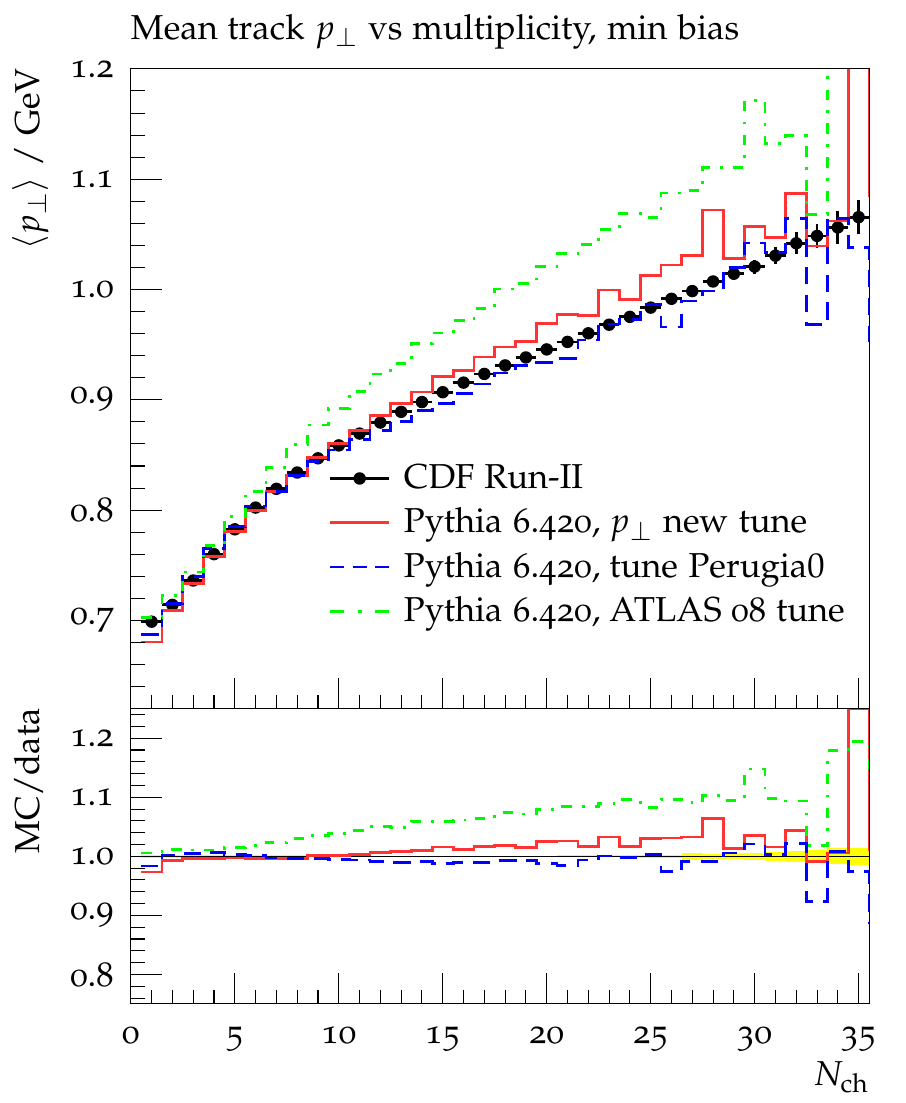}
    \captionof{figure}{The upper plots show the \PZ \pT{} distribution as
      measured by CDF~\cite{Affolder:1999jh} compared to different tunes of
      the virtuality-ordered shower with the old MPI model (left) and the
      \pT{}-ordered shower with the interleaved MPI model (right). Except for
      tune~A, all tunes describe this observable; the fixed version of
      tune~A, called AW, is basically identical to DW. The lower plots show
      the average track \pT{} as a function of the charged multiplicity in
      minimum bias events~\cite{Aaltonen:2009ne}. This observable is quite
      sensitive to colour reconnection. Only the recent tunes hit the data
      here (except for \ATLAS.)}
    \label{fig:tune-ue-1}
  \end{center}

  \clearpage
  \enlargethispage{2\baselineskip}
  \vspace*{-1em}
  \begin{center}
    \vspace*{2em}
    \includegraphics[width=0.45\textwidth]{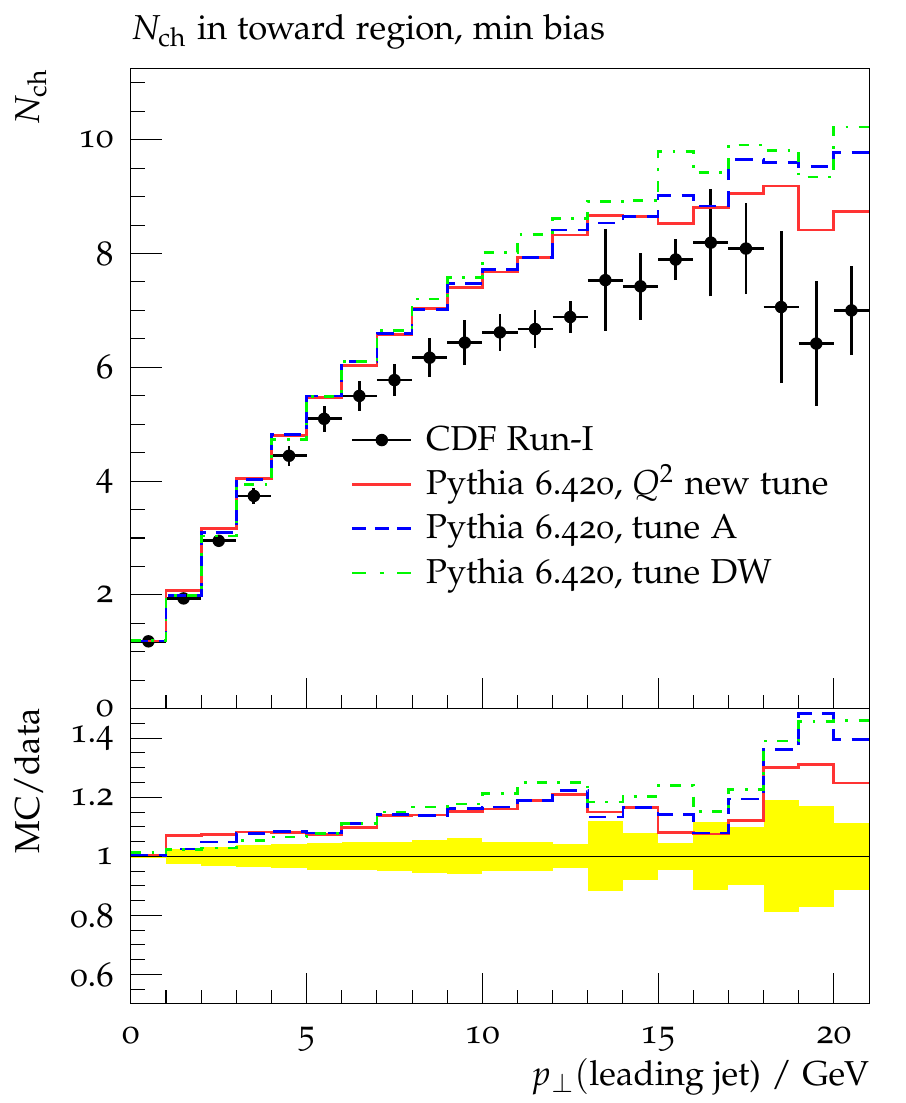}
    \includegraphics[width=0.45\textwidth]{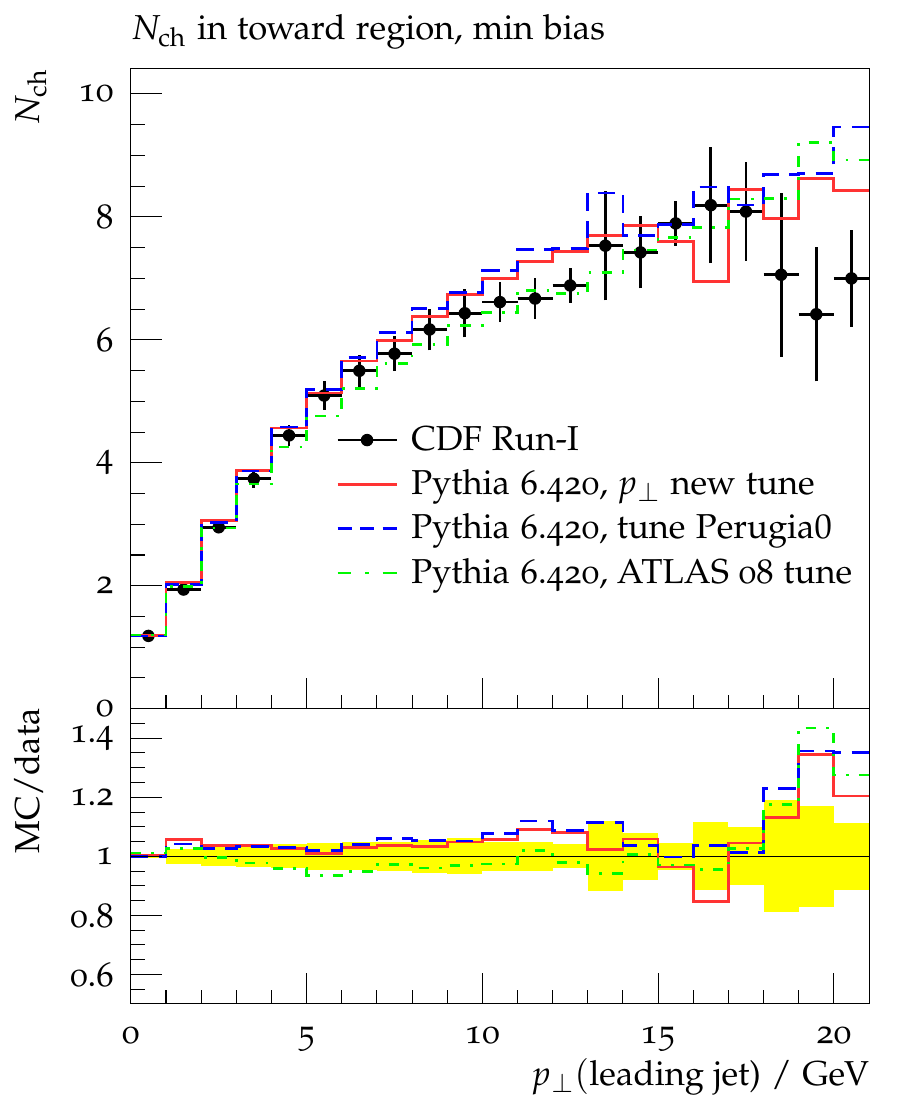}\\
    \vspace*{2em}
    \includegraphics[width=0.45\textwidth]{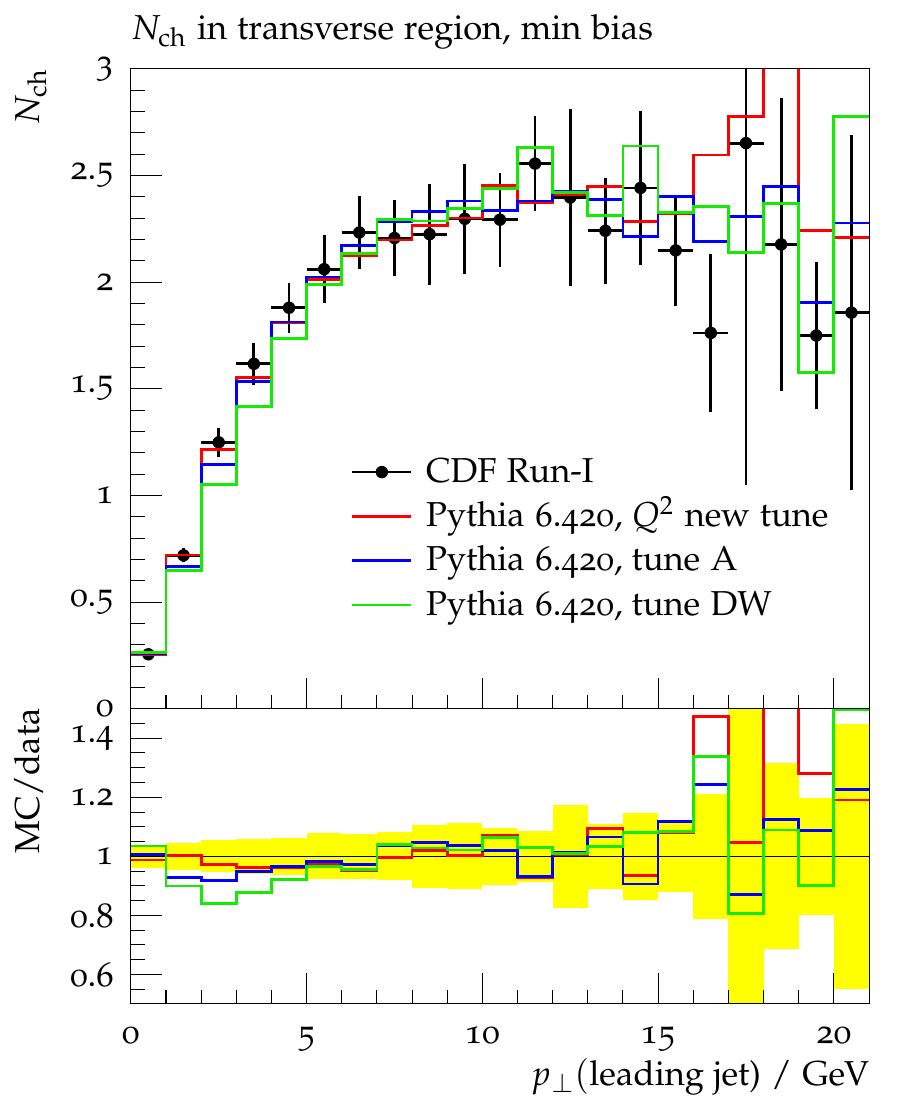}
    \includegraphics[width=0.45\textwidth]{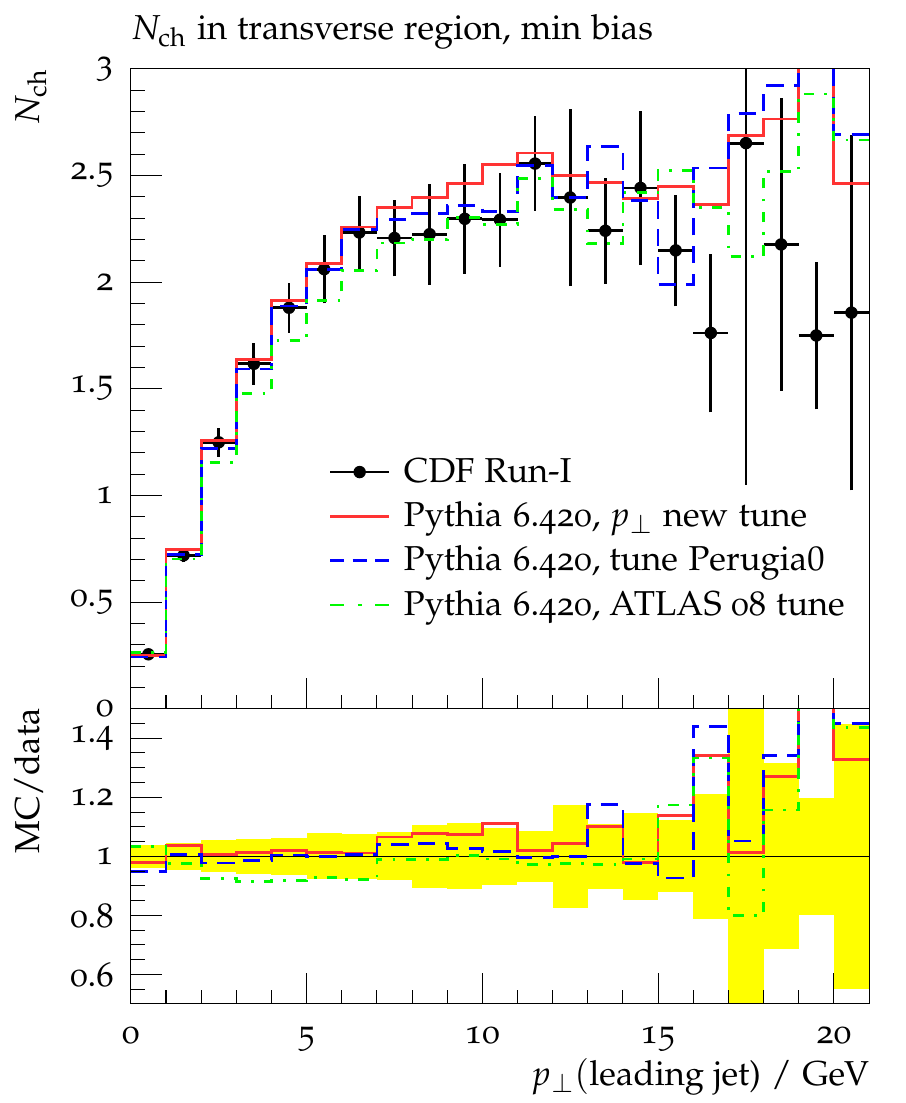}
    \captionof{figure}{These plots show the average charged multiplicity in
      the toward and transverse regions as function of the leading jet \pT{}
      in minimum bias events~\cite{Affolder:2001xt}. The left side shows tunes of
      the virtuality-ordered shower with the old MPI model, while on
      the right side the \pT{}-ordered shower with the interleaved MPI model
      is used. The old model is known to be a bit too ``jetty'' in the toward
      region, which can be seen in the first plot. Other than this, all tunes
      are very similar.}
    \label{fig:tune-ue-2}
  \end{center}

  \clearpage
  \enlargethispage{2\baselineskip}
  \vspace*{-1em}
  \begin{center}
    \vspace*{2em}
    \includegraphics[width=0.45\textwidth]{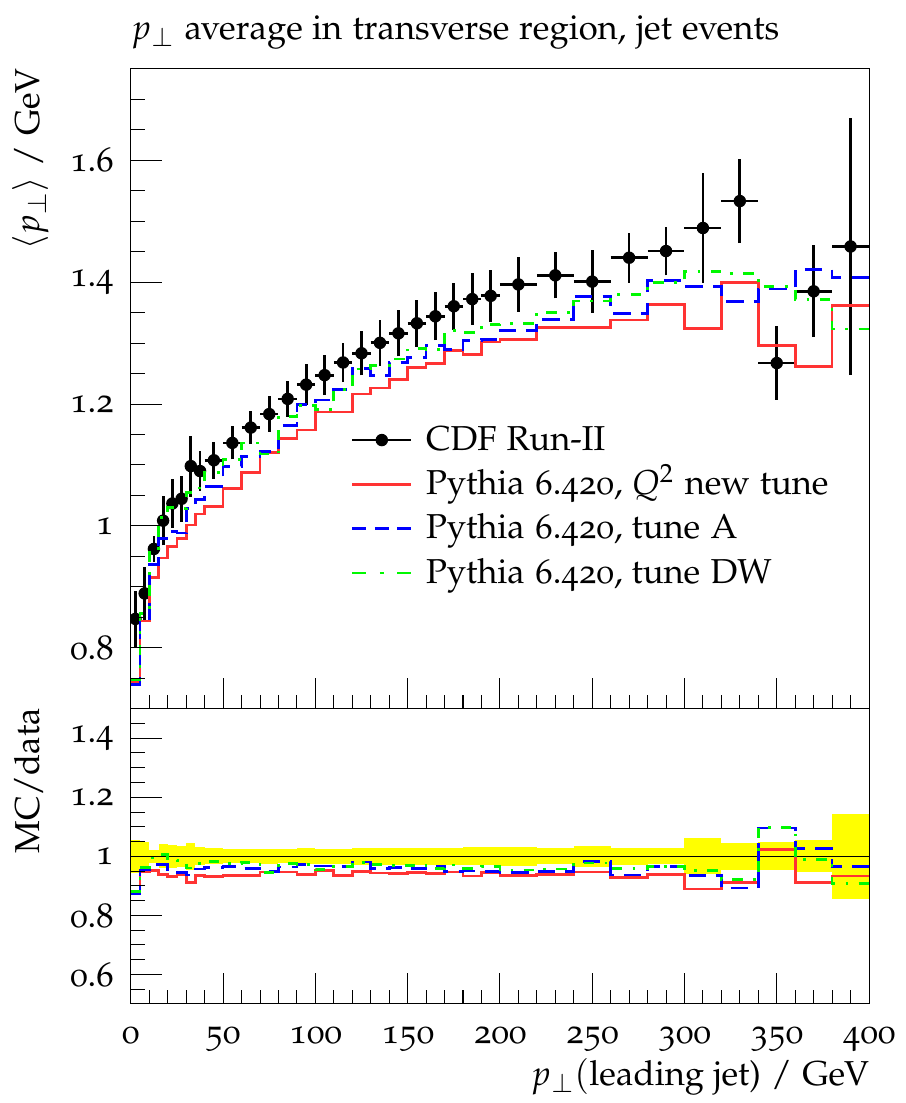}
    \includegraphics[width=0.45\textwidth]{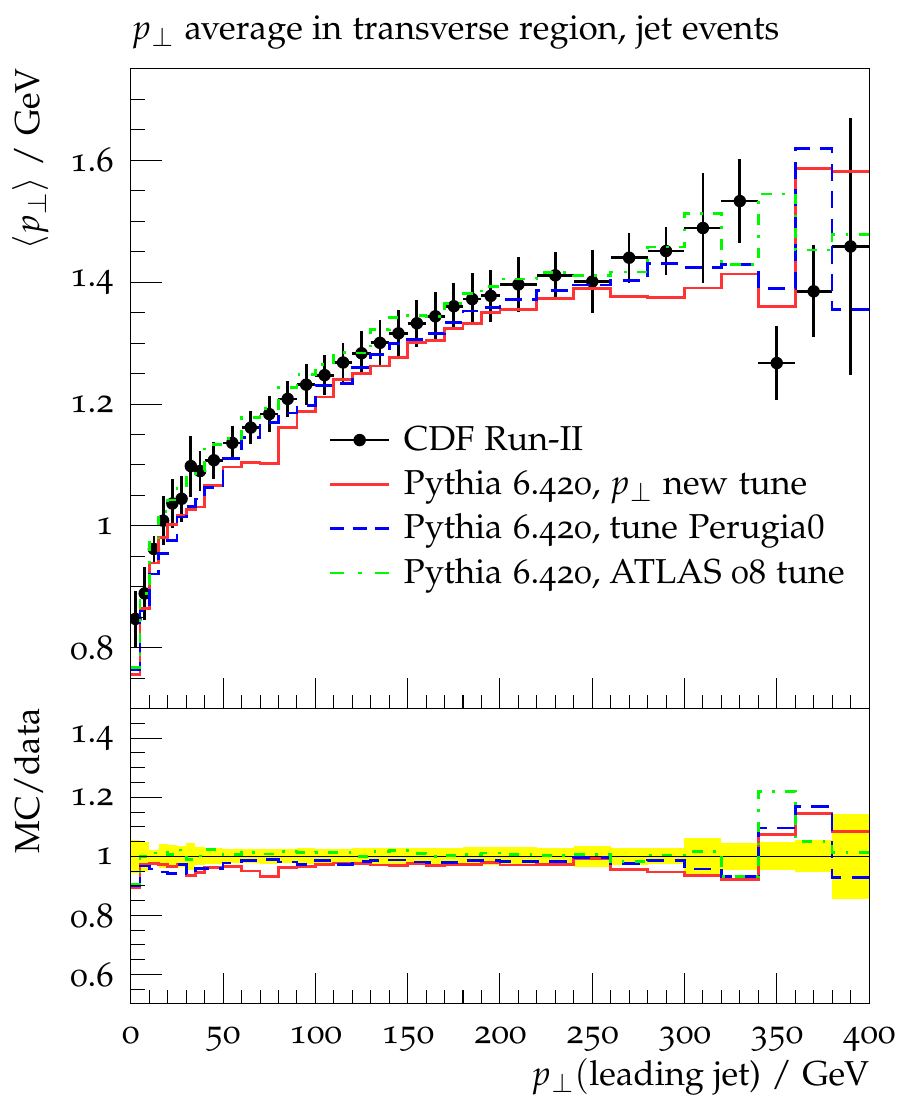}\\
    \vspace*{2em}
    \includegraphics[width=0.45\textwidth]{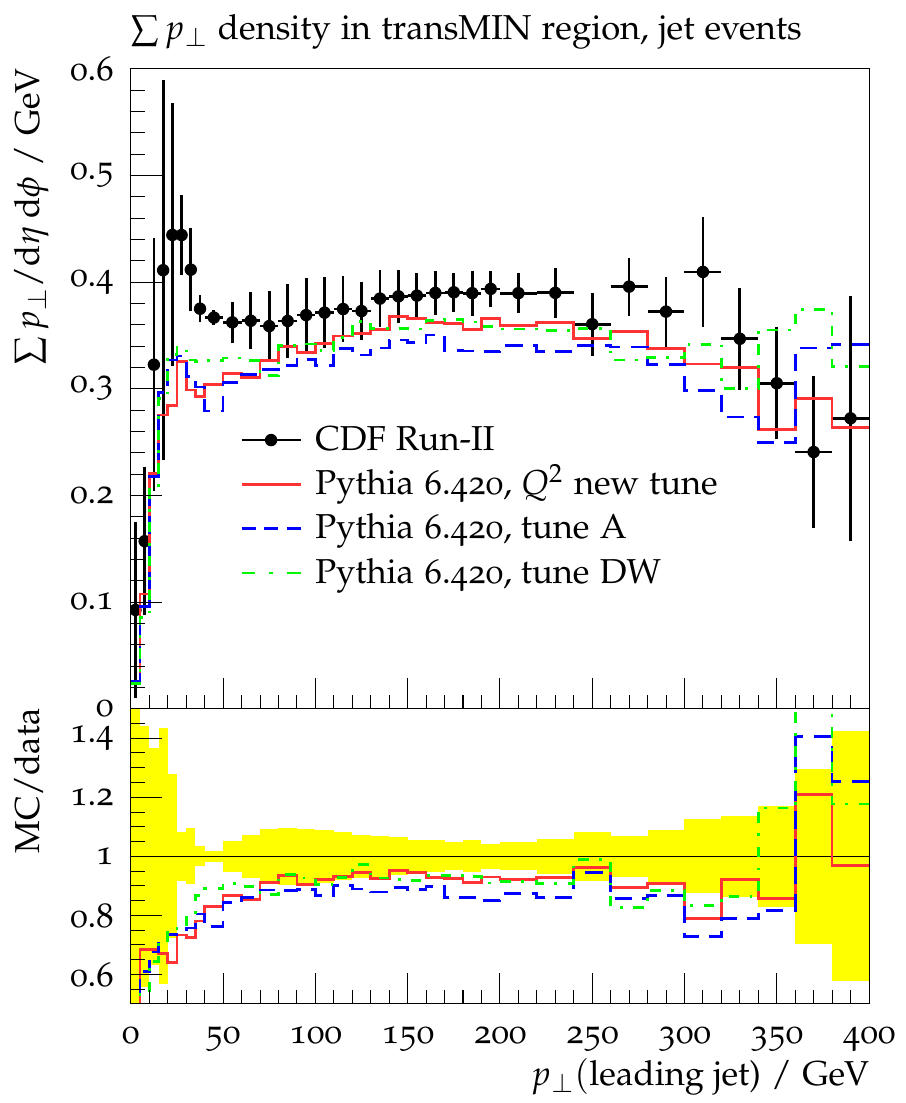}
    \includegraphics[width=0.45\textwidth]{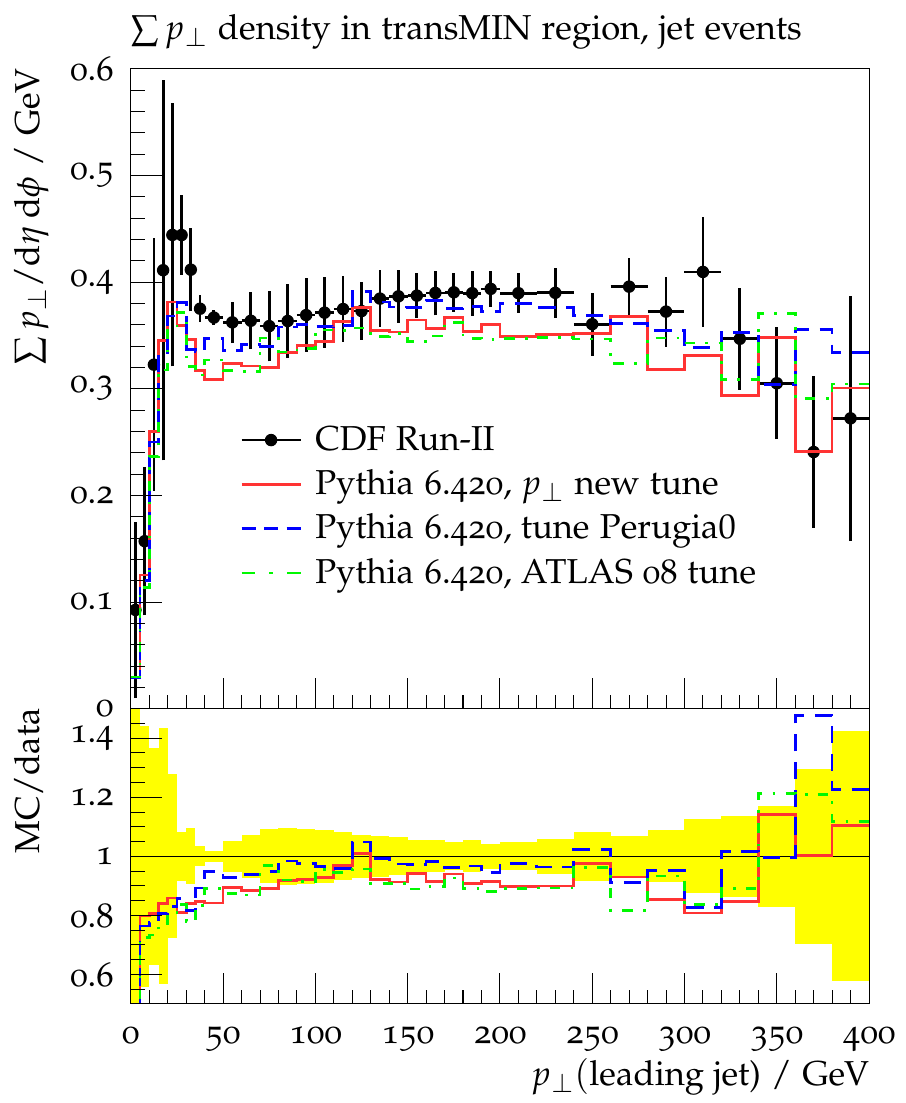}
    \captionof{figure}{These plots show the average track \pT{} in the
      transverse region (top) and the $\sum \pT$ density in the transMIN
      region (bottom) in leading jet events~\cite{cdf-leadingjet}. The new
      model (on the right) seems to have a slight advantage over the
      virtuality-ordered shower with the old MPI model shown on the left, both
      in the turn-on hump and in overall activity.}
    \label{fig:tune-ue-3}
  \end{center}

  \clearpage
  \enlargethispage{2\baselineskip}
  \vspace*{-1em}
  \begin{center}
    \vspace*{2em}
    \includegraphics[width=0.45\textwidth]{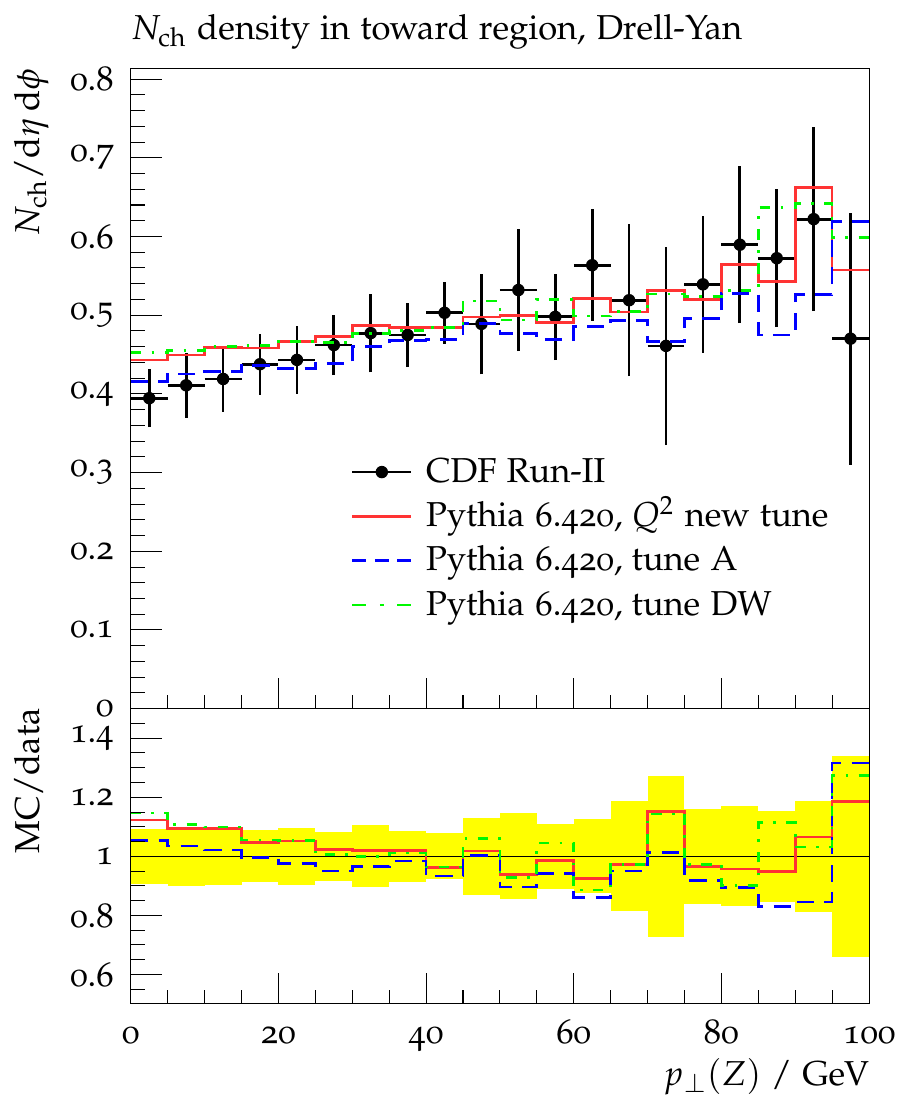}
    \includegraphics[width=0.45\textwidth]{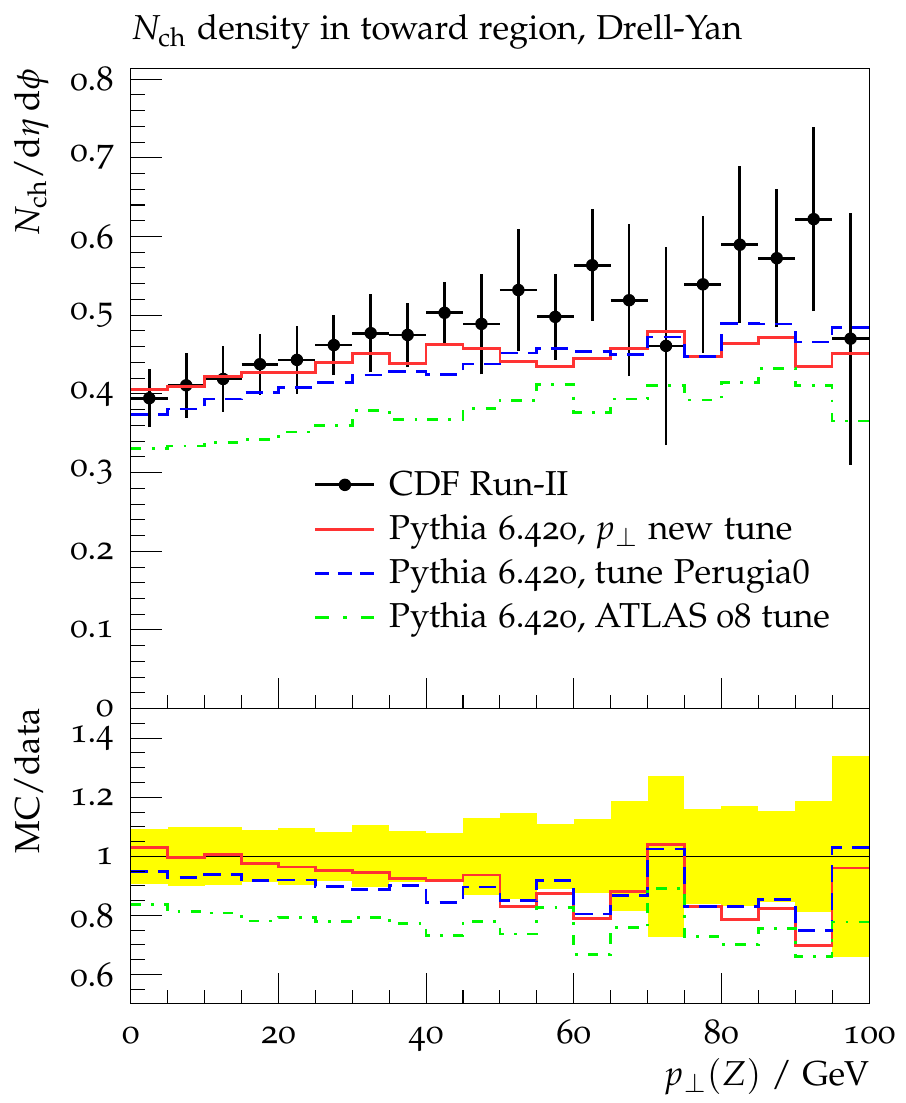}\\
    \vspace*{2em}
    \includegraphics[width=0.45\textwidth]{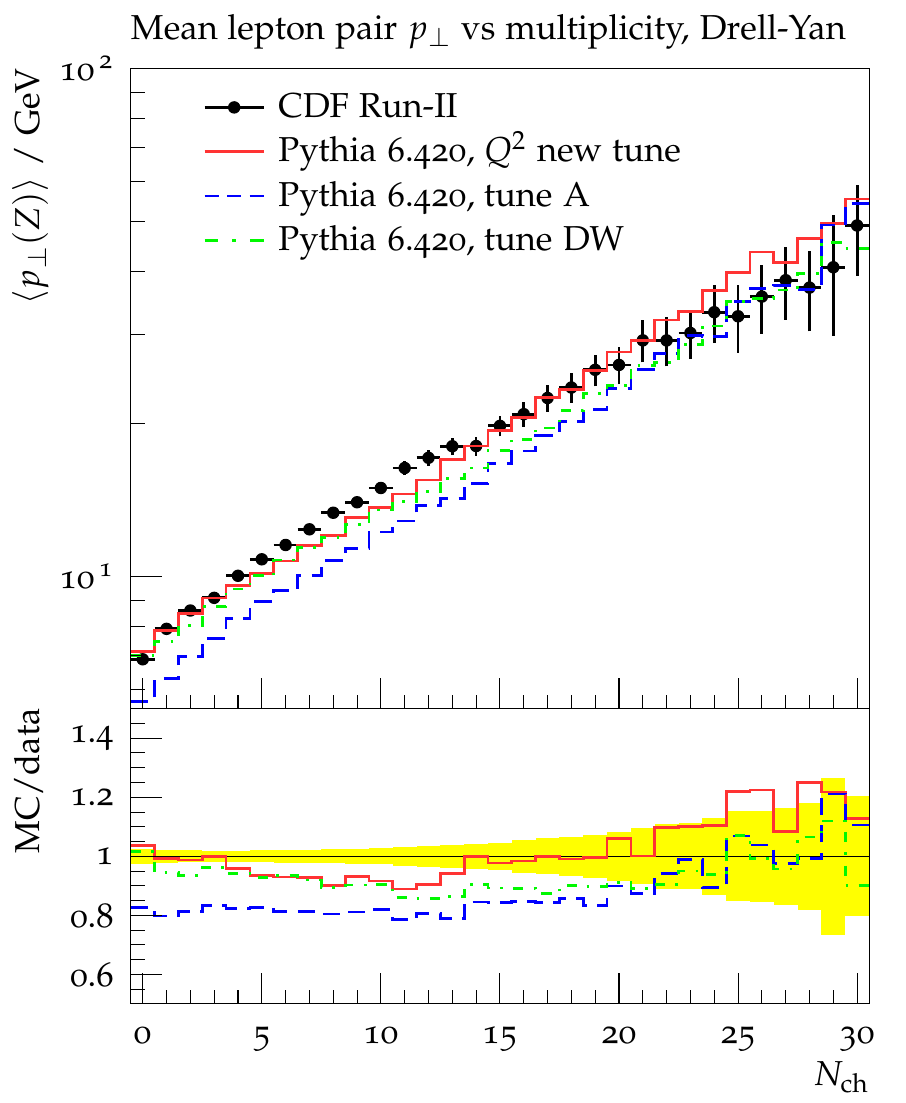}
    \includegraphics[width=0.45\textwidth]{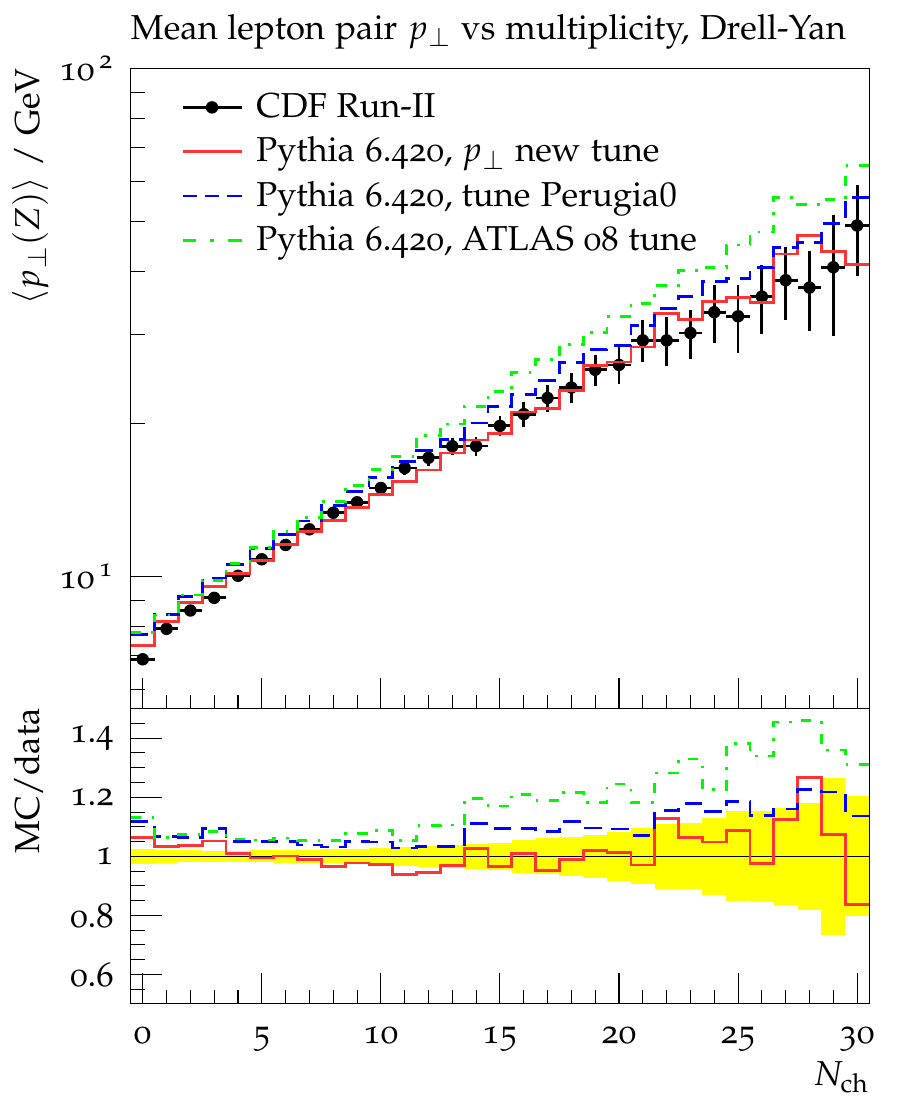}
    \captionof{figure}{In Drell-Yan~\cite{cdf-note9351} the new MPI model
      consistently produces less underlying event activity than for the old model (top plots.) 
      This underestimation is particularly pronounced for the \ATLAS tune.
      Nevertheless, most of the recent tunes are able to describe the
      multiplicity dependence of the \PZ \pT (bottom plots.)}
    \label{fig:tune-ue-4}
  \end{center}

  \clearpage
  \enlargethispage{2\baselineskip}
  \vspace*{-1em}
  \begin{center}
    \vspace*{2em}
    \includegraphics[width=0.45\textwidth]{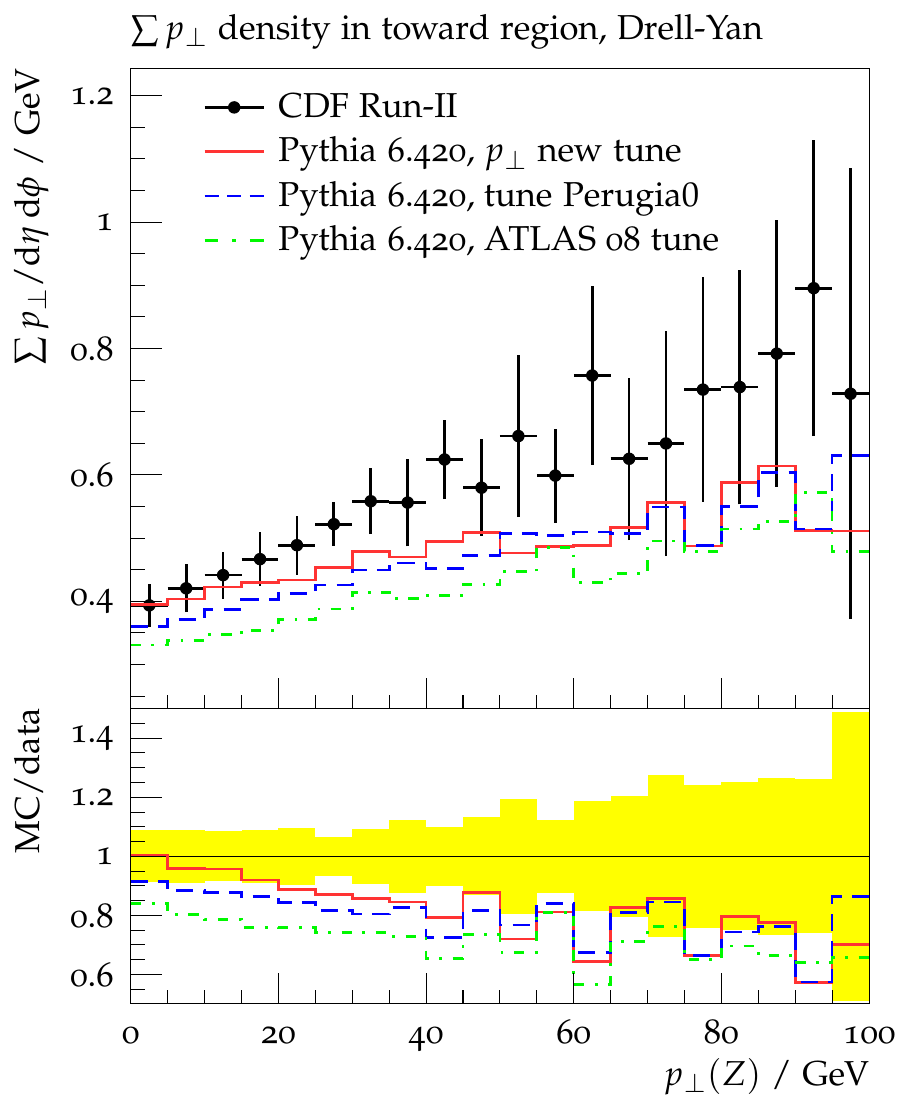}
    \includegraphics[width=0.45\textwidth]{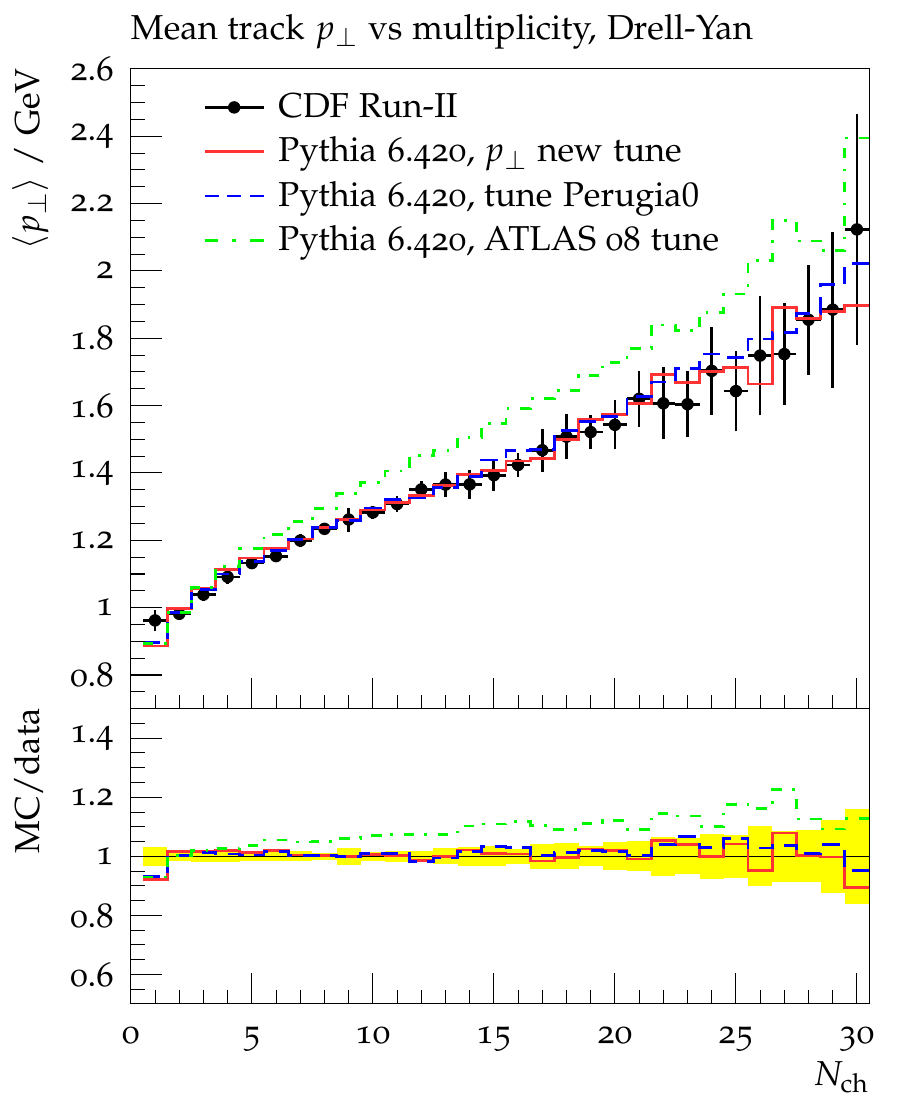}\\
    \vspace*{2em}
    \includegraphics[width=0.45\textwidth]{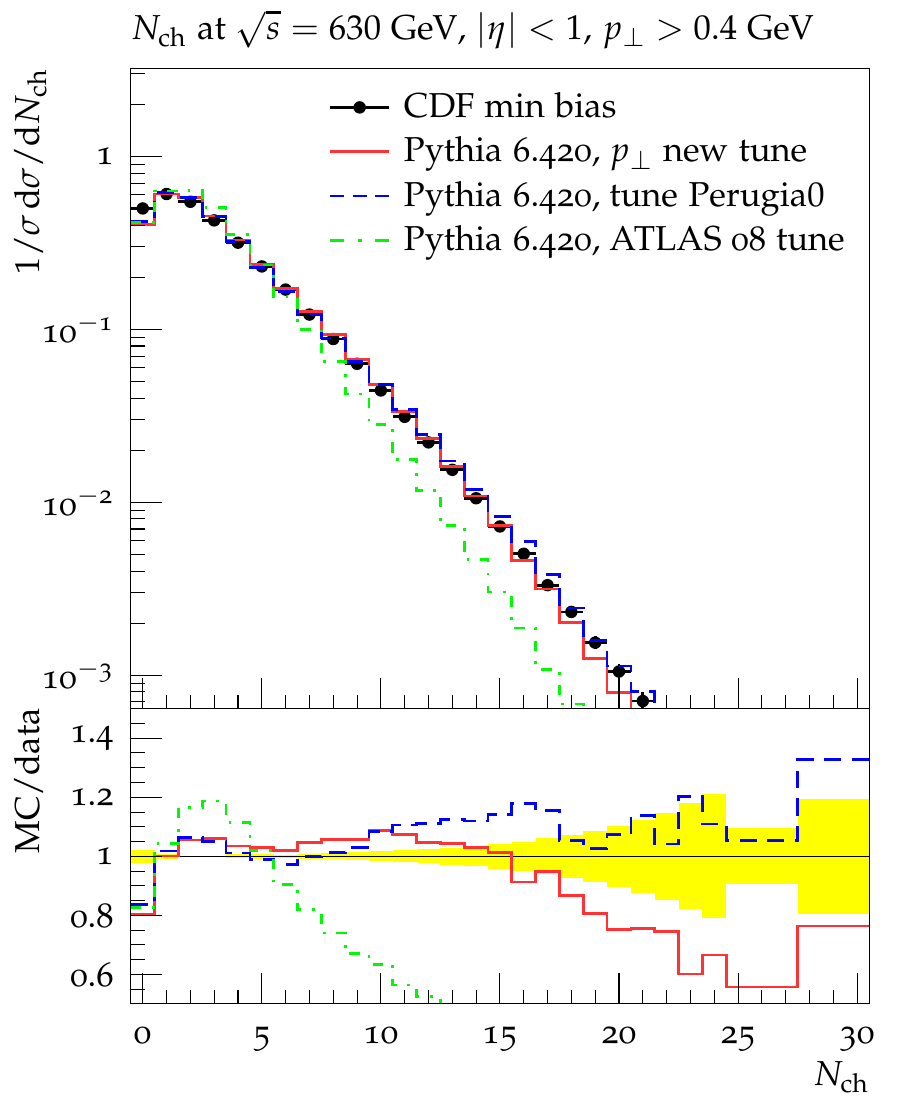}
    \includegraphics[width=0.45\textwidth]{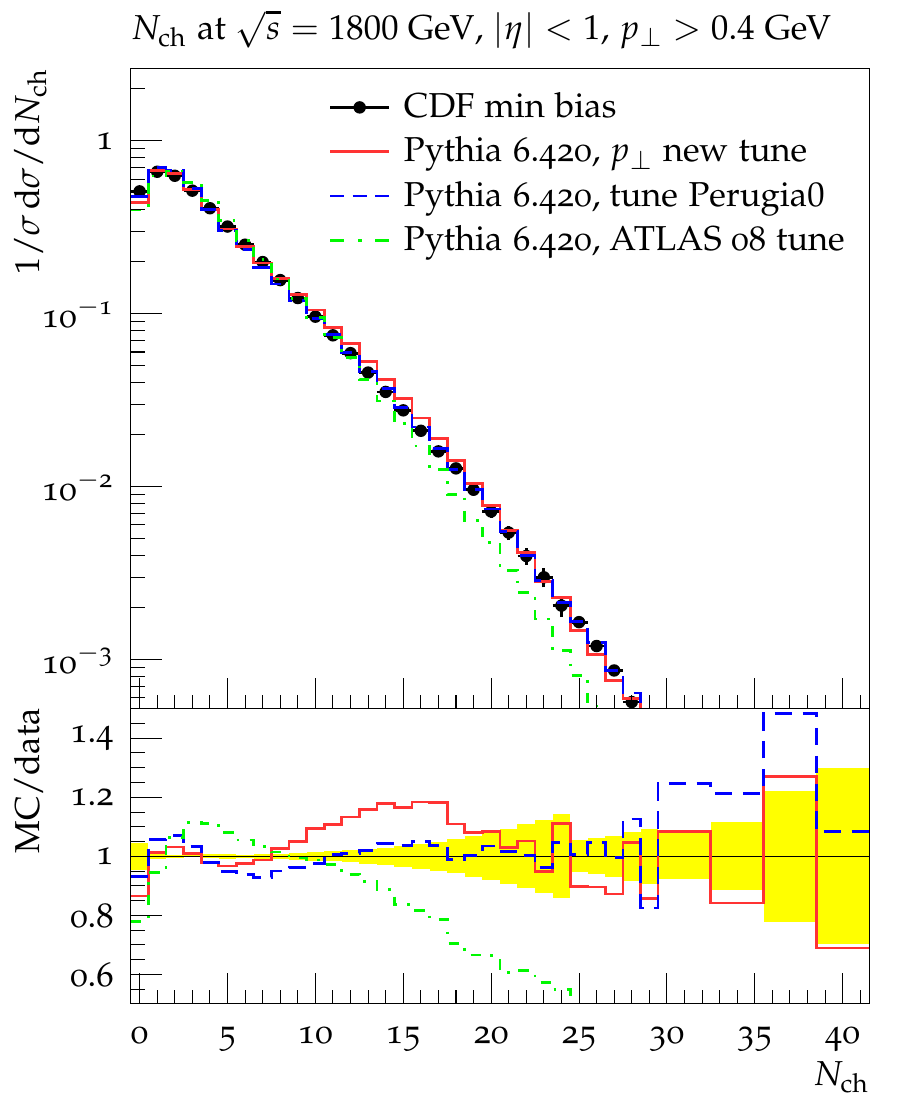}
    \captionof{figure}{Some more plots showing the behaviour of the
      interleaved MPI model and the \pT{}-ordered shower. The two upper plots
      focus on the underlying event in Drell-Yan~\cite{cdf-note9351}. On the
      left we see again that the new model underestimates the activity in
      Drell-Yan events (like in \FigRef{fig:tune-ue-4}.)  Regardless of that,
      the top right plot shows that the average track \pT{} as function of the
      charged multiplicity is described well, except by the \ATLAS tune. The
      \ATLAS tune also shows strong disagreement with the multiplicity distribution in
      minimum bias events, even at the reference energy of \unit{1800}{\GeV}, 
      as shown in the lower two plots~\cite{Acosta:2001rm}.}
    \label{fig:tune-ue-5}
  \end{center}

\end{appendix}

\end{document}